\documentclass[fleqn,usenatbib]{mnras}

\usepackage{newtxtext,newtxmath}

\usepackage[T1]{fontenc}

\usepackage{graphicx}   
\usepackage{amsmath}    
\usepackage{epsfig}
\usepackage{xcolor}
\usepackage{tabularx}
\usepackage{threeparttable}
\usepackage{hyperref}
\hypersetup{colorlinks=true, citecolor=blue, linkcolor=blue}
\usepackage{comment}
\usepackage{xspace}
\let\oldAA\AA
\renewcommand{\AA}{\text{\oldAA}\xspace}

\newcommand{\redtxt}[1]{\textcolor{black}{#1}}

\newcommand{\target}{GS\_3073}

\newcommand{\hii}{H\,{\sc ii}}
\newcommand{\hei}{He\,{\sc i}}
\newcommand{\heii}{He\,{\sc ii}}
\newcommand{\neiii}{[Ne\,{\sc iii}]}
\newcommand{\nev}{[Ne\,{\sc v}]}

\newcommand{\oiii}{[O\,{\sc iii}]}
\newcommand{\oiiis}{O\,{\sc iii}]}
\newcommand{\oii}{[O\,{\sc ii}]}
\newcommand{\sii}{[S\,{\sc ii}]}
\newcommand{\cii}{C\,{\sc ii}}
\newcommand{\ciii}{C\,{\sc iii}]}
\newcommand{\civ}{C\,{\sc iv}}
\newcommand{\feii}{Fe\,{\sc ii}}
\newcommand{\feiv}{[Fe\,{\sc iv}]}
\newcommand{\fevi}{[Fe\,{\sc vi}]}
\newcommand{\fevii}{[Fe\,{\sc vii}]}
\newcommand{\fexiv}{[Fe\,{\sc xiv}]}
\newcommand{\nv}{N\,{\sc v}}
\newcommand{\niv}{N\,{\sc iv}]}
\newcommand{\niii}{N\,{\sc iii}]}
\newcommand{\nii}{[N\,{\sc ii}]}
\newcommand{\cav}{[Ca\,{\sc v}]}
\newcommand{\ariv}{[Ar\,{\sc iv}]}
\newcommand{\lya}{Ly$\alpha$}

\title[An extremely nitrogen-loud galaxy at $z\sim 5.55$]{GA-NIFS: An extremely nitrogen-loud and chemically stratified galaxy at $z\sim 5.55$}

\author[Ji et al.]{Xihan Ji,$^{1,2}$\thanks{E-mail: \href{mailto:xj274@cam.ac.uk}{xj274@cam.ac.uk}}
Hannah \"{U}bler,$^{1,2}$
Roberto Maiolino,$^{1,2,3}$
Francesco D'Eugenio,$^{1,2}$
Santiago Arribas,$^{4}$
\newauthor
Andrew J. Bunker,$^{5}$
Stéphane Charlot,$^{6}$
Michele Perna,$^{4}$
Bruno Rodr\'iguez Del Pino,$^{4}$
\newauthor
Torsten B\"{o}ker,$^{7}$
Giovanni Cresci,$^{8}$
Mirko Curti,$^{9}$
Nimisha Kumari,$^{10}$
Isabella Lamperti$^{4}$
\\
$^{1}$Kavli Institute for Cosmology, University of Cambridge, Madingley Road, Cambridge, CB3 0HA, UK\\
$^{2}$Cavendish Laboratory, University of Cambridge, 19 JJ Thomson Avenue, Cambridge, CB3 0HE, UK\\
$^{3}$Department of Physics and Astronomy, University College London, Gower Street, London WC1E 6BT, UK\\
$^{4}$Centro de Astrobiolog\'{\i}a (CAB), CSIC-INTA, Ctra. de Ajalvir km 4, Torrej\'on de Ardoz, E-28850, Madrid, Spain\\
$^{5}$University of Oxford, Department of Physics, Denys Wilkinson Building, Keble Road, Oxford OX13RH, UK\\
$^{6}$Sorbonne Universit\'e, CNRS, UMR 7095, Institut d'Astrophysique de Paris, 98 bis bd Arago, 75014 Paris, France\\
$^{7}$European Space Agency, c/o STScI, 3700 San Martin Drive, Baltimore, MD 21218, USA\\
$^{8}$INAF - Osservatorio Astrofisico di Arcetri, largo E. Fermi 5, 50127 Firenze, Italy\\
$^{9}$European Southern Observatory, Karl-Schwarzschild-Strasse 2, 85748 Garching, Germany\\
$^{10}$AURA for European Space Agency, Space Telescope Science Institute, 3700 San Matin Drive, Baltimore, MD 21218, USA
}

\begin{document} 
\label{firstpage}
\pagerange{\pageref{firstpage}--\pageref{lastpage}}
\maketitle

\begin{abstract}
    {We report the chemical abundance pattern of GS\_3073, a galaxy hosting an overmassive active black hole at $z=5.55$, by leveraging observations from JWST/NIRSpec and VLT/VIMOS. Based on the rest-frame UV emission lines, which trace high-density ($\sim 10^5~{\rm cm^{-3}}$) and highly ionized gas, we derive $\rm \log(N/O) = 0.42^{+0.13}_{-0.10}$. At an estimated metallicity of $0.2~Z_{\odot}$, this is the most extreme nitrogen-rich object found by JWST thus far. In comparison, the relative carbon abundance derived is $\rm \log(C/O) = -0.38^{+0.13}_{-0.11}$, which is not significantly higher than those in local galaxies and stars with similar metallicities. We also found potential detection of \fevii$\lambda 6087$ and \fexiv$\lambda 5303$, both blended with \cav. We inferred a range of Fe abundances compatible with those in local stars and galaxies. Overall, the chemical abundance pattern of GS\_3073 is compatible with enrichment by supermassive stars with $M_* \gtrsim 1000~M_\odot$, asymptotic giant branch (AGB) stars, or Wolf-Rayet (WR) stars. Interestingly, when using optical emission lines which trace lower density ($\sim 10^3~{\rm cm}^{-3}$) and lower ionization gas, we found a sub-solar N/O ratio, consistent with local galaxies at the same metallicity. We interpret the difference in N/O derived from UV lines and optical lines as evidence for a stratified system, where the inner and denser region is both more chemically enriched and more ionized. Our results suggest that nitrogen loudness in high-$z$ galaxies might be confined to the central, dense, and highly ionized regions of the galaxies, while the bulk of the galaxies evolves more normally.}
\end{abstract}

\begin{keywords}
galaxies: abundances -- galaxies: active -- galaxies: evolution -- galaxies: high-redshift
\end{keywords}
%

\section{Introduction}

The chemical abundances of galaxies are imprints of their past evolution \citep{dave2011,lilly2013}.
Observational studies of the chemical abundances of galaxies in the local Universe have found tight correlations between the relative abundances of different elements.
The scaling relations between elemental abundances strongly indicate a common chemical evolution pathway in typical galaxies \citep[cf.,][and references therein]{maiolino2019}.
Among the established scaling relations between different elemental abundances, one of the most widely studied is the relation between the oxygen-to-hydrogen abundance, O/H, and the nitrogen-to-oxygen abundance, N/O.
Previous observations have shown a correlation between O/H and N/O in local \hii\ regions and in the interstellar medium (ISM) of star-forming (SF) galaxies \citep[e.g.,][]{vilacostas1993,vanzee1998,pilyugin2012,andrews2013,berg2016,berg2019,berg2020,nicholls2017,belfiore2017,Hayden-Pawson22}.
At low oxygen abundances, or ``metallicities'', the abundance ratio of N/O asymptotically approaches a constant value.
In contrast, in the intermediate to high metallicity regime, N/O roughly scales quadratically with O/H \citep{nicholls2017}.
In theory, this behavior is a reflection of the different enrichment mechanisms of oxygen and nitrogen.
The oxygen in the ISM is promptly enriched by
core-collapse supernovae (CCSNe, from massive stars with stellar masses of $M_* \gtrsim 8~M_\odot$ and short lifetimes). In contrast, the enrichment of nitrogen has another path available through stellar winds during the Asymptotic
Giant Branch (AGB) phase of intermediate-mass stars with stellar masses of $2~M_\odot < M_* < 8~M_\odot$.
Due to the CNO cycle at work in the intermediate-mass stars, the nitrogen production is boosted in the metal-rich stars, leading to the scaling relation between O/H and N/O.
Since the enrichment by AGB stars lags behind the enrichment by CCSNe, there is expected to be an intrinsic scatter around the scaling relation \redtxt{\citep[e.g.,][]{andrews2013}}.
In addition, the inflow of low-metallicity gas, as well as variations in the star formation efficiency, can also lead to scatter in the N/O versus O/H relation \citep{vincenzo2016,kumari2018,schaefer2020,luo2021}.

Despite the intrinsic scatter of the N/O versus O/H relation, it is rare to observe galaxies with an N/O deviating from the locally established relation by more than 0.5 dex.
One interesting outlier, for example, is \redtxt{a local} compact dwarf galaxy Mrk 996, which has a rather high nitrogen abundance of $\rm log(N/O)=-0.76$ (or $\rm [N/O]\equiv log(N/O) - log(N/O)_\odot = 0.10$) considering its extremely low metallicity of $\rm 12+log(O/H) = 7.15$ (or $\rm [O/H] = -1.54$; \citealp{thuan1996,berg2016}).
More recently, by reanalyzing the UV emission lines of Mrk 996 with photoionization models, \citet{senchyna2023} and \citet{isobe2023} inferred an even higher abundance ratio of $\rm [N/O] \sim 0.86$.
\redtxt{According to \citet{james2009}, however, the emission-line spectrum of Mrk 996 has two kinematic components, indicative of two distinct regions with different gas densities and abundance patterns. The low-density region has a density of $n_{\rm e} = 170\pm 40~{\rm cm^{-3}}$ and a lower nitrogen abundance of $\rm \log(N/O) = -1.43\pm 0.14$. In contrast, the high-density region has a density of $\log (n_{\rm e}/{\rm cm^{-3}}) = 7.25^{+1.25}_{-0.75}$ and a higher nitrogen abundance of $\rm \log(N/O) = -0.13\pm 0.28$.}
The ``nitrogen-loudness'' of this target potentially comes from the peculiar enrichment from Wolf-Rayet \citep[WR; e.g.,][]{crowther2007} stars, where the powerful stellar winds enriched with light elements, including nitrogen, boost N/O even though the metallicity is low.
Due to the relatively short time scale of the WR phase, the enhanced N/O would be quickly lowered again by enrichment from CCSNe of massive stars a few Myr after the initial burst of star formation \citep{limongi2018,watanabe2024}.
Thus, it is expected that such a peculiar abundance pattern should be rarely encountered in observations.
\redtxt{As another example of nitrogen enhanced star-forming system, a young super star cluster with Lyman continuum leakage (LyC) in the gravitationally lensed Sunburst Arc at $z = 2.37$ was recently found to show both a pressure (and density) stratification and a nitrogen abundance stratification \citep{pascale2023}. According to \citet{pascale2023}, the high-pressure and high-density region in the LyC cluster has an enhanced nitrogen abundance of $\rm \log(N/O) = -0.21^{+0.10}_{-0.11}$ at a low metallicity of $\rm 12 + log(O/H) = 8.03\pm 0.06$.}

Interestingly, previous observations have revealed a population of active galactic nuclei (AGNs) potentially having enhanced nitrogen abundances, which are called ``nitrogen-loud'' AGNs \citep{baldwin2003}.
Nitrogen-loud AGNs are a rare subclass of AGNs ($\sim 1\%$ in SDSS quasars with $1.6\lesssim z\lesssim 4$; \citealp{bentz2004a,bentz2004,jiang2008}).
The nitrogen loudness refers to the strong rest-frame UV nitrogen lines observed in these AGNs.
These AGNs typically show clear broad components in the strong UV nitrogen lines, and the abundance estimations based on the nitrogen lines have suggested significantly supersolar metallicities \citep[e.g.,][]{hamann1992,dietrich2003a,baldwin2003,nagao2006a}.
There are a number of longstanding questions about nitrogen-loud AGNs, including whether elements other than nitrogen are also enriched, whether the nitrogen-loudness is only limited to the broad-line regions (BLRs), and so forth \citep{collin1999,baldwin2003,jiang2008,wang2011,araki2012,batra2014,matsuoka2011,matsuoka2017,villarmartin2020}.
There are also discussions on the potential systematic uncertainties of abundance estimations based on broad emission lines \citep{batra2014,temple2021}.
For the enrichment mechanisms of these nitrogen-loud AGNs, depending on the actual abundance pattern and the site of the enrichment, it is possible to have fast enrichment within the small nuclear region \redtxt{\citep{collin1999,wang2011,cantiello2021,huang_agndiscsf_2023,ali_agndiscsf_2023,chen_agndiscsf_2024}} or wide-spread enrichment during a specific stage of galaxy evolution \citep{hamann1993,baldwin2003}.
To provide more constraints on the true nature of nitrogen-loud AGNs, it is thus important to probe the chemical abundances of different elements as well as within different ionization zones in AGNs.
It is also vital to find evidence of nitrogen enrichment in the nitrogen-loud AGNs at early times.

Based on the local N/O versus O/H relation, one might expect high redshift galaxies to populate the low-metallicity end of the relation.
Intriguingly, with the advent of the James Webb Space Telescope \citep[JWST;][]{jwst0,jwst1}, an abundance pattern with enhanced N/O has recently been confirmed in a galaxy, GN-z11, at $z=10.603$ \citep{bunker2023,cameron2023}.
\redtxt{It is noteworthy that the flux ratios between the semi-forbidden transitions of \niii\ as well as the flux ratio between [\niv\ and \niv\ in the JWST spectrum of GN-z11 indicate a high density \citep{maiolino2023}, typical of the BLRs of AGNs \citep{netzer1990}.}
Based on flux measurements of UV emission lines in GN-z11, \citet{cameron2023} inferred a fiducial abundance ratio of $\rm [N/O] > 0.61$. This puts GN-z11 in the category of nitrogen-loud AGNs previously found at lower redshifts and raises the question about the enrichment site of nitrogen in this galaxy \citep{cameron2023,maiolino2023}.
Thus far, several nitrogen-loud galaxies at redshifts of $z > 6$ have been confirmed spectroscopically by JWST \redtxt{\citep{bunker2023,cameron2023,senchyna2023,isobe2023,topping2024,castellano_ghz2_2024,schaerer_nloud_2024}}.
These nitrogen-loud galaxies are characterized by strong rest-frame UV emission from high ionization species of nitrogen, including \niii$\lambda 1746$-$1754$ and \niv$\lambda \lambda 1483,1486$, with fluxes considerably higher than that of the UV oxygen emission from \oiiis$\lambda \lambda 1661,1666$.
The discovery of these sources seems to suggest a higher fraction of nitrogen-loud galaxies in the early Universe.
Still, most of the high redshift galaxies currently observed are not bright enough for reliable constraints on the nitrogen abundance \citep{isobe2023}.
To explain the peculiar abundance pattern of these systems, especially given their age, various enrichment mechanisms have been proposed including WR stellar winds with direct-collapse WR stars, winds from very-massive stars (VMSs, with $M_* > 100~M_\odot$) and super-massive stars (SMSs, with $M_* > 1000~M_\odot$), and tidal disruption events (TDEs) \citep{cameron2023,charbonnel2023,nagele2023,senchyna2023,vink2023,kobayashi2024,nandal2024,watanabe2024}. 
It remains unclear whether these enrichment mechanisms have any connections with AGN activity.

With JWST observations, we can now explore more details on the abundance pattern of nitrogen-loud galaxies/AGNs in the early Universe.
One particular target of interest for exploring the nature of nitrogen loudness in high-$z$ galaxies is GS\_3073 (RA: $\rm 3^h32^m18.93^s$, DEC: $\rm -27^{\circ}53^{\prime}2.96^{\prime \prime}$) found in the Cosmic Assembly Near-infrared Deep Extragalactic Legacy Survey/Great Observatories Origins Deep Survey field centered on the Chandra Deep Field South \citep[CANDELS/GOODS-S,][]{candels}.
The target was previously observed by ground-based spectroscopy and was confirmed to have a redshift of $z\sim 5.55$ \citep{vanzella2010,mclure2018,grazian2020}.
Based on emission line ratio diagnostics, \citet{grazian2020} and \citet{barchiesi2023} speculated that there is AGN activity in GS\_3073.
Recent observations taken with JWST/Near InfraRed Spectrograph (NIRSpec) integral field unit (IFU) confirmed it is a Type-1 AGN (specifically type 1.8) showing prominent broad emission lines in the rest-frame optical \citep{ubler2023a}.
The super-massive black hole (SMBH) in GS\_3073 appears ``overmassive'', with its mass lying above the local black hole mass versus stellar mass relation
\citep{ubler2023a}.
Interestingly, the JWST/NIRSpec PRISM observation of GS\_3073 reveals strong nitrogen lines in the rest-frame UV spectrum, which potentially put GS\_3073 among the nitrogen-loud Type-1 AGNs observed at the highest redshifts \citep{ubler2023a}. At the same time, with the NIRSpec coverage of the rest-frame optical regime of GS\_3073, forbidden transitions of \nii\ are also detected. Thus, it is possible to explore the nitrogen abundance in different density regimes.
In this work, we investigate the chemical abundance pattern of GS\_3073, combining ground-based spectroscopic data from the Very Large Telescope/VIsible Multi-Object Spectrograph (VLT/VIMOS) and space-based spectroscopic data primarily from the JWST/NIRSpec PRISM.

The layout of the manuscript is as follows.
In Section~\ref{sec:data}, we describe the observational data and data reduction.
In Section~\ref{sec:spec_fit}, we describe the method we used to fit the observed spectra of GS\_3073.
In Section~\ref{sec:diagnostics}, we derive the chemical abundances of GS\_3073 based on our emission-line measurements.
We discuss the enrichment mechanism as well as the enrichment site in this galaxy in Section~\ref{sec:discussion} and draw our conclusions in Section~\ref{sec:conclude}.

Throughout the manuscript, we assume a flat $\rm \Lambda CDM$ cosmology with $H_0 = 70~{\rm km/s/Mpc}$, $\Omega _{\Lambda} = 0.7$, and $\Omega _{m} = 0.3$. We assume solar abundances provided in \citet{grevesse2010} (with $\rm 12+log(O/H)_\odot = 8.69$, $\rm 12+log(N/H)_\odot = 7.83$, and $\rm 12+log(C/H)_\odot = 8.43$) unless otherwise specified.

\section{Observational data}
\label{sec:data}

In this work, we used both space- and ground-based spectra for GS\_3073.
The space-based spectra were obtained with the IFU of JWST/NIRSpec
\citep{jakobsen2022,boker2022,boker2023} and the ground-based spectrum was obtained with VIMOS \citep{lefevre2003}, formerly at VLT.
Below we summarize these observations.

\subsection{JWST/NIRSpec observation and data processing}

The JWST/NIRSpec observations of GS\_3073 discussed here are from the GTO Program, Galaxy Assembly with NIRSpec IFS (GA-NIFS; Program ID 1216; PI: Nora L\"{u}tzgendorf).
The configuration of the observation is detailed in \citet{ubler2023a}.
In brief, two dispersers were used, resulting in IFU cubes with different spectral resolutions.
The high-resolution IFU cube was obtained using the G395H grating, and has a total integration time of 5 hr. The final G395H spectra have a spectral resolution of $R\sim 1900-3600$, and covers the spectral range $2.85-4.85~{\rm \mu m}$, which corresponds to \redtxt{$4400-7400~{\rm \AA}$} in the rest frame of GS\_3073.
The low-resolution IFU cube was obtained using the PRISM disperser, and has a total integration time of 1.1 hr.
The final PRISM spectra have a spectral resolution of $R\sim 30-300$, covering a spectral range of $0.6-5.3~{\rm \mu m}$ in the observed frame and \redtxt{$900-8100~{\rm \AA}$} in the rest frame of GS\_3073.

In this work, we utilize the NIRSpec PRISM observations of GS\_3073. Raw data files were downloaded from the Barbara A.~Mikulski Archive for Space Telescopes (MAST) and processed with the {\it JWST} Science Calibration pipeline\footnote{\url{https://jwst-pipeline.readthedocs.io/en/stable/jwst/introduction.html}} version 1.11.1 under the Calibration Reference Data System (CRDS) context jwst\_1143.pmap. 
We made some modifications to the default reduction steps to produce the data cube \citep[see][for details]{Perna23}.
The final cube was combined using the ``drizzle'' method. The analysis in this paper is based on the combined cube with a pixel scale of $0.05''$.
To subtract the background (which is assumed to be uniform across the IFU field of view), we constructed an inverse-variance weighted average spectrum from the spaxels outside the central $1''\times 1''$ region, where no apparent emission line feature can be found.
We then subtracted this average background spectrum from the central spaxels.
We tested different background extractions with excluded central areas ranging from $0.6''\times 0.6''$ to $1.4''\times 1.4''$ and the results are essentially the same.


The low spectrum resolution of the PRISM spectrum leads to blending of several important rest-frame UV emission lines, such as \lya\ and \nv, and [N\,{\sc iv}]$\lambda 1483$ and \niv$\lambda 1486$.
To deblend these lines, we used the publicly available VLT/VIMOS spectrum of GS\_3073 that overlaps with the PRISM spectrum in the observational wavelength range of $\rm 0.6~\mu m<\lambda <1.03~\mu m$.
The VLT/VIMOS spectrum can also provide a consistency check for emission-line fluxes measured from the PRISM spectrum.

\subsection{VLT/VIMOS observation}

The VLT/VIMOS spectrum we used was extracted from the observations by the multi-object spectroscopy survey VANDELS, with an integration time of 20 hr \citep[ESO program ID 194-A-2003;][]{mclure2018}.
This source was assigned to a slit
with a width of $1^{\prime \prime}$ \citep{mclure2018}, which gives a nominal spectral resolution of $R\sim 580$.
The slit width is larger than the size of GS\_3073, which is around $0.15^{\prime \prime}-0.30^{\prime \prime}$ from the \textsc{galfit} results of previous observations \citep{vanzella2010,vanderwel2012} as well as from the inspection of the rest-frame UV fluxes from the JWST/NIRSpec PRISM IFU cube \citep{ubler2023a}.
Despite the compact size of GS\_3073, 
the typical full width at half maximum (FWHM) of the point spread function (PSF) of VANDELS observations is $\sim 0.6''-1.0''$ \citep{mclure2018}, meaning the actual spectral resolution of the data could vary in a range of $580 \lesssim R \lesssim 970$.
We used the 1D spectrum reduced by \citet{grazian2020} with the reduction pipeline described in \citet{grazian2018}.

The VIMOS spectrum covers a spectral range of $4800-10300$ \AA\ in the observed frame and $700-1600$ \AA\ in the rest frame of GS\_3073.
Within this spectral regime, detection of emission lines including O\,{\sc vi}$\lambda \lambda 1032,1038$, Ly$\alpha$, \nv$\lambda \lambda 1238,1242$, C\,{\sc ii}$\lambda \lambda 1334,1335$, and \niv$\lambda \lambda 1483,1486$ was previously reported \citep{grazian2020,barchiesi2023}.
We note that previous spectroscopic observations of GS\_3073 with the FOcal Reducer and low dispersion Spectrograph (FORS2) at VLT \citep{vanzella2010} and X-Shooter at VLT (Program IDs: 384.A-0886 and 089.A-0679) cover a similar spectral regime.
While Ly$\alpha$ and \niv$\lambda \lambda 1483,1486$ are also detected in the FORS2 spectrum and X-Shooter spectrum \citep{raiter2010,grazian2020},
the VIMOS spectrum has the highest signal-to-noise ratio (S/N) and is the only one showing the clear detection of all the aforementioned emission lines.

In the next section, we describe the method we adopted to fit the JWST/NIRSpec and VLT/VIMOS spectra and show the fitting results.

\section{Spectral fitting}
\label{sec:spec_fit}

\begin{figure*}
    \centering\includegraphics[width=0.95\textwidth]{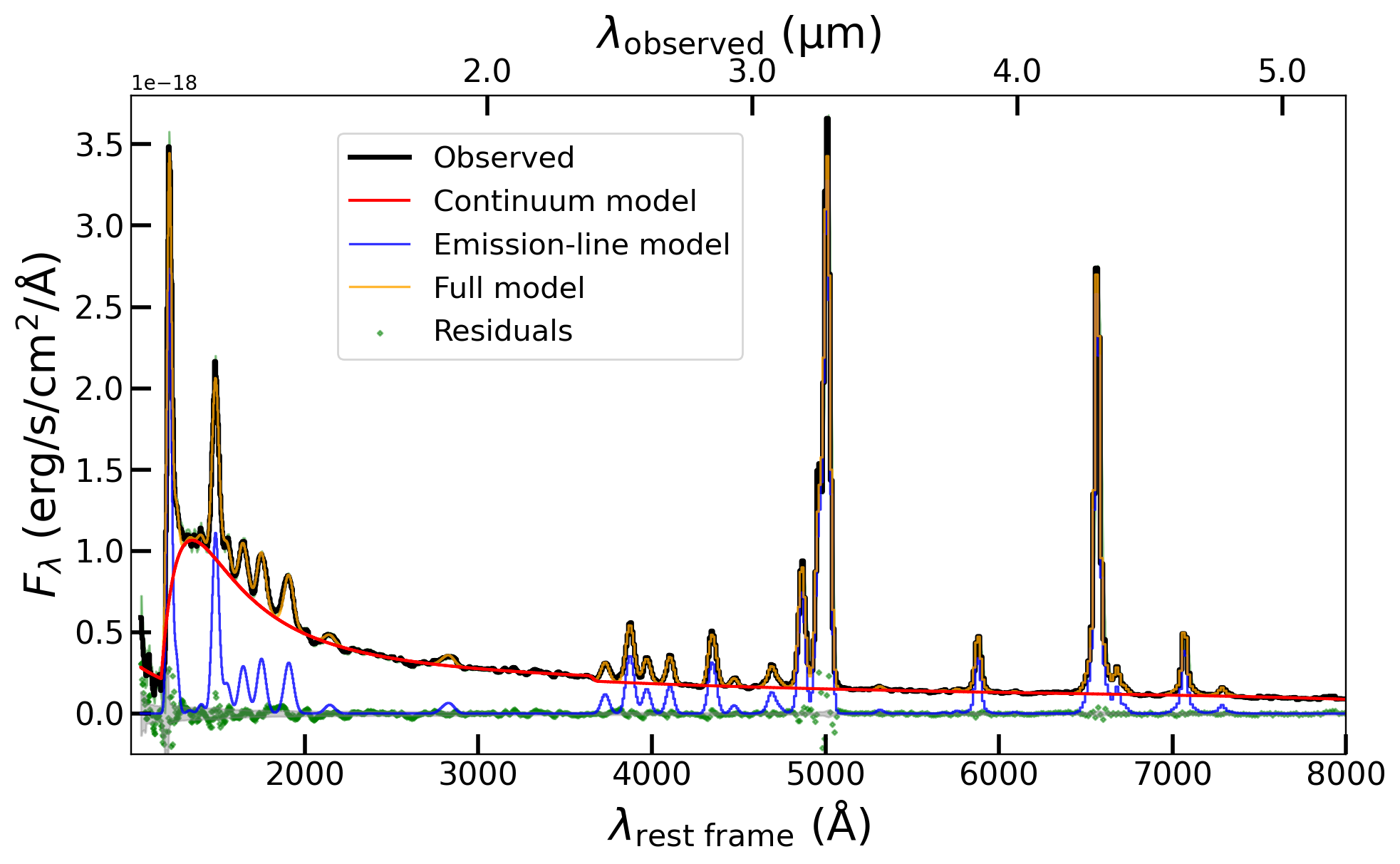}

    \includegraphics[width=0.48\textwidth]{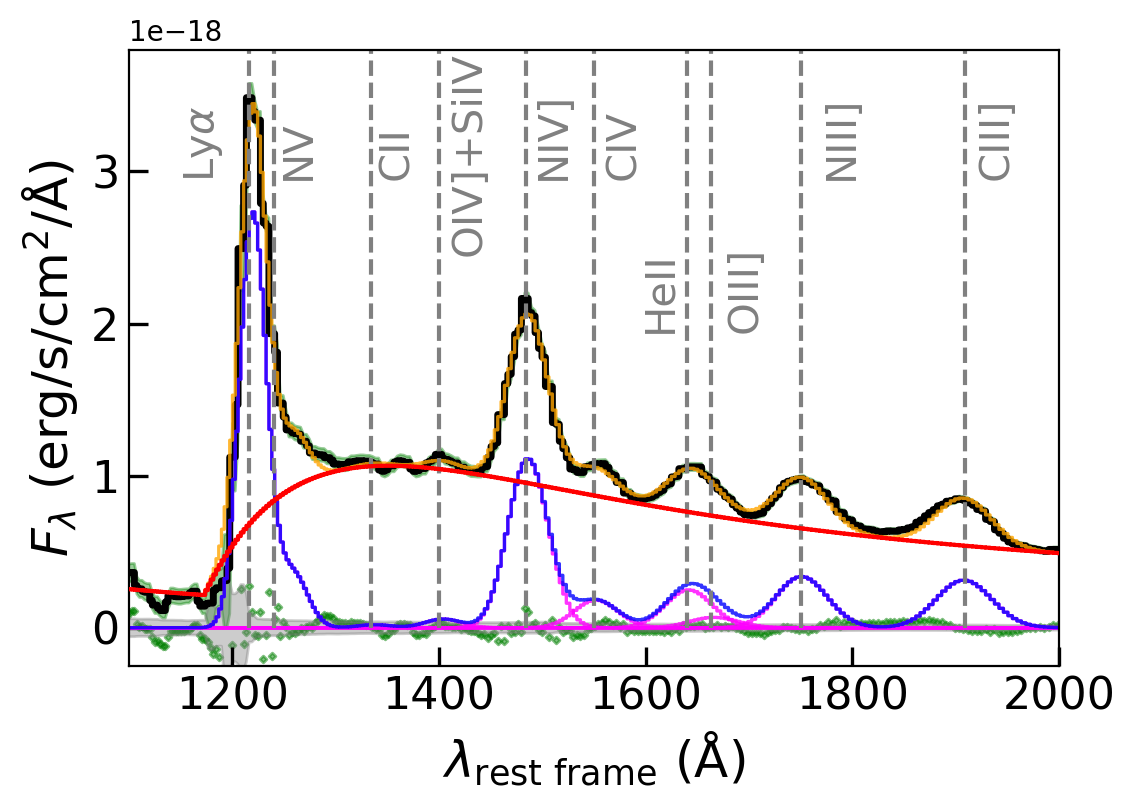}
    \includegraphics[width=0.455\textwidth]{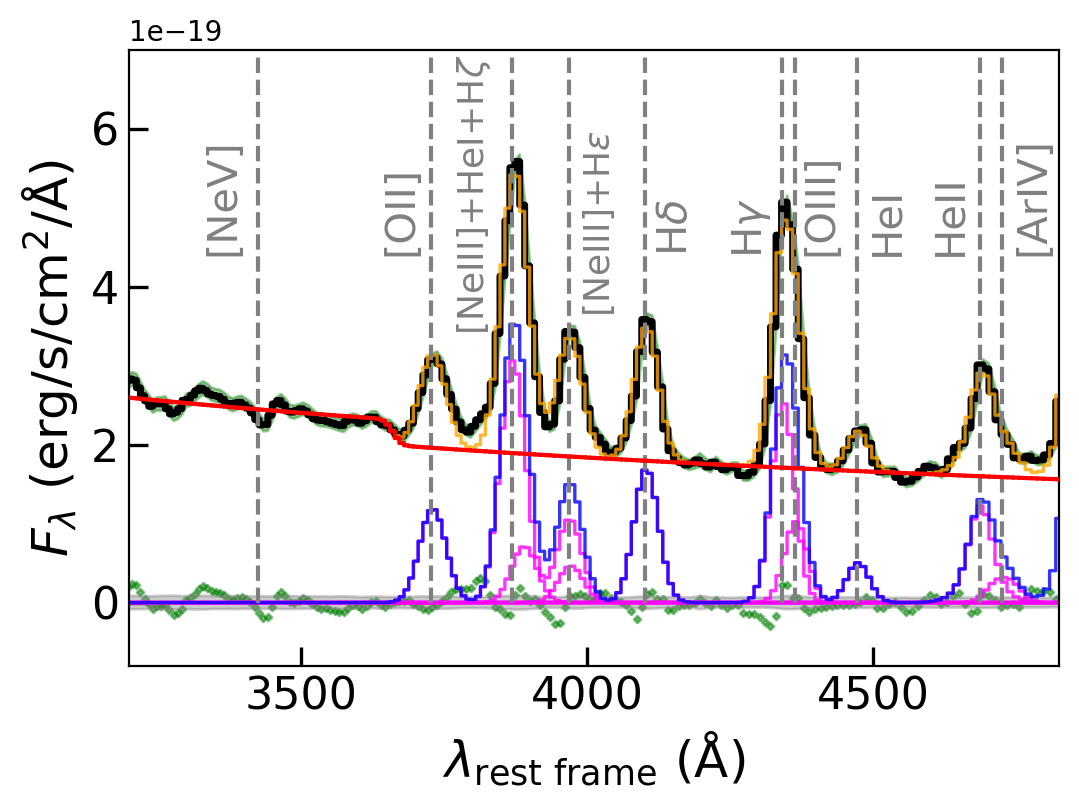}
    
    \caption{Best-fit spectral model for the JWST/NIRSpec PRISM spectrum integrated from the central $0.3^{\prime \prime} \times 0.3^{\prime \prime}$ region of GS\_3073.
    The solid black line is the observed spectrum rebinned by \textsc{pPXF} and the green shaded region indicates the rescaled $1\sigma$ uncertainty of the flux density.
    The solid red line is the continuum model.
    The solid orange line is the emission line model on top of the continuum model.
    The solid blue line is the emission-line only model.
    The green diamonds represent the residual of the fit.
    The grey shaded region indicates the rescaled $1\sigma$ uncertainty of the residual.
    \textit{Top:} Fitting result for the PRISM spectrum across the full spectral range of $1100-8000$ \AA\ in the rest frame.
    \textit{Bottom left:} Zoom-in view of the fit around the rest-frame UV lines.
    \textit{Bottom right:} Zoom-in view of the fit around some relatively weak rest-frame optical lines.
    Fitted models for individual lines are colored in magenta.
    The dashed grey lines mark the locations of emission lines included in the model.
    We caution that the dashed grey lines do not mean the marked lines are detected.}
    \label{fig:prism_fit}
\end{figure*}

\begin{figure}
    \centering\includegraphics[width=0.48\textwidth]{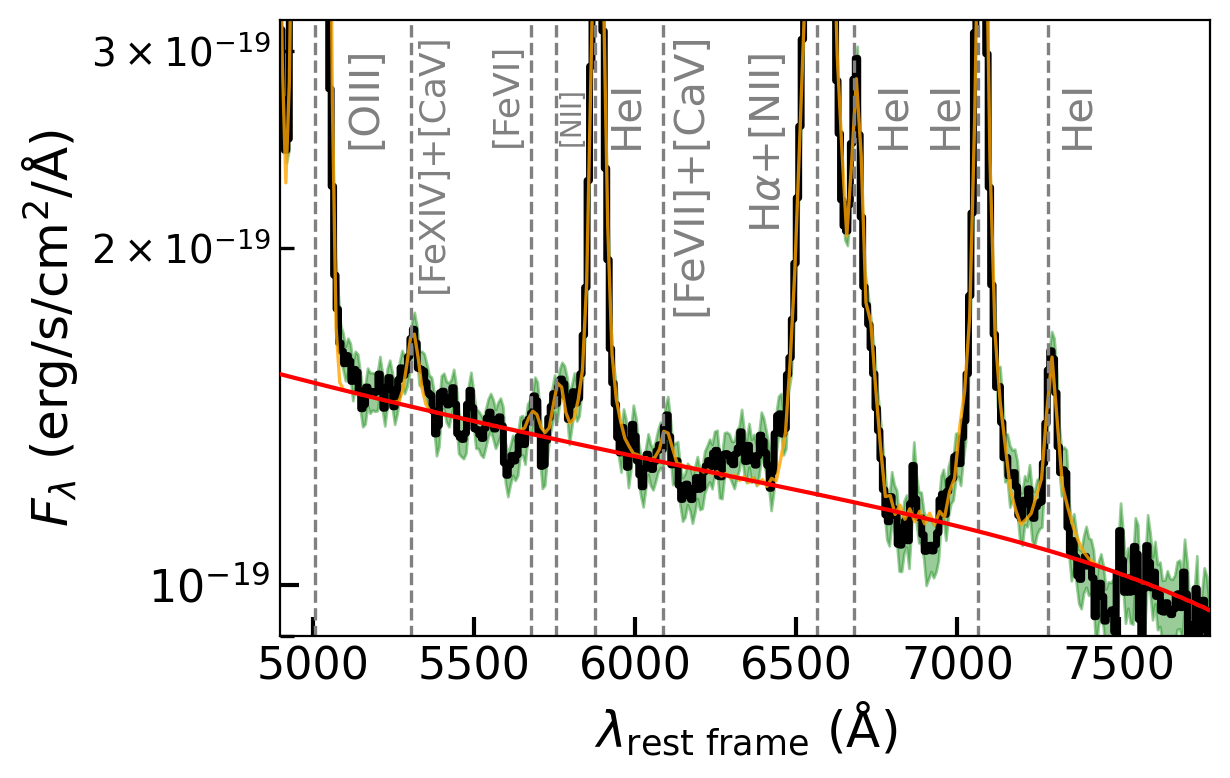}
    
    \caption{Zoom-in view of the best-fit spectral model for the JWST/NIRSpec PRISM spectrum integrated from the central $0.3^{\prime \prime} \times 0.3^{\prime \prime}$ region of GS\_3073.
    The flux density is displayed in log to highlight the weak lines.
    The solid black line is the observed spectrum and the green shaded region indicates the rescaled $1\sigma$ uncertainty of the flux density.
    The solid red line is the continuum model.
    The solid orange line is the emission line model on top of the continuum model.
    The dashed grey lines mark the locations of emission lines included in the model.
    }
    \label{fig:prism_fe}
\end{figure}

\begin{figure*}
    \centering\includegraphics[width=0.95\textwidth]{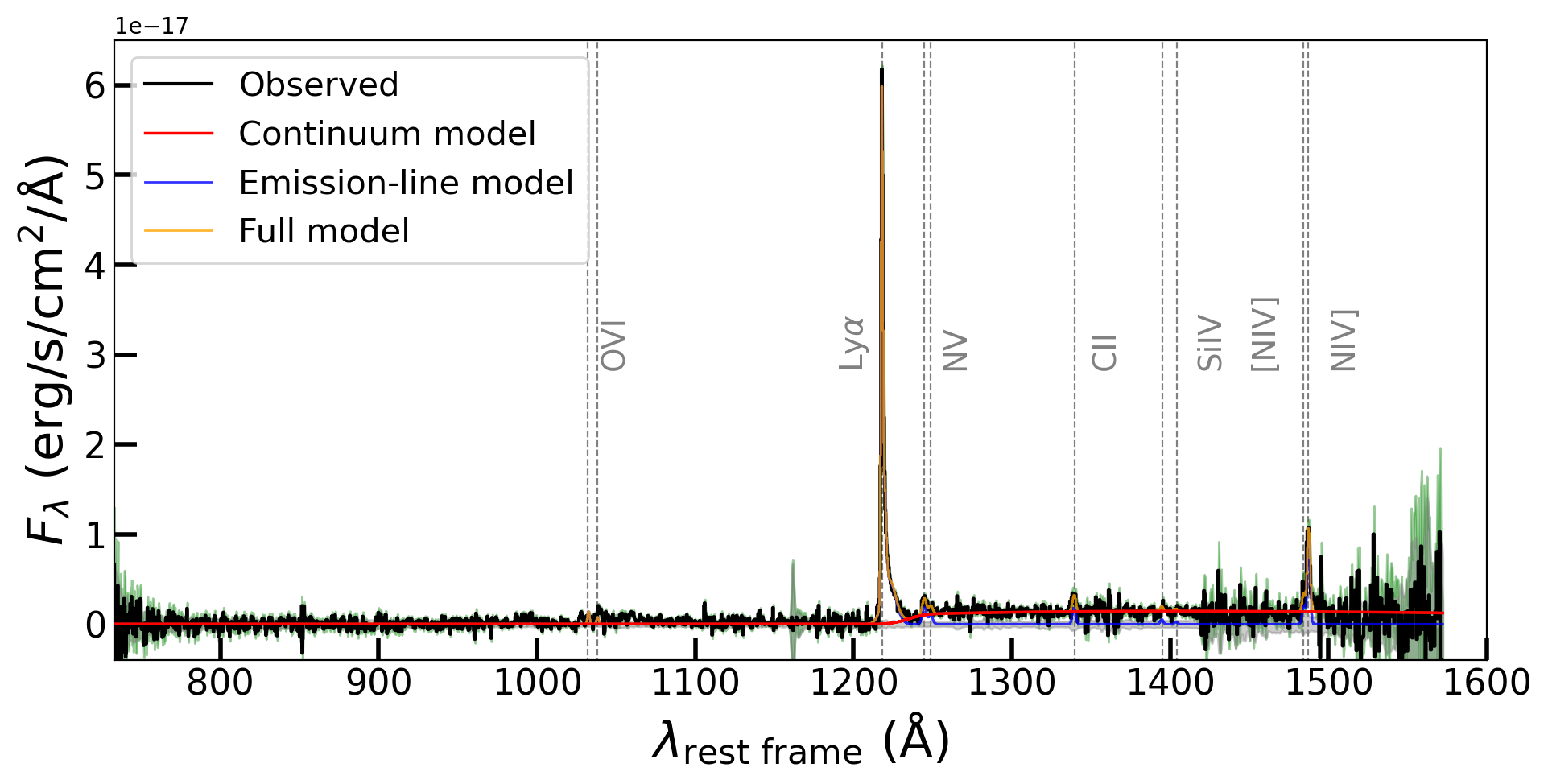}

    \includegraphics[width=0.47\textwidth]{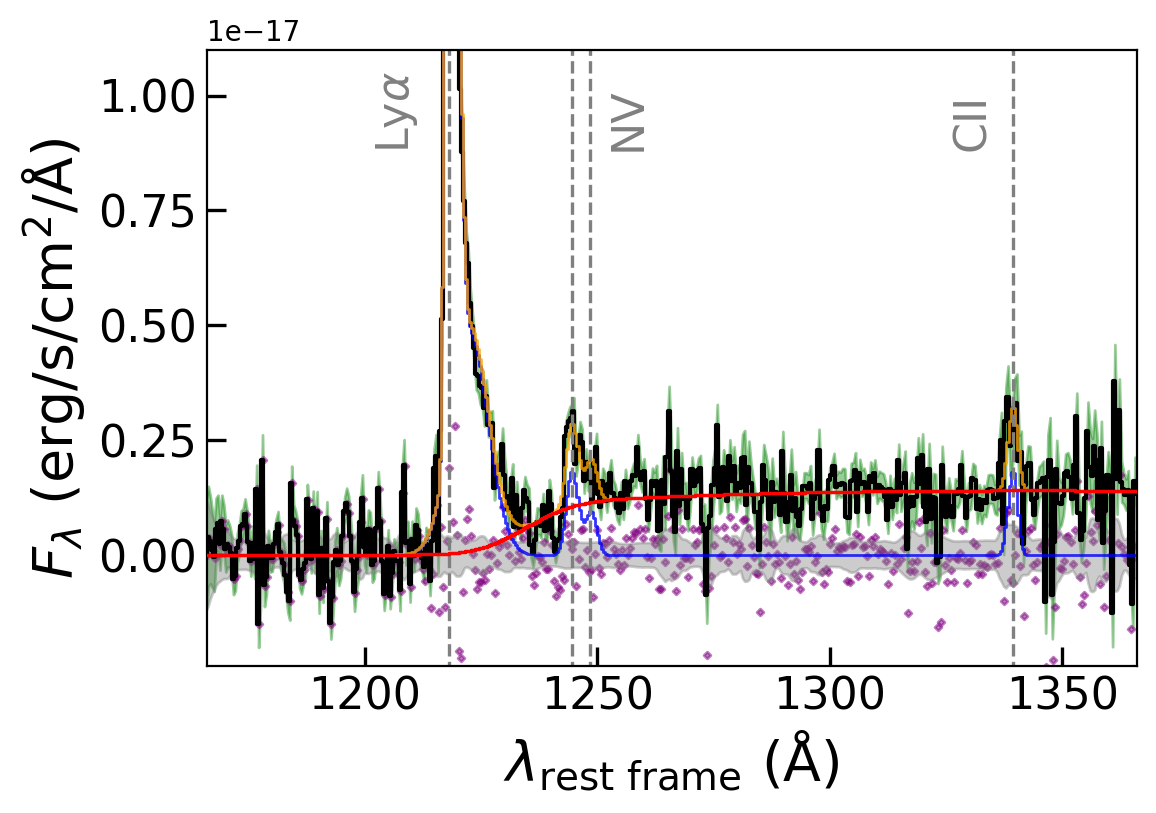}
    \includegraphics[width=0.47\textwidth]{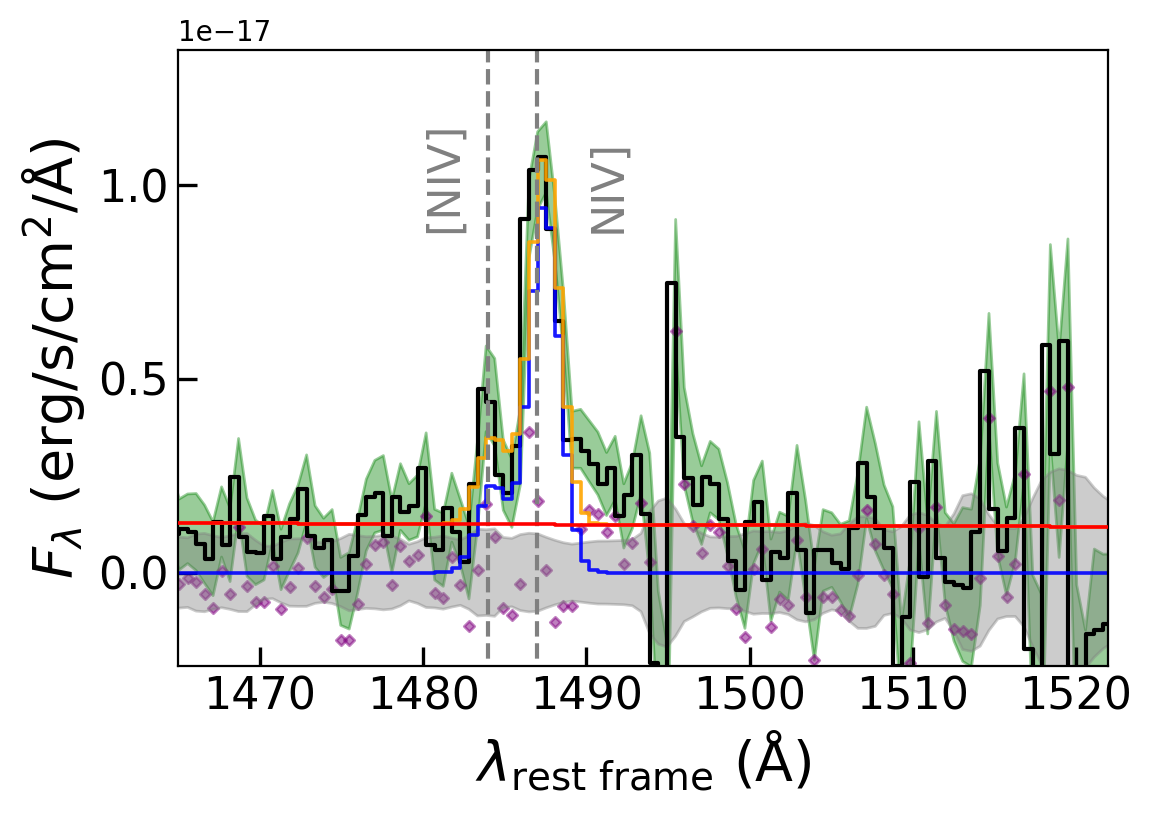}
    
    \caption{Best-fit spectral model for the VLT/VIMOS spectrum measured from a $1^{\prime \prime}$-wide slit. The solid black line is the observed spectrum rebinned by \textsc{pPXF} and the green shaded region indicates the $1\sigma$ uncertainty of the flux density.
    The solid red line is the continuum model.
    The solid orange line is the emission line model on top of the continuum model.
    The solid blue line is the emission-line only model.
    The purple diamonds represent the residual of the fit.
    The grey shaded region indicates the $1\sigma$ uncertainty of the residual.
    The dashed grey lines mark the locations of emission lines included in the model.
    \textit{Top:} Fitting result for the VIMOS spectrum across the full spectral range of $700-1600$ \AA\ in the rest frame.
    \textit{Bottom left:} Zoom-in view of the fit around Ly$\alpha$, \nv, and C\,{\sc ii}.
    \textit{Bottom right:} Zoom-in view of the fit around \niv.
    }
    \label{fig:vimos_fit}
\end{figure*}

\begin{table*}
	\caption{Total fluxes of emission lines (in unit of $10^{-18}~{\rm erg/s/cm^2}$) measured from different spectra for GS\_3073.}
	\label{tab:fluxes}
	\begin{tabular}{lcccc} 
		\hline
		\hline
		Emission line & PRISM ($0.3^{\prime \prime} \times 0.3^{\prime \prime}$) & PRISM ($1^{\prime \prime} \times 1^{\prime \prime}$) & G395H ($0.3^{\prime \prime} \times 0.3^{\prime \prime}$) & VIMOS
		\\
  	  O\,{\sc vi}$\lambda \lambda 1032, 1038$ & - & - & - & $3.6\pm 0.3^{\rm d}$ 
		  \\
        \redtxt{Ly$\alpha ~{\rm (total)^a}$} & $103.3\pm 2.8$ & $166.1\pm 4.0$ & - & $159.6 \pm 1.3$
		  \\
    	  \nv$\lambda \lambda 1238,1242^{\rm a}$ & - & - & - & $9.2\pm 1.8$
		  \\
    	  C\,{\sc ii}$\lambda 1335^{\rm a}$ & $1.0\pm 1.6$ & $11.9\pm 1.9$ & - & $4.8\pm 0.6$
		  \\
    	  O\,{\sc iv}]$\lambda 1400$+Si\,{\sc iv}$\lambda 1402$ & $2.8\pm 0.9$ & $5.4\pm 1.8$ & - & $1.8\pm 0.4$
		  \\
    	  \niv$\lambda \lambda 1483, 1486$ & $54.1\pm 0.9$ & $65.1\pm 1.7$ & - & $4.7\pm 0.8~(\lambda 1483)$
		  \\
             &  & & & $22.0\pm 0.9~(\lambda 1486)$
		  \\
    	  C\,{\sc iv}$\lambda \lambda 1548,1550$ & $9.4\pm 1.1$ & $13.3\pm 1.4$ & - & $<37^{\rm c}$
		  \\
    	  \heii$\lambda 1640^{\rm b}$ & $13.9\pm 0.9$ & $16.4\pm 1.2$ & - & -
		  \\
    	  O\,{\sc iii}]$\lambda \lambda 1661,1666^{\rm b}$ & $3.8\pm 0.9$ & $<4.4^{\rm c}$ & - & -
		  \\
    	  \niii$\lambda$1746-1754 & $20.7\pm 0.8$ & $22.0\pm 1.0$ & - & -
		  \\
    	  C\,{\sc iii}]$\lambda \lambda 1906,1908$ & $21.0\pm 0.4$ & $27.6\pm 0.9$ & - & -
		  \\
    	  N\,{\sc ii}]$\lambda 2142$ & $3.9\pm 0.4$ & $5.9\pm 0.8$ & - & -
		  \\
    	  Mg\,{\sc ii}]$\lambda \lambda 2795,2802$ & $1.9\pm 0.3$ & $2.9\pm 0.9$ & - & -
		  \\
    	  $[$Fe\,{\sc iv}]\redtxt{$\lambda \lambda 2829, 2835$} & $3.6\pm 0.4$ & $3.9\pm 2.7$ & - & -
		  \\
    	  $[$Ne\,{\sc v}]$\lambda 3426$ & $<0.62^{\rm c}$ & $<2.0^{\rm c}$ & - & -
		  \\
    	  $[$O\,{\sc ii}]$\lambda \lambda 3726,3729$ & $7.0\pm 0.2$ & $11.4\pm 0.3$ & - & -
		  \\
    	  $[$Ne\,{\sc iii}]$\lambda 3869$+\hei$\lambda 3889$+H$\zeta$ & $20.4\pm 0.2$ & $28.5\pm 0.5$ & - & -
		  \\
    	  $[$Ne\,{\sc iii}]$\lambda 3968$+H$\epsilon$ & $9.8\pm 0.2$ & $15.8\pm 0.4$ & - & -
		  \\
    	  H$\delta$ & $9.32\pm 0.14$ & $13.9\pm 0.5$ & - & -
		  \\
    	  H$\gamma ^{\rm a}$ & $13.1\pm 0.2$ & $19.1\pm 0.4$ & - & -
            \\
    	  $[$O\,{\sc iii}]$\lambda 4363^{\rm a}$ & $5.3\pm 0.2$ & $7.9\pm 0.4$ & $4.4\pm0.6$ & -
		  \\
    	  \hei$\lambda 4472$ & $2.67\pm 0.13$ & $4.0\pm 0.3$ & $1.4\pm0.4$ & -
		  \\
    	  \heii$\lambda 4686 ^{\rm a}$ & $7.6\pm 0.3$ & $12.7\pm 0.5$ & $4.92\pm 0.58$ & -
		  \\
    	  $[$Ar\,{\sc iv}]$\lambda \lambda 4711,4740$+\hei$\lambda 4713^{\rm a}$ & $2.0\pm 0.2$ & $0.9\pm 0.3$ & 
          $1.9\pm0.4$ & - \\
          & - & - & \redtxt{$0.39\pm 0.12$~\ariv$(\lambda 4711)$} & -
		  \\
    & - & - & \redtxt{$0.39\pm 0.08$~\ariv$(\lambda 4740)$} & -
		  \\
       	H$\beta$ & $37.9\pm 0.2$ & $52.5\pm 0.5$ & $30.86\pm 0.65$ & -
		  \\
    	  $[$O\,{\sc iii}]$\lambda 5007$ & $146.0\pm 0.2$ & $222.6\pm 0.3$ & $143.63\pm 0.43$ & -
            \\
    	  $[$Fe\,{\sc xiv}]$\lambda 5303$+$[$Ca\,{\sc v}]$\lambda 5309$ & $1.1\pm 0.1$ & $2.0\pm 0.2$ & - & -
		  \\
    	  $[$Fe\,{\sc vi}]$\lambda 5677$ & $0.30\pm 0.09$ & $<1.7^{\rm c}$ & - & -
		  \\
    	  $[$N\,{\sc ii}]$\lambda 5755$ & $0.7\pm 0.1$ & $1.3\pm 0.2$ & - & -
		  \\
    	  \hei$\lambda 5876$ & $16.4\pm 0.2$ & $21.2\pm 0.4$ & $13.7\pm0.5$ & -
		  \\
    	  $[$Fe\,{\sc vii}]$\lambda 6087$\redtxt{+\cav$\lambda 6087$} & $0.39\pm 0.09$ & $0.31\pm 0.17$ & - & -
		  \\
    	  H$\alpha ^{\rm a}$ & $123.7\pm 0.3$ & $182.1\pm 0.7$ & $99.5\pm 1.3$ & -
		  \\
    	  $[$N\,{\sc ii}]$\lambda 6583 ^{\rm a}$ & $<1.4^{\rm c}$ & $<2.6^{\rm c}$ & $2.95\pm 0.27$ & -
		  \\
    	  $[$S\,{\sc ii}]$\lambda \lambda 6716,6731$ & $1.26\pm 0.14$ & $2.3\pm 0.3$ & $0.42\pm 0.08$ & -
		  \\
     & - & - & \redtxt{$0.17\pm 0.07~(\lambda 6716)$} & -
		  \\
    & - & - & \redtxt{$0.25\pm 0.07~(\lambda 6731)$} & -
		  \\
    	  \hei$\lambda 6678$ & $7.6\pm 0.3$ & $12.4\pm 0.7$ & $7.3\pm0.5$ & -
		  \\
    	  \hei$\lambda 7065$ & $21.0\pm 0.4$ & $31.9\pm 0.6$ & $17.0\pm0.7$ & -
		  \\
    	  \hei$\lambda 7281$ & $3.6\pm 0.3$ & $8.7\pm 0.5$ &  $1.8\pm0.5$ & -
		  \\
		\hline
	\end{tabular}
	\begin{tablenotes}
        \small
        \item $\bf Notes.$
        \item $^{\rm a}$ Blended with the adjacent line in the PRISM spectrum.
        \item $^{\rm b}$ While these two lines are blended in the PRISM spectrum, the blended line profile peaks near the central wavelength of \heii.
        \item $^{\rm c}$ The value listed is the $3\sigma$ upper limit.
        \item $^{\rm d}$ Potentially large systematic uncertainties in the continuum not taken into account.
    \end{tablenotes}
\end{table*}

We used the Penalized PiXel-Fitting code (\textsc{pPXF}; \citealp{cappellari2004,cappellari2017}) to perform spectral fitting.

\subsection{JWST/NIRSpec PRISM spectrum}

Our main focus is on emission line fluxes.
For the JWST/NIRSpec PRISM IFU observations, we extract the central $0.3^{\prime \prime} \times 0.3^{\prime \prime}$ (or equivalently $\rm 1.8~kpc\times 1.8~kpc$) to match the fluxes measured from the JWST/NIRSpec high-resolution spectrum in \cite{ubler2023a}.
We used a combination of power-law functions and an analytical function for an optically thin Balmer continuum \citep{grandi1982} to fit the rest-frame UV-to-optical continuum simultaneously with emission lines.
\redtxt{Specifically, we used three power laws, one has a slope of $-2.33$ following the \citet{ss_disk_1973} (SS73) accretion disk, the other two have slopes of $-2.66$ and $0.5$, respectively.
The fitting results are relatively insensitive to the values of the latter two slopes and their role is to account for deviation from the SS73 disk.
The Balmer continuum component, on the other hand, is motivated by the clear Balmer jump feature in the rest-frame near UV spectrum of \target.
}
To account for variations in the continuum shape and the effect of dust attenuation, we included $\rm 8^{th}$-order multiplicative polynomials when fitting the continuum.
We also tried $\rm 4^{th}$-order multiplicative polynomials and our results are unaffected.
\redtxt{Since our main focus is on emission-line fluxes, we did not attempt to interpret the physical origin of the continuum based on spectral fitting.}

As shown by \citet{ubler2023a}, the G395H spectrum of GS\_3073 reveals several kinematic components in optical emission lines.
In the G395H spectrum, there is detection of a broad component in the Balmer lines and optical \hei\ and \heii\ lines with an FWHM of $3370.4\pm 41.4$ km/s as well as an outflow component with an FWHM of $1306.7\pm 23.7$ km/s.
In contrast, for forbidden lines in the optical, \citet{ubler2023a} only found a narrow component and an outflow component.

\redtxt{We assumed Gaussian profiles for emission lines.}
During the fit of emission lines, we included two different kinematic components for Balmer lines as well as optical He\,{\sc i} and \heii\ lines to represent the narrow components and the broad components.
\redtxt{We tied the kinematics of all narrow lines and broad lines, respectively.}
Since the spectral resolution of the PRISM spectrum ($\rm R\sim 30-300$) is generally not high enough to separate the two sets of components, we fixed the velocity dispersion of the narrow component to the values measured from the G395H spectrum in \cite{ubler2023a} convolved with the nominal resolution of the PRISM spectrum.
For the broad component, however, we did not fix its velocity dispersion \redtxt{to the value derived by \citet{ubler2023a}} in order to account for potential mismatch between the nominal spectral resolution and the actual spectral resolution \citep{jakobsen2022,degraaff2023}.
Following the results of \citet{ubler2023a}, for each line, if there are both narrow and broad components, we set the velocities of the two components the same.
For optical forbidden lines, only the narrow components were considered.
Finally, we tied the kinematics of all narrow optical lines and those of all broad optical lines, respectively.
While \cite{ubler2023a} identified an outflow component for optical lines in the G395H spectrum,
to avoid overfitting of the low-resolution spectrum, we did not include a third kinematic component.
The outflow contribution to the flux as fitted by \cite{ubler2023a} is in general small for strong lines and we discuss its potential impact when analyzing specific lines.
For UV lines, we simply considered a single kinematic component due to the much lower spectral resolution.
\redtxt{This means the UV line fluxes we measured should be the total fluxes including both broad line fluxes and narrow line fluxes. During our analyses later in the manuscript, we discuss the effects of such blending on our results.}
We tied the kinematics of all UV lines except Ly$\alpha$ due to the complex nature of its intrinsic line profile.
\redtxt{While \nv, \cii, and \civ\ are also resonant or have a resonant component, the spectral resolution of the PRISM spectrum at their locations are relatively low and insufficient to discern their potentially different kinematics. As a sanity check, we performed a fit where the kinematics of the above lines are set free.
The final fluxes we obtained are consistent with our current fitting results within $1\sigma$ of the measurement uncertainties.
}

To account for the spatial covariance of the flux errors, we first calculated the integrated errors assuming all spaxels are independent. We then performed an initial spectral fit assuming the true errors equal to the integrated errors.
Next, we rescaled the integrated errors using the standard deviation obtained from the residuals of the initial fit. To do this, we calculated standard deviations within several spectral windows without apparent emission features.
We inspected five spectral regions in the spectrum centered around 2400 \AA, 3300 \AA, 5600 \AA, 6300 \AA, and 7700 \AA\ in the rest frame, and we tested different widths for the spectral regions\footnote{We caution, however, that the spectral regime in the rest-frame UV might be contaminated by \feii\ emission}.
The standard deviations obtained from these regions are in general $2-3$ times larger than the integrated errors and thus we scaled up all integrated errors by a factor of 3 as a conservative estimation.
The final flux uncertainty for each emission line was calculated by adding the noise to the original spectrum and refitting.
We repeated the above procedure 500 times and took the standard deviation of the flux distribution divided by $\sqrt{2}$ as the final uncertainty (since the effective flux uncertainties have been enlarged by a factor of $\sqrt{2}$ after adding the noise).
The uncertainty calculated this way is broadly consistent with the formal error reported by \textsc{pPXF}.

Figure~\ref{fig:prism_fit} shows the fitting result of our extracted PRISM spectrum.
The continuum model suggests the existence of a Balmer continuum.
Meanwhile, the UV continuum appears to feature a few bumps spanning from $2200~\AA$ to $3600~\AA$.
We speculate the existence of \feii\ emission in this regime and will present the analyses in a future paper.
In the optical, one can see a clear broad component in H$\alpha$ even at the low resolution of the PRISM spectrum.
Our fit gives $\rm FWHM_{H\alpha} \sim 4.3\times 10^3$ km/s for the broad component.
This is larger than the value measured by \citet{ubler2023a} from the high resolution G395H spectrum, which is $\rm FWHM_{H\alpha} = 3370.4 \pm 41.4$ km/s.
The difference is likely due to our adoption of a canonical spectral resolution for the PRISM spectrum, which could be different from the actual spectral resolution in observations.
Judging from the residuals, the emission lines are overall well fitted.
The bottom panels of Figure~\ref{fig:prism_fit} and Figure~\ref{fig:prism_fe} show the zoom-in views around several UV lines and weak optical lines.
Specifically, we plotted the zoom-in view around H$\alpha$ on a logarithmic scale to highlight the potential detection of weak coronal lines of Fe.

We summarize the fluxes of emission lines in Table~\ref{tab:fluxes}.
We also list the emission line fluxes measured from the high resolution spectrum by \citet{ubler2023a}.
These fluxes are measured within the central $0.3^{\prime \prime} \times 0.3^{\prime \prime}$ region of the G395H IFU cube.
For each line, we simply summed all its kinematic components.
While the \oiii$\lambda 5007$ flux from the PRISM matches that from the G395H grating, other strong lines such as H$\alpha$, H$\beta$, and He\,{\sc i}$\lambda 7065$ have their PRISM fluxes higher than the G395H fluxes by $\sim 20\%$.
We note that the flux differences might come from a mix of effects from the flux calibration of the PRISM spectrum \redtxt{\citep{arribas2023,jades_dr3}} and different data reductions adopted for the PRISM cube we used and the G395H cube analyzed by \citet{ubler2023a}.
Also, the new reduction we adopted generated a spaxel size of $0.05'' \times 0.05''$ for the PRISM cube, different from the previous reduction that produced a spaxel size of $0.1'' \times 0.1''$ for the G395H cube.
This potentially causes an offset between our extracted area and that in \citet{ubler2023a}.
In addition, even if the center of the two extracted areas match exactly, the spaxels we selected near the boundary of the area would be partially outside the extracted area of \citet{ubler2023a}.
We performed a check by shrinking the size of the extracted aperture in the PRISM cube by one pixel and added one-pixel displacement into different directions; the resulting fluxes of H$\alpha$ and H$\beta$ are more consistent with those measured from the G395H spectrum but the \oiii$\lambda 5007$ flux is lowered by $\sim 15-20\%$.
Regardless, most of our analyses rely on relative fluxes within the PRISM spectrum or the G395H spectrum, respectively.
The only major impact of the flux difference is on the flux ratio of \nii$\lambda 6583$/\oii$\lambda \lambda 3726,3729$, where the narrow \nii$\lambda 6583$ is only measured in G395H and the \oii$\lambda \lambda 3726,3729$ is only measured in PRISM.
To account for the systematics associated with the flux mismatch, we included an extra 20\% systematic uncertainty for the flux ratio.
As we show later, this uncertainty is smaller than the potential systematics associated with the outflowing component in these lines.

A number of lines are blended in the PRISM spectrum.
In the UV, \heii$\lambda 1640$ is blended with O\,{\sc iii}]$\lambda \lambda 1661, 1666$.
However, the blend of these lines clearly peaks close to the central wavelength of \heii$\lambda 1640$ as shown in Figure~\ref{fig:prism_fit}, suggesting its flux dominates the blend\footnote{We note that the wavelength sampling adopted by the data reduction pipeline we used is 50 \AA\ in the observed frame and 7.6 \AA\ in the rest frame at $z=5.55$. We also tried resampling the spectrum to another wavelength grid as adopted in the reduction of MSA data.
Despite the latter having a range of wavelength intervals from 26 \AA\ to 122 \AA\ in the observed frame, our emission line measurements including the decomposition of blended lines remain largely unchanged.
}.
The same situation applies to \heii$\lambda 4686$, [Ar\,{\sc iv}]$\lambda \lambda 4711, 4740$, and \hei$\lambda 4713$, where \heii$\lambda 4686$ likely dominates the blend.
While it appears that \nv$\lambda \lambda 1238, 1242$ is present in the PRISM spectrum, it might be blended with the red wing of Ly$\alpha$.
In the next subsection, we used the VIMOS spectrum to check the existence of this line.

Finally, besides the $0.3^{\prime \prime} \times 0.3^{\prime \prime}$ extraction, we also fit the PRISM spectrum extracted from a larger aperture of $1^{\prime \prime} \times 1^{\prime \prime}$ in order to match the slit width of the VIMOS observation.
The integrated line fluxes in general reach the maximum value with this extraction but have larger uncertainties due to the inclusion of the noisy background.

\subsection{VLT/VIMOS spectrum}
\label{subsec:vimos_measurement}

The VIMOS spectrum has a much higher spectral resolution compared to the PRISM spectrum.
Meanwhile, the noise level of the VIMOS spectrum is higher than that of the PRISM spectrum and thus the broad component of emission lines could potentially be lost in the noise.

We used \textsc{pPXF} to fit the VIMOS spectrum.
Again, we assumed the continuum is a combination of power laws.
Due to the high noise level of the VIMOS spectrum, we used the best-fit continuum model from the fitting of the PRISM spectrum to construct the input continuum template.
Specifically, we fixed the relative shape of the continuum redwards of Ly$\alpha$ to that of the best-fit PRISM continuum model, while setting the continuum bluewards of Ly$\alpha$ a flat line.
This choice is motivated by the fact that the shape of the damped Lyman continuum is uncertain but it has an intrinsically low flux level.
Furthermore, we did not include any polynomials during the fit in order to avoid overfitting the noise.
Thus, our final continuum model is the weighted sum of a best-fit PRISM continuum and a flat continuum in the blue.

We note that the choice of the continuum is vital for the subsequent flux measurements.
For example, by assuming the underlying continuum has a stellar origin, \citet{barchiesi2023} obtained flux measurements different from ours especially for weak lines (e.g., their flux for \nv\ is only $\sim 1/3$ of ours).
Figure~\ref{fig:vimos_fit} shows the VIMOS spectrum and the emission lines we identified.
Among the lines, Ly$\alpha$ clearly shows an asymmetric line profile.
Meanwhile, \nv\ exhibits a P-Cygni profile, which is attributed to a $\sim 3$ Myr-old stellar population by \citet{barchiesi2023}.
\redtxt{\citet{barchiesi2023} argue the feature cannot be due to absorption in the AGN continuum since GS\_3073 was not confirmed to be a type-1 AGN before JWST observations and no clear absorptions in other lines were found.}
It is \redtxt{now} clear from the high resolution optical spectrum that GS\_3073 is a type-1 AGN \citep{ubler2023a}.
\redtxt{Still, we note that the rest-frame UV continua of many JWST-identified type-1 AGNs could have significant stellar light contributions \citep[e.g.,][]{kocevski2023,harikane2023,greene2024,matthee2024,juodzbalis_absagn_2024}.
For young stellar populations at low metallicities, the UV continua are also well approximated by featureless power laws \citep[e.g.,][]{stanway2018}.
The only relevant difference is whether the emission at the location of \nv\ has a stellar continuum origin or a purely nebular origin (ionized either by stellar radiation or AGN radiation).
According to the fit done by \citet{barchiesi2023}, a full stellar UV continuum model would still leave part of the \nv\ emission nebular, indicative of the presence of highly ionized gas.
We emphasize that our main conclusions are largely unchanged no matter whether we adopt a pure nebular origin for the \nv\ emission or a partially nebular origin for the \nv\ emission based on the results of \citet{barchiesi2023}.
Even without including \nv\ in our nebular diagnostics, we obtained similar abundance patterns, as shown in Section~\ref{subsec:nitrogen}.
}


In the VIMOS spectrum, there is also detection of C\,{\sc ii}$\lambda \lambda 1334,1335$, although it is unclear whether both lines within the doublet are present.
Interestingly, while C\,{\sc ii}$\lambda 1334$ is also a resonant line similar to \nv, no blueshifted absorption is visible in the spectrum.
In addition, there is \niv$\lambda \lambda 1483,1486$, and potentially O\,{\sc vi}$\lambda \lambda 1032,1038$ and Si\,{\sc iv}$\lambda \lambda 1393,1402$ in the spectrum.
While the spectrum reaches the location of \civ$\lambda \lambda 1548,1551$, the noise level becomes very high in this regime and there is no clear detection of this doublet.

Due to the complex nature of different UV lines, we adopted the following fitting strategy.
First, we used three Gaussian components to approximate the asymmetric profile of Ly$\alpha$, where one narrower component is used to fit the peak of the line profile, and the other two broader components are used to capture the red wing of the line profile and thus their centroids are allowed to be redshifted with respect to that of the narrower component.
\redtxt{We note that this multi-Gaussian fitting approach is only for estimating the total flux of Ly$\alpha$, whose profile indicates impacts from radiative transfer, and adding more Gaussians do not significantly improve the fit.
}
Second, we tied the kinematics of most UV lines to that of \niv$\lambda \lambda 1483,1486$, which is the strongest line besides Ly$\alpha$.
The kinematics of Ly$\alpha$, \nv$\lambda \lambda 1238, 1242$, and C\,{\sc ii}$\lambda \lambda 1334,1335$ are allowed to vary with respect to those of \niv$\lambda \lambda 1483,1486$ as these lines can be resonantly scattered and absorbed.
\redtxt{Indeed, from the VIMOS spectrum, we found the peak of Ly$\alpha$ is redshifted with respect to that of the narrow \niv$\lambda \lambda 1483,1486$ by roughly 250 km/s.
Also, \nv\ and \cii\ show velocity shifts of roughly 1200 km/s and 750 km/s, respectively.
}
We did not include broad components for emission lines due to the noise.
We plotted the fitting result in Figure~\ref{fig:vimos_fit} and summarized the emission line fluxes in Table~\ref{tab:fluxes}.

Comparing the VIMOS fluxes with the $1^{\prime \prime}$-PRISM fluxes listed in Table~\ref{tab:fluxes}, the total fluxes of \lya\ measured from these spectra are roughly consistent.
However, the rest of the lines, especially the strong \niv$\lambda \lambda 1483,1486$ exhibits a clear mismatch, where the PRISM flux is significantly higher.
We suggest that the difference originates from a broad (BLR) component of these lines
(i.e., associated with the BLR component seen in the hydrogen Balmer lines and helium lines), but undetected in the VIMOS spectrum because of the higher spectral resolution and larger noise.
To check this, we convolved the VIMOS spectrum to the nominal spectrum resolution of the PRISM spectrum and resampled the VIMOS spectrum at the wavelength grid of the PRISM spectrum.
We then refitted the convolved and resampled VIMOS spectrum, obtaining the total flux of the \niv\ doublet to be $\rm 56.3\pm 3.5 \times 10^{-18}~erg/s/cm^{-2}$, which is higher than the flux obtained from the original VIMOS spectrum but close to the flux obtained from the PRISM spectrum (c.f., Table~\ref{tab:fluxes}).
Still, we caution that the flux measurement can suffer from the following uncertainties. First, in the rebinned VIMOS spectrum, the broad wing of \niv\ merges with the blue wing of \civ\ and is truncated at the red end of the spectrum, leading to a potential systematic associated with deblending and continuum determination.
\redtxt{Second, the imperfect aperture matching and the relative flux calibrations could lead to further systematic uncertainties.
By directly taking the total flux ratio of \lya+\nv\ and \niv, the VIMOS result is roughly 15\% lower than that of $1''$-PRISM.
However, as we show later, this systematic would only bias the abundance patterns lower, leaving our main conclusions unchanged.
}

Finally, as noted by \citet{ubler2023a}, there is a companion or an outflow component to the northwest and another one to the east of GS\_3073, which would be included if the extracted size of the aperture exceeds $\sim 0.3^{\prime \prime}$.
Based on the publicly available NIRCam images from JADES shown in Appendix~\ref{appendix:a}, the northwest and east targets are separated from the central source.
The separation is clear in photometric bands including F200W, which essentially measures the UV continuum flux in the rest frame of GS\_3073.
Given the presence of continuum emission, these targets are likely satellites of the central source rather than outflows.
Both the VIMOS spectrum and the $1^{\prime \prime}$-PRISM spectrum can potentially have contamination from the light of these satellites.

In the following analyses, we use all the emission-line measurements listed in Table~\ref{tab:fluxes}.
Below we summarize the limitations of different measurements.
\begin{enumerate}
    \item The PRISM spectrum has a spectral coverage from the rest-frame UV to the rest-frame optical. But several emission lines are blended. The G395H spectrum and the VIMOS spectrum only cover the rest-frame optical and the rest-frame UV, respectively.
    \item There is potentially a systematic difference between the PRISM flux and the grating flux, which varies as a function of the wavelength.
    The difference can be caused by imperfect flux calibrations and different IFU cube constructions.
    \item The VIMOS spectral measurements can suffer from an uncertain continuum model and the loss of the broad component due to the noise.
    \item Spectra extracted from an aperture with a size larger than $0.3^{\prime \prime} \times 0.3^{\prime \prime}$ can have contamination from the nearby companions.
\end{enumerate}

Despite these limitations, our conclusions are not significantly affected as most of our analyses do not involve flux ratios between the PRISM measurements and the G395H measurements and thus are insensitive to flux calibration issues.
In addition, in the following analyses we discuss results from different extracted apertures of the PRISM cube when appropriate, and consider the potential loss of the broad component in the VIMOS spectrum.

\section{Nebular diagnostics}
\label{sec:diagnostics}

\begin{figure*}
    \centering\includegraphics[width=0.98\textwidth]{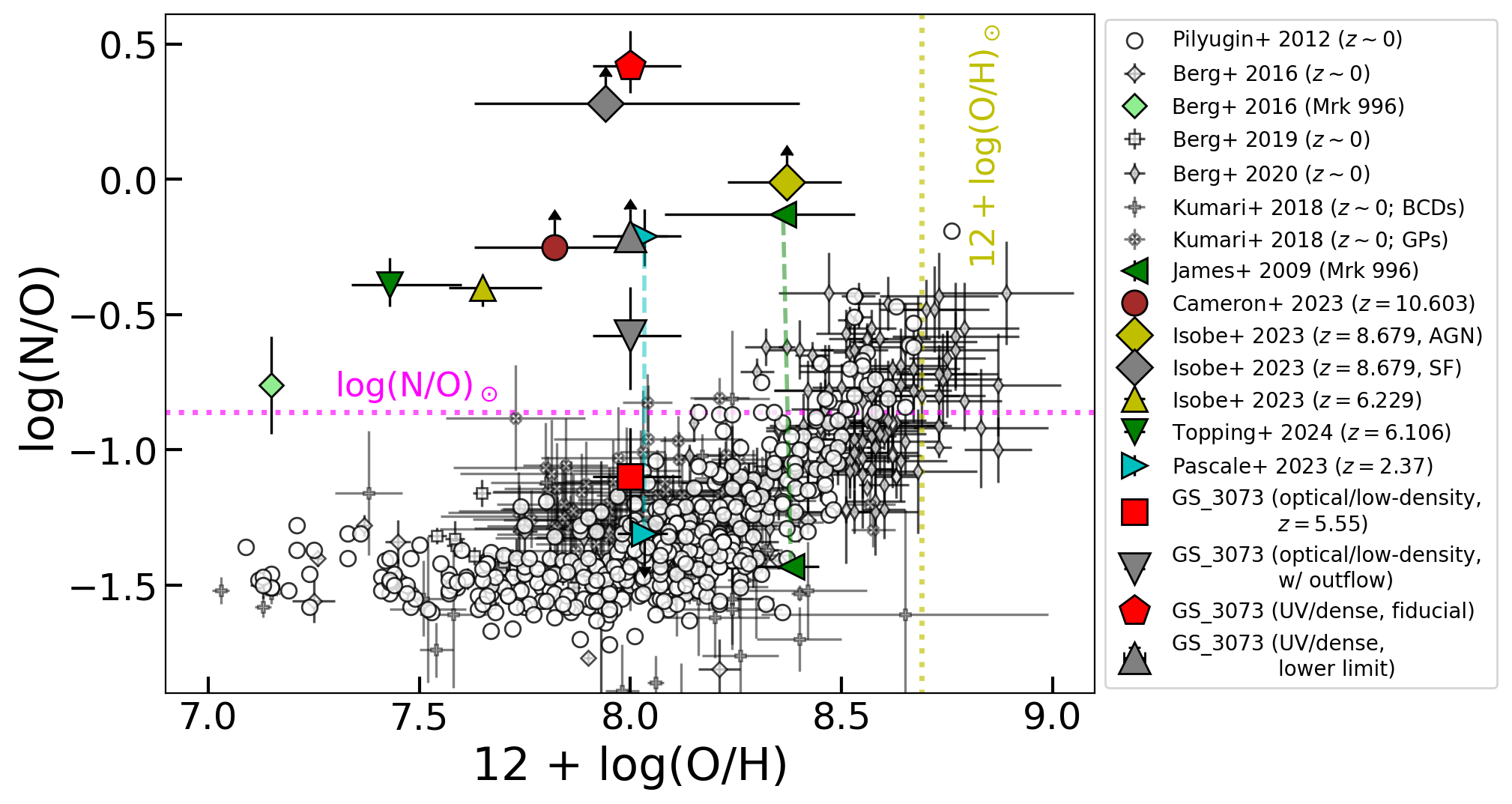}
    
    \caption{Relation between N/O and O/H measured from different systems.
    For clarity, we used open symbols to represent all $z\sim 0$ sources except Mrk 996.
    The open circles, diamonds, and squares are local \hii\ regions and galaxies with abundances determined by \citet{pilyugin2012}, \citet{berg2016}, \citet{berg2019}, and \citet{berg2020}.
    The open plus and cross symbols represent the abundances of a sample of local blue compact dwarf galaxies (BCDs) and green pea galaxies compiled by \citet{kumari2018}.
    The compilation includes abundances derived by \citet{cairos2009}, \citet{james2010,james2013}, \citet{perez-montero2011}, \citet{westmoquette2013}, \citet{lagos2014,lagos2016}, \citet{kehrig2016}, and \citet{kumari2018}.
    The abundances of the WR galaxy, Mrk 996, derived by \citet{berg2016} are indicated by the large lightgreen diamond.
    \redtxt{We also show the results from \citet{james2009} for two kinematically different gas components in Mrk 996 as two left-pointing triangles connected by the dashed green line.}
    \redtxt{Another source within the Sunburst Arc at $z=2.37$ also showing two different densities and chemical abundance patterns as derived by \citet{pascale2023} is shown as two cyan right-pointing triangles connected by the dashed cyan line.}
    The green triangle, yellow triangle, and yellow diamond are nitrogen-loud galaxies at $z = 6.106$, $z = 6.229$, and $z=8.679$, respectively, with their abundances measured by \citet{isobe2023} (also by \citealp{marques-chaves2024}) and \citet{topping2024}.
    The grey diamond is the estimation by \citet{isobe2023} assuming an SF-dominated scenario rather than an AGN-dominated scenario.
    The brown circle is an AGN candidate, GN-z11, at $z=10.603$ with abundances estimated by \citet{cameron2023}.
    The red square is our estimation for GS\_3073 using only the optical lines.
    The grey inverted triangle is also estimated based solely on optical lines, but with the assumption that a fraction of the \oii$\lambda \lambda3726,3729$ is actually in the outflow.
    The red pentagon represents the fiducial abundances we estimated for GS\_3073, where N/O is estimated from UV lines originating in a dense region.
    The grey triangle corresponds to the lower limit of N/O based on UV lines, where the flux of the whole \heii$\lambda 1640$ + \oiiis$\lambda \lambda 1661,1666$ is assigned to \oiiis$\lambda \lambda 1661,1666$.
    The dotted lines mark the solar abundances from \citet{grevesse2010}.
    }
    \label{fig:no_relation}
\end{figure*}

\begin{figure*}
    \centering\includegraphics[width=0.98\textwidth]{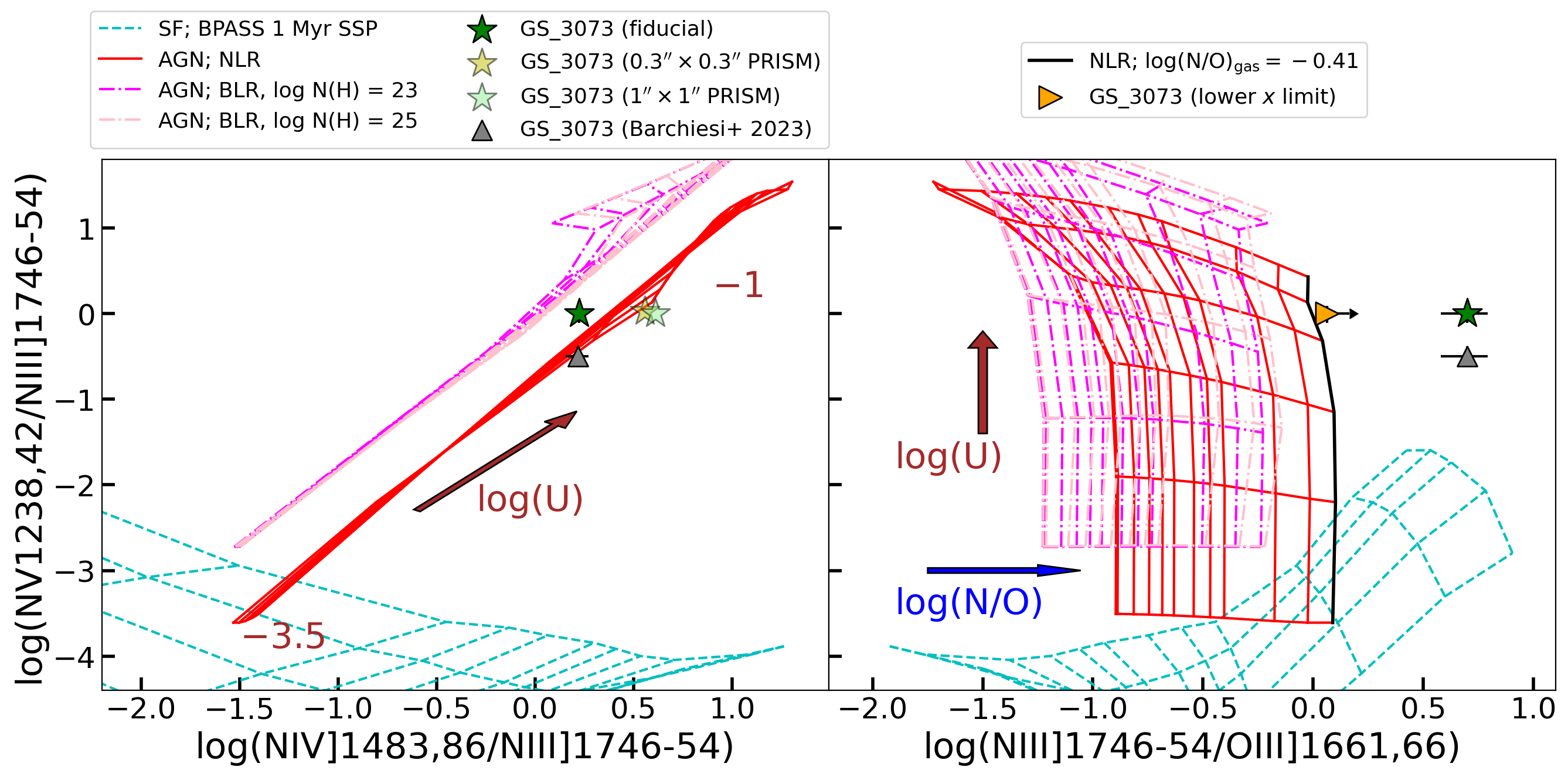}
    
    \caption{\textit{Left:} Line ratios predicted by different photoionization models together with the observed line ratios of GS\_3073 in the \nv/\niii\ versus \niv/\niii\ diagram. The dashed cyan grid, solid red grid, and the dotted-dashed magenta and pink grids represent photoionization models for SF regions, NLRs of AGNs, and BLRs of AGNs.
    All model grids span the same range of metallicity and ionization parameter, with $Z/Z_\odot \sim 5\times 10^{-4}-2$ and $U \sim 10^{-3.5}-10^{-1}$. The arrow indicates the increasing direction of $U$ for the models.
    The green star is the fiducial measurements we adopted for GS\_3073.
    The yellow and light green stars replace the narrow \niv\ measured from VIMOS with the \niv\ measured from the PRISM with different extracted apertures.
    The grey triangle replaces \nv\ and \niv\ with the measurements by \citet{barchiesi2023}, which assume a stellar origin for the UV continuum.
    \textit{Right:} Line ratios predicted by different photoionization models together with the observed line ratios of GS\_3073 in the \nv/\niii\ versus \niii/O\,{\sc iii}] diagram.
    The arrows indicate the increasing directions of $U$ and N/O for the models.
    The solid black line represents NLR models with a gas-phase nitrogen abundance of $\rm log(N/O)_{gas} = -0.41$.
    The yellow triangle represents a lower limit for N/O set by replacing the flux of O\,{\sc iii}] with the summed flux of \heii$\lambda 1640$ and O\,{\sc iii}].
    The rest of the symbols are the same as the left panel.
    }
    \label{fig:diagrams_n}
\end{figure*}

\begin{figure*}
    \centering\includegraphics[width=\columnwidth]{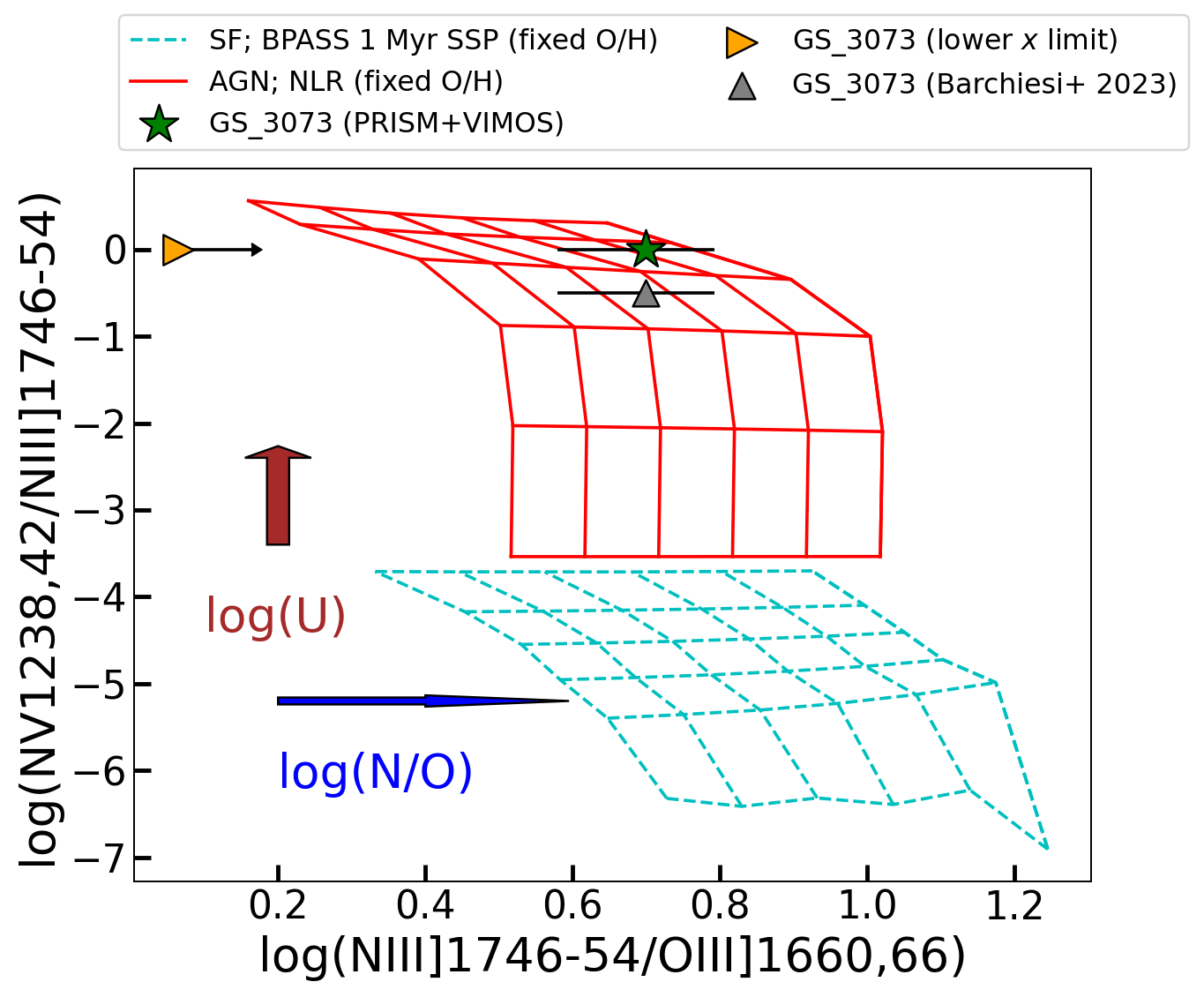}
    \includegraphics[width=\columnwidth]{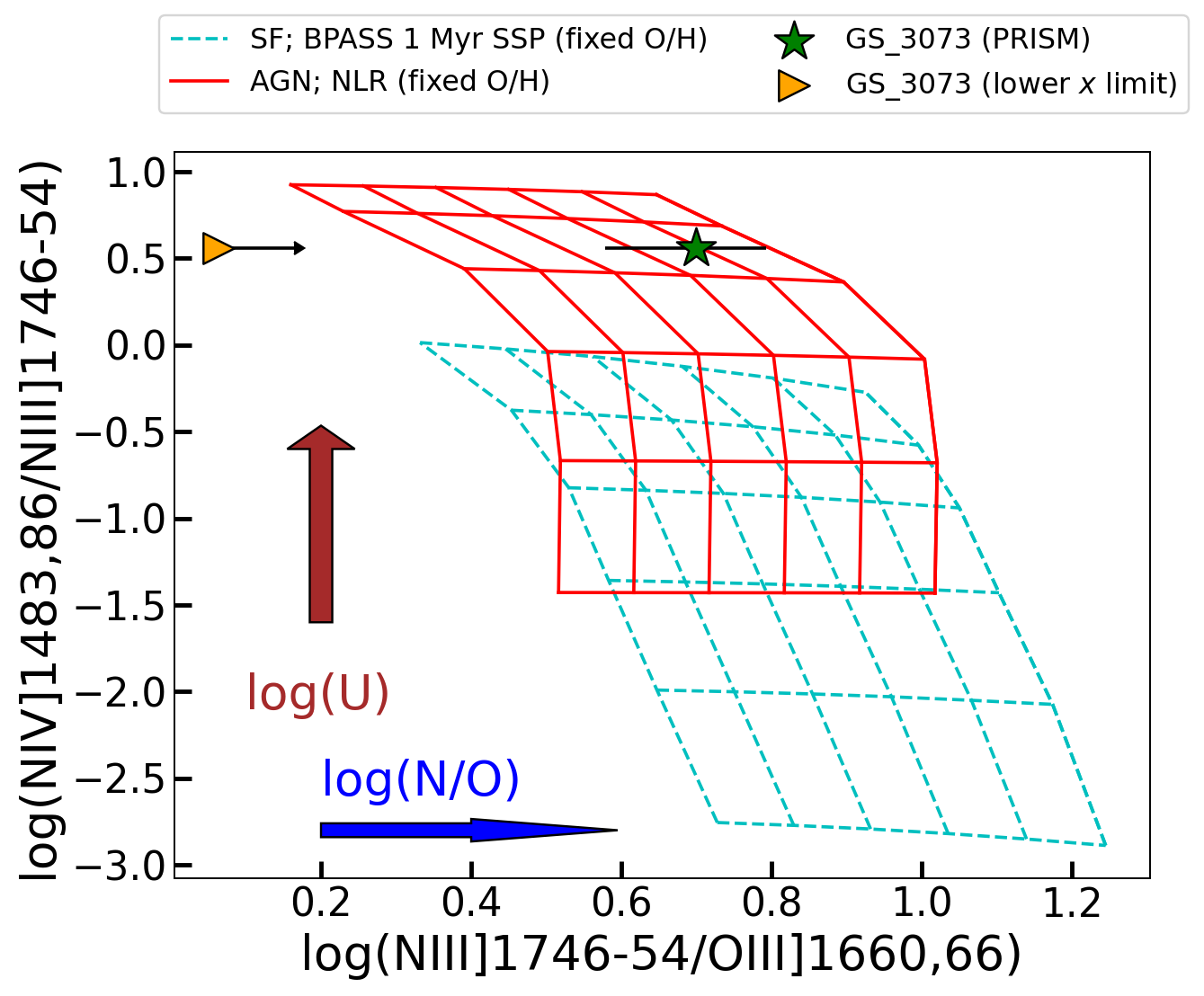}
    
    \caption{\textit{Left:} Diagnostic diagram composed of \nv/\niii\ and \niii/\oiiis.
    The photoionization models shown have a fixed metallicity of $\rm 12+\log(O/H) = 8.0$ and log(N/O) ranging from 0.03 dex to 0.53 dex.
    The rest of the model parameters are the same as the corresponding models in Figure~\ref{fig:diagrams_n}.
    The green star represents the fiducial line ratios we measured for GS\_3073.
    The grey triangle replaces the \nv\ flux with that measured by \citet{barchiesi2023} assuming the UV continuum is stellar dominated.
    The orange triangle set the lower $x$-axis limit by assigning the whole \heii$\lambda 1640$ + \oiiis$\lambda \lambda 1661,1666$ flux to \oiiis$\lambda \lambda 1661,1666$.
    \textit{Right:} Diagnostic diagram composed of \niv/\niii\ and \niii/\oiiis.
    The models shown are the same as the ones on the left panel.
    The green star and the orange triangle represent measurements from the $0.3^{''}$-PRISM spectrum only.
    }
    \label{fig:diagram_n_no}
\end{figure*}

\begin{figure*}
    \centering\includegraphics[width=.95\columnwidth]{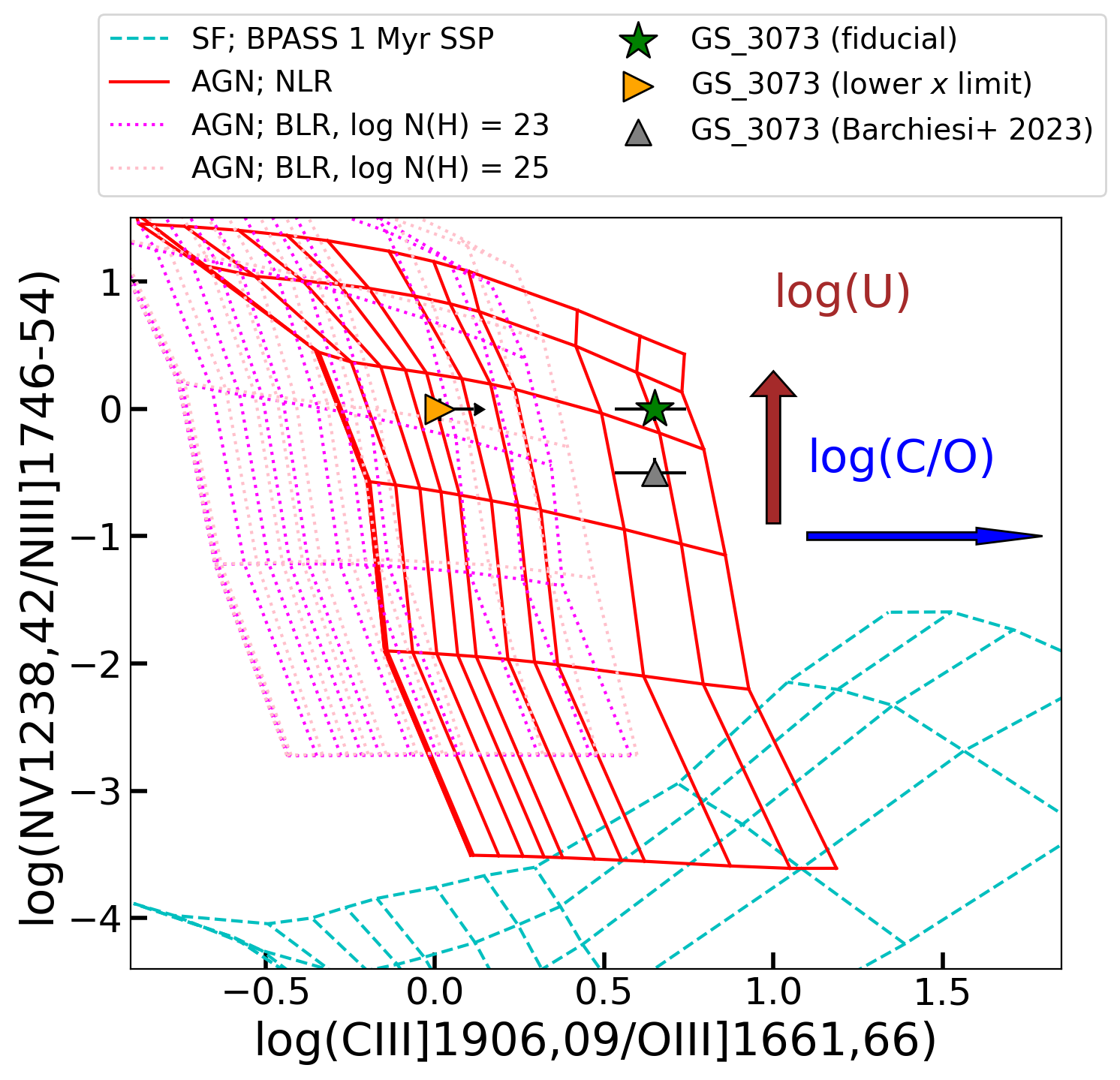}
    \includegraphics[width=1.05\columnwidth]{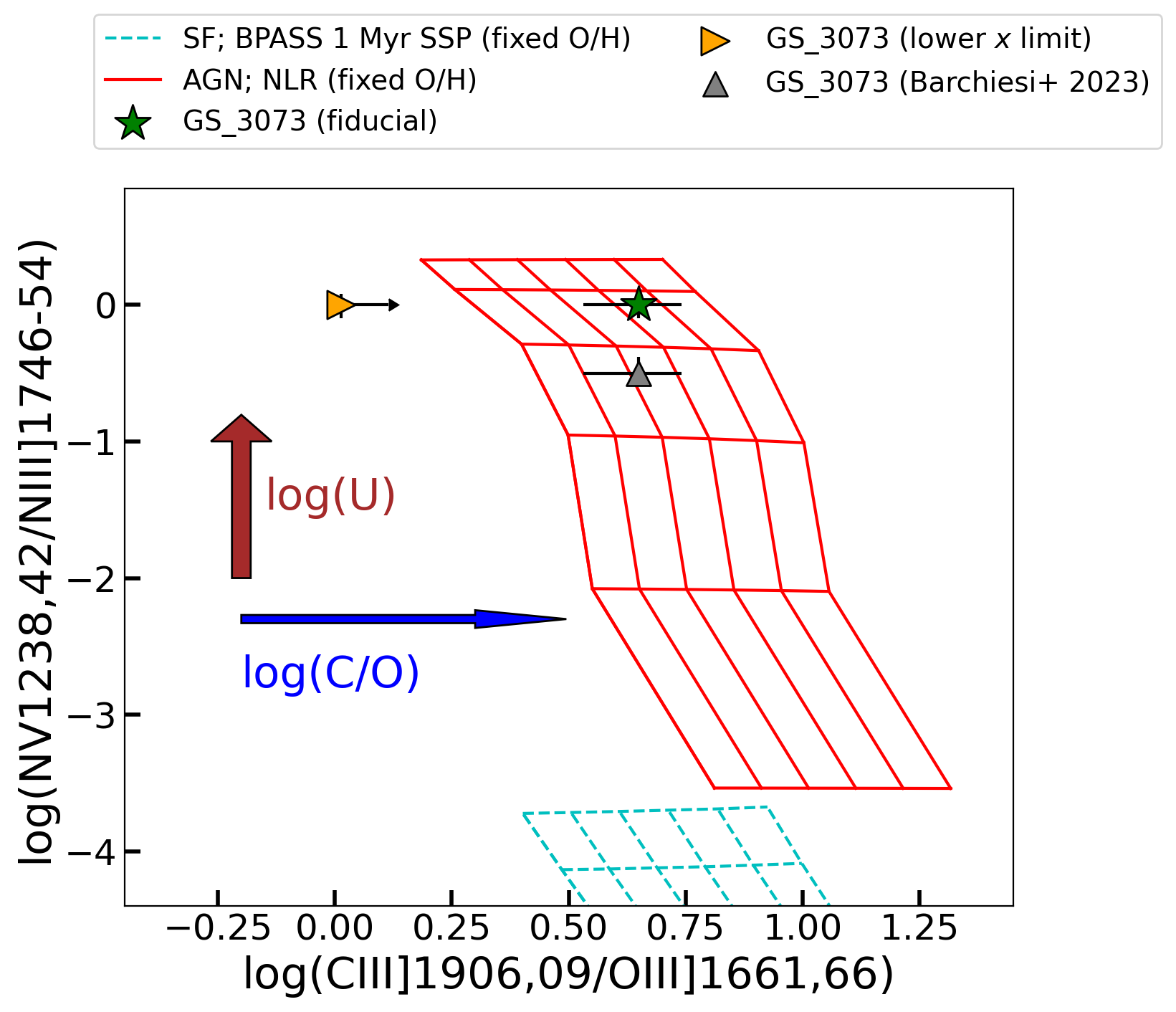}
    \includegraphics[width=1.05\columnwidth]{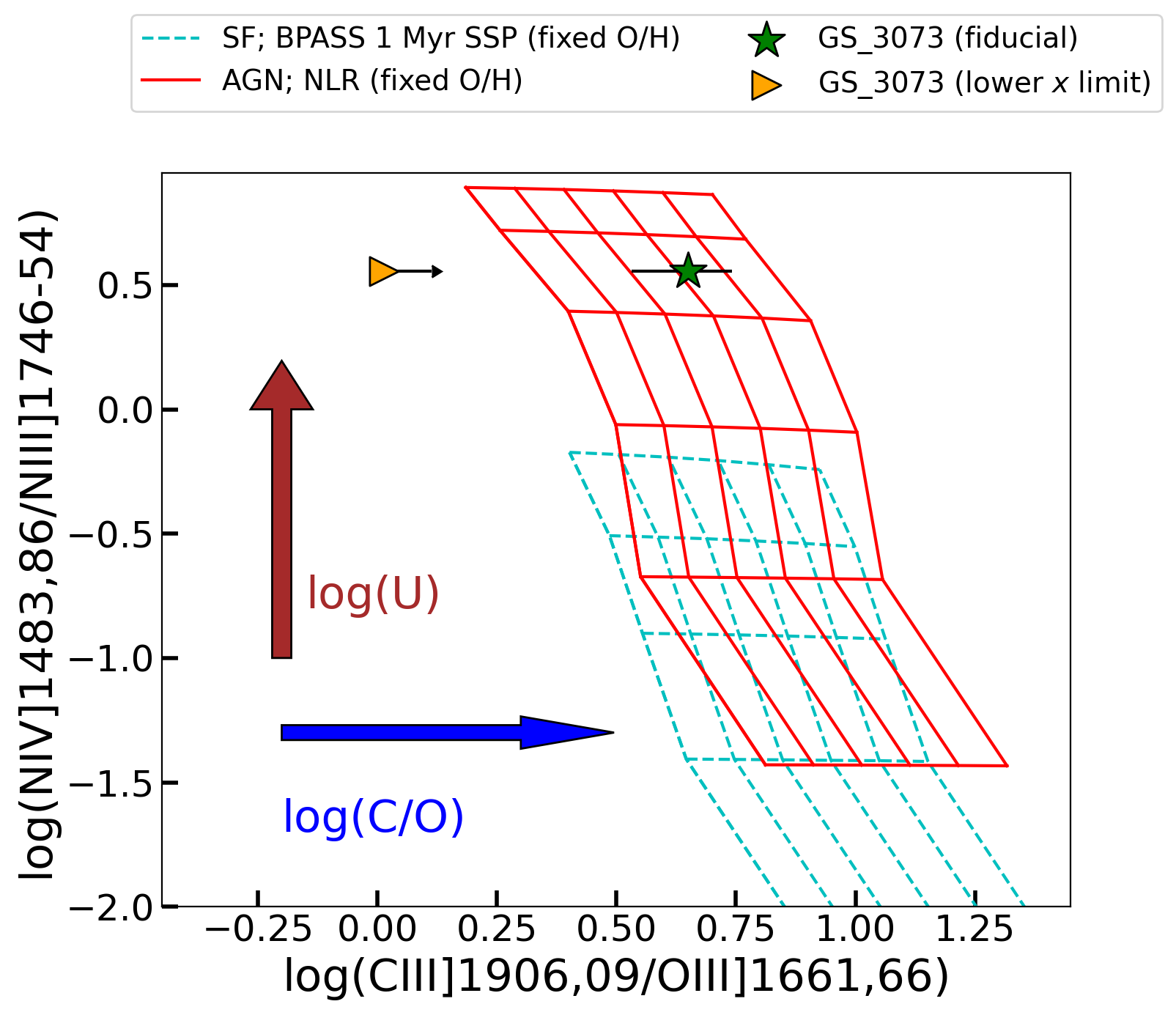}
    
    \caption{\redtxt{\textit{Upper left:}}
    Diagnostic diagram composed of \nv/\niii\ and \ciii/\oiiis.
    The photoionization models are the same as those in Figure~\ref{fig:diagrams_n}.
    The green star represents the fiducial line ratios we measured for GS\_3073.
    The grey triangle replaces the \nv\ flux with that measured by \citet{barchiesi2023} assuming the UV continuum is stellar dominated.
    The orange triangle set the lower limit for C/O by assigning the whole \heii$\lambda 1640$ + \oiiis$\lambda \lambda 1661,1666$ flux to \oiiis$\lambda \lambda 1661,1666$. The rightmost two vertical model lines of the NLR grid correspond to $\rm log(C/O)_{gas} = -0.32$ and $\rm log(C/O)_{gas} = -0.21$, respectively.
    \redtxt{\textit{Upper right:} The models shown have a fixed metallicity of $\rm 12+log(O/H) = 8.0$, a fixed nitrogen abundance of $\rm log(N/O) = 0.43$, and a range of carbon abundances of $\rm -0.5\leq log(C/O) \leq 0$. \textit{Bottom:} Diagnostic diagram composed of \niv/\niii\ and \ciii/\oiiis. The models shown are the same as those in the upper right panel.
    The green star and the orange triangle represent measurements from the $0.3^{''}$-PRISM spectrum only.}
    }
    \label{fig:diagram_c}
\end{figure*}

\begin{figure*}
    \centering\includegraphics[width=0.98\textwidth]{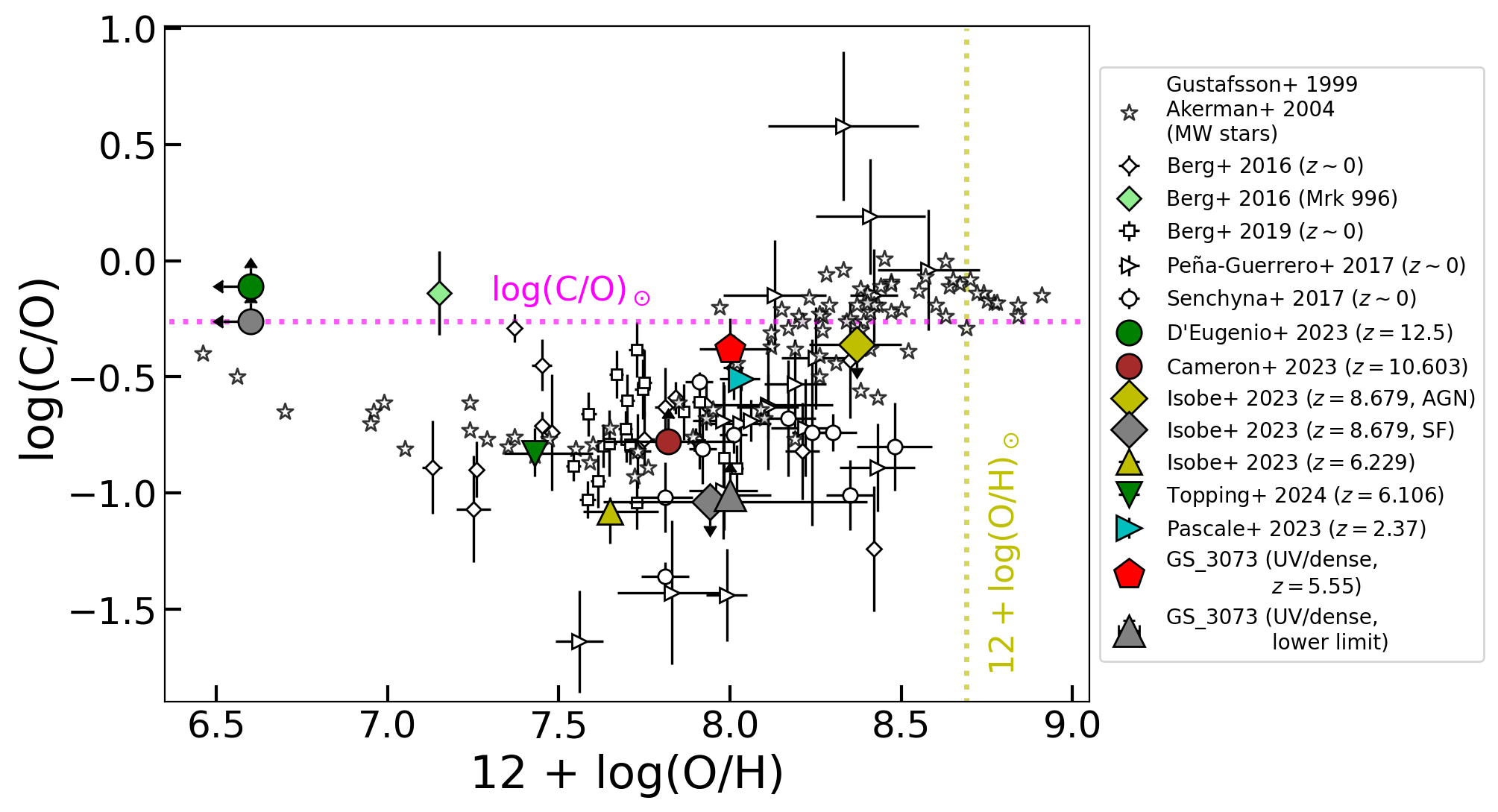}
    
    \caption{Relation between C/O and O/H measured from different systems.
    For clarity, we used open symbols to represent all $z\sim 0$ sources except Mrk 996.
    The open stars are Milky Way (WM) stars with abundances determined by \citet{gustafsson1999} and \citet{akerman2004}.
    The small open diamonds, squares, triangles, and circles are local \hii\ regions and galaxies with abundances determined by \citet{berg2016}, \citet{berg2019}, \citet{pena-guerrero2017}, and \citet{senchyna2017}.
    The large lightgreen diamond is Mrk 996, a metal-poor blue compact dwarf galaxy showing WR features.
    \redtxt{The right-pointing cyan triangle represents the abundances of gas surrounding a super-star cluster in the Sunbusrt Arc at $z=2.37$ measured by \citet{pascale2023}.}
    The large green triangle, large yellow triangle, and large yellow diamond are nitrogen-loud galaxies at $z = 6.106$, $z = 6.229$, and $z=8.679$, respectively, with their abundances measured by \citet{isobe2023} (also by \citealp{marques-chaves2024}) and \citet{topping2024}.
    The grey diamond is the estimation by \citet{isobe2023} assuming an SF-dominated scenario rather than an AGN-dominated scenario.
    The brown circle is a nitrogen-loud AGN candidate, GN-z11, at $z=10.603$ with abundances estimated by \citet{cameron2023}.
    The large green and grey circles corresponds to a carbon rich galaxy, GS-z12, at $z=12.5$, with abundances constrained by \citet{deugenio2023} with different assumptions.
    The red pentagon represents the fiducial abundances we estimated for GS\_3073, where C/O is estimated from UV lines.
    The grey triangle corresponds to the lower limit of C/O based on UV lines, where the flux of the whole \heii$\lambda 1640$ + \oiiis$\lambda \lambda 1661,1666$ is assigned to \oiiis$\lambda \lambda 1661,1666$.
    The dotted lines mark the solar abundances from \citet{grevesse2010}.
    }
    \label{fig:co_relation}
\end{figure*}

In this section we summarize the emission-line diagnostics we performed to derive the physical conditions of the ionized gas in GS\_3073.
Our analyses are largely based on \textsc{PyNeb} \citep{luridiana2015} and \textsc{Cloudy} \citep{ferland2017}.
For calculations with \textsc{PyNeb}, we used \redtxt{atomic data from} the CHIANTI database\footnote{\redtxt{While the version of \textsc{Cloudy} we used has an older version of CHIANTI (v7) compared to the latest version (v10) we used during the \textsc{PyNeb} calculations, we performed consistency checks to confirm the two data versions do not lead to noticeable difference in the predicted line ratios.}}\citep{chianti0,chianti1} 
\redtxt{
for all relevant atoms and ions
except for $\rm O^{2+}$, $\rm N^{2+}$, and $\rm C^{2+}$. 
For $\rm O^{2+}$, we took its transition probabilities from \citet{oiii_as_tz17} and collisonal strengths from CHIANTI.
For $\rm N^{2+}$, we took its transition probabilities from \citet{niii_as_g98} and collisonal strengths from CHIANTI.
For $\rm C^{2+}$, we took its transition probabilities from \citet{ciii_as_n78} and \citet{ciii_as_t99}, and its collisional strengths from \citet{ciii_coll_b85} and \citet{ciii_coll_k92}.
}

\subsection{Dust attenuation}
\label{subsec:dust_att}

Our analyses are broadly separated between the rest-frame UV regime and the rest-frame optical regime.
Within each spectral regime, we mostly used emission lines with similar wavelengths to perform nebular diagnostics.
However, to evaluate the consistency between UV and optical diagnostics, we need to combine lines with very different wavelengths, which are therefore subject to the effect of potential dust attenuation.

In \cite{ubler2023a}, dust attenuation is not taken into account due to the small deviation of the Balmer decrement from the typical values found in dust-free AGNs and the analyses are limited to the optical regime.
To evaluate the impact of dust on UV-to-optical line ratios, we used \heii$\lambda 1640$ and \heii$\lambda 4686$.
These two lines are recombination lines of He and their intrinsic ratio is insensitive to variations in temperature and density.
The typical intrinsic ratio of \heii$\lambda 1640$/\heii$\lambda 4686$ is 7 -- 8, no matter whether the lines originate in \hii\ regions, narrow-line regions (NLRs), BLRs, or stellar winds \citep{maschmann2024}.
Even with a considerable contribution from collisional excitation, as can happen in shocks, the expected intrinsic ratio of \heii$\lambda 1640$/\heii$\lambda 4686$ would still be around 6 -- 7 \citep{alarie2019}.
Intriguingly, the measured \heii$\lambda 1640$/\heii$\lambda 4686$ from the $0.3^{\prime \prime}$-PRISM spectrum is $1.87\pm 0.05$, significantly smaller than the intrinsic value.
The ratio is even smaller in the $\sim 1^{\prime \prime}$ PRISM spectrum.
Even if we consider the uncertainty due to blending with O\,{\sc iii}]$\lambda \lambda 1661,1666$, the measured flux ratio of (\heii$\lambda 1640$+O\,{\sc iii}]$\lambda \lambda 1661,1666$)/\heii$\lambda 4686$ is $2.49\pm 0.06$.
Finally, the ratio becomes $2.8\pm 0.4$ if we take the \heii$\lambda 1640$ flux from PRISM and the \heii$\lambda 4686$ flux from G395H while ignoring the potential systematic uncertainty, still incompatible with the intrinsic ratio.
The observed \heii\ ratio reveals the dusty nature of this AGN, which is, however, not apparent from the Balmer decrement.

To estimate the magnitude of dust attenuation, we adopted an intrinsic value of \heii$\lambda 1640$/\heii$\lambda 4686 \approx 7$, typical for ionized gas with an electron temperature of $T_e \approx 1.5\times 10^4$ K and an electron density of $n_e < 10^6~{\rm cm^{-3}}$.
We assumed the shape of the attenuation curve to follow the shape of the average SMC extinction curve derived by \cite{gordon2003} with $R_V = 3.1$, which can approximately describe the nebular attenuation in some galaxies at $z\sim 1-3$ \citep{reddy2023}.
While we cannot separate the broad and narrow components in \heii$\lambda 1640$, the intrinsic \heii\ ratios are similar in both the NLR and the BLR.
Therefore, we further assumed the same column of dust is in the foreground of both the BLR and the NLR, and used the integrated flux for each \heii\ line.
According to \cite{ubler2023a}, in the R2700 spectrum, the flux of the broad \heii$\lambda 4686$ is $\sim 2.6$ times that of the narrow \heii$\lambda 4686$.
If the actual attenuation is different for the NLR and the BLR, our last assumption likely makes the resulting attenuation a lower limit for the BLR.
With these assumptions, we obtained $A_V = 0.49\pm 0.01$, or $A_V = 0.39\pm 0.01$ if we used a more conservative estimate by treating the whole \heii$\lambda 1640$+O\,{\sc iii}]$\lambda \lambda 1661,1666$ blend as \heii$\lambda 1640$.
The corresponding attenuated flux ratio of H$\alpha$/H$\beta$ would be $\sim 3.4$ assuming an intrinsic Case B ratio of 2.86. This value is slightly lower than the ratio found in the broad Balmer lines, which is $3.87\pm 0.22$ according to \cite{ubler2023a}.
Given the potential collisional excitation of the Balmer lines in the BLR, which would enhance the Balmer decrement, we considered the observed Balmer ratio consistent with both attenuation values derived from the helium lines\footnote{The Balmer decrement would be $\sim 3.6$ if we assume instead an intrinsic H$\alpha$/H$\beta$ of 3.0 possible for AGNs \citep{netzer1982}.}.

\redtxt{One might wonder whether the SMC extinction curve we adopted is appropriate for dereddening the emission lines of GS\_3073. In Appendix~\ref{appendix:dust}, we compare the measured ratio of line fluxes to the predictions by different extinction curves. In summary, if we took the total emission line fluxes measured from PRISM, which in principle correspond to the attenuation averaged between the BLR and the NLR, the \heii\ ratio and the Balmer decrement combined favor the SMC extinction. However, if one only uses Balmer line ratios, a combination of H$\alpha$/H$\beta$ and H$\delta$/H$\beta$ would favor an average MW extinction curve given by \citet{fitzpatrick1999} with $R_V = 3.1$. In comparison, H$\gamma$ appears more attenuated, possibly affected by blending with \oiii$\lambda 4363$. Separating the attenuation of the BLR and the NLR is uncertain and leads to more possibilities. Regardless, in all cases, the resulting corrections for \nii/\oii\ have relatively small impacts on the derived N/O, which is the abundance ratio affected most by the attenuation correction.}
We adopted an average value of $A_V = 0.44\pm 0.05$ \redtxt{and the SMC extinction curve} in the following calculations unless otherwise specified.
\redtxt{We also discuss the case where narrow lines are actually not attenuated.}

\subsection{Electron temperature and density}

As shown by \cite{ubler2023a}, the density sensitive lines \sii$\lambda \lambda 6716,6731$ and \ariv$\lambda \lambda 4711,4740$ are resolved in the G395H spectrum of JWST/NIRSpec.
Meanwhile, \oiii$\lambda 4363$ is partially covered in the G395H spectrum (as it lies on the edge of the spectral range), whose flux is constrained by tying its kinematics with other narrow emission lines.
Using \oiii$\lambda 4363$, \oiii$\lambda 5007$, and \sii$\lambda \lambda 6716,6731$, \cite{ubler2023a} derived an electron temperature of \redtxt{$T_{e} = 1.42^{+0.13}_{-0.15}\times 10^4$ K} and an electron density of \redtxt{$n_{e} \sim 2\times 10^3~{\rm cm^{-3}}$}. In comparison, the electron density inferred from the \ariv\ doublet \redtxt{(with \hei$\lambda 4713$ fitted and subtracted in G395H)} is \redtxt{$n_{e} \sim 3\times 10^3~{\rm cm^{-3}}$}.
We rederived the temperature using \textsc{PyNeb} and obtained $T_e = 1.45\pm 0.13 \times 10^4$ K \redtxt{at the low-density limit for the $\rm O^{2+}$ zone, which translates to a temperature of $T_e = 9.4\pm 0.1 \times 10^3$ K for the low ionization zone following the calibrations of \citet{dors2020} based on photoionization modeling of AGN NLRs.}
\redtxt{The slight difference in the temperature is due to different choices of the atomic database as \citet{ubler2023a} used the collisional strengths from \citet{oiii_coll_p12}.}
\redtxt{For consistency, we also rederived the densities.
For \sii, the probability distribution of the density has a median of $2.9\times 10^3~{\rm cm^{-3}}$, a $\rm 16^{th}$-percentile value of $7\times 10^2~{\rm cm^{-3}}$, and an $\rm 84^{th}$-percentile limit of $> 10^4~{\rm cm^{-3}}$.
For \ariv, the probability distribution of the density has a median of $3.6\times 10^3~{\rm cm^{-3}}$, a $\rm 16^{th}$-percentile limit of $< 10^2~{\rm cm^{-3}}$, and an $\rm 84^{th}$-percentile value of $1.3\times 10^4~{\rm cm^{-3}}$.
Besides the slightly higher S/N of \ariv, its higher density sensitivity at high densities compared to \sii\ might have allowed it to better constrain the high density end. The temperature uncertainty induced by the $1\sigma$ uncertainty in the density traced by \ariv\ is negligible compared to the uncertainty caused by the measurement uncertainty of \oiii$\lambda 4363$.}

With the JWST/NIRSpec PRISM spectra, one can estimate the electron temperature and density probed by the rest-frame UV lines.
Specifically, O\,{\sc iii}]$\lambda \lambda 1661,1666$ can be combined with \oiii$\lambda 5007$ to estimate the temperature and \niv$\lambda \lambda 1483,1486$ can be used to estimate the density.
However, the caveat on this approach is that O\,{\sc iii}]$\lambda \lambda 1661,1666$ is blended with \heii$\lambda 1640$ in the PRISM spectra and is the weaker line in the blend.
In addition, the ratio between O\,{\sc iii}]$\lambda \lambda 1661,1666$ and \oiii$\lambda 5007$ is sensitive to dust attenuation.
Aware of these caveats, we estimated the electron temperature from the $0.3^{\prime \prime}$-PRISM spectrum.
Without correcting for dust attenuation, O\,{\sc iii}]$\lambda \lambda 1661,1666$/\oiii$\lambda 5007$ gives a temperature $T_e = 1.33^{+0.05}_{-0.11}\times 10^4$ K at the low density limit.
In contrast, applying the dust attenuation drives the UV-inferred temperature to $T_e \sim 1.76-1.93\times 10^4$ K depending on the adopted $A_V$.
The discrepancy between the temperatures inferred from O\,{\sc iii}]$\lambda \lambda 1661,1666$/\oiii$\lambda 5007$ and [O\,{\sc iii}]$\lambda 4363$/\oiii$\lambda 5007$ can be resolved by either having an NLR less dusty than the BLR, an overestimation of the O\,{\sc iii}]$\lambda \lambda 1661,1666$ flux due to blending, or a considerable contribution from a broad component in O\,{\sc iii}]$\lambda \lambda 1661,1666$.
It is also possible that O\,{\sc iii}]$\lambda \lambda 1661,1666$ probes a different temperature/density zone in the NLR compared to \oiii$\lambda \lambda 4959,5007$ due to its higher critical density and energy level \citep{peimbert1967}.
Finally, we note another possibility is the underestimation of the \oiii$\lambda 4363$ flux as part of the line profile is not covered by the G395H spectrum \citep{ubler2023a}.

With the UV coverage, we can also estimate the electron density using the \niv$\lambda \lambda 1483, 1486$ doublet, whose narrow component is resolved in the VLT/VIMOS spectrum.
Assuming $T_e \approx 1.5\times 10^4$ K, the density derived from [\niv$\lambda 1483$/\niv$\lambda 1486$ is $n_e = 5.0\pm 1.0 \times 10^5~{\rm cm^{-3}}$, consistent with the limit obtained by \citet{raiter2010}.
Increasing $T_e$ would increase the estimate for $n_e$ as well. For example, the electron density becomes $n_e = 6.1\pm 1.3 \times 10^5~{\rm cm^{-3}}$ at $T_e \approx 2\times 10^4$ K.
The density probed by \niv\ lines is two orders of magnitudes higher than that probed by \ariv\ and \sii\ lines, implying the narrow components of the high ionization UV lines likely originate from a more central and denser region in the NLR.
We note that there is an uncertainty associated with the continuum model for the VIMOS spectrum.
By adopting a stellar continuum model, \citet{barchiesi2023} obtained a higher ratio for [\niv$\lambda 1483$/\niv$\lambda 1486$, which corresponds to a density of $n_e = 2.5^{+1.0}_{-0.7} \times 10^5~{\rm cm^{-3}}$ at $T_e \approx 1.5\times 10^4$ K.
Still, in our case the underlying continuum is likely better constrained as it is constructed from the high S/N PRISM spectrum.

\redtxt{One might wonder whether assuming the higher density derived from \niv\ can be an alternative explanation for the different temperatures derived from \oiiis$\lambda \lambda 1661,1666$/\oiii$\lambda 5007$ and \oiii$\lambda 4363$/\oiii$\lambda 5007$. Assuming a uniform density of $n_e = 5\times 10^5~{\rm cm^{-3}}$, the attenuation corrected \oiiis$\lambda \lambda 1661,1666$/\oiii$\lambda 5007$ would correspond to a temperature of $T_e \sim 1.53-1.65\times 10^4$ K, with an additional uncertainty of roughly 1000 K due the attenuation correction.
However, since the critical density of \oiii$\lambda 5007$ ($\sim 10^6~{\rm cm^{-3}}$) is lower than that of \oiii$\lambda 4363$ ($\sim 3\times 10^7~{\rm cm^{-3}}$), at $n_e = 5\times 10^5~{\rm cm^{-3}}$, the derived temperature based on these two lines would be $T_e = 1.05_{-0.08}^{+0.06}\times 10^4$ K, leading to an even larger discrepancy.
Thus, the more natural explanation would still be \oiiis$\lambda \lambda 1661,1666$ partly originating in a denser and/or hotter region, possibly the BLR or the inner region of the NLR.
}

\subsection{Oxygen abundance}
\label{subsec:oxygen}

With the inferred temperature from $\rm O^{2+}$, one can estimate the oxygen abundance using narrow optical lines including [O\,{\sc iii}]$\lambda 5007$, \oii$\lambda \lambda 3726,3729$, and H$\beta$, assuming the temperature for $\rm O^{+}$ scales with that of $\rm O^{2+}$ following the scaling relation in \cite{dors2020}, which is calibrated from photoionization models.
According to \cite{ubler2023a}, the narrow components of \heii\ and He\,{\sc i} lines in the optical indicate the ionization correction factor (ICF) for $\rm O^{2+}$ is $\sim 1.017$, meaning the part of the NLR that optimally emits the optical lines barely has any higher ionization species of oxygen.
As we show in the next subsection, this assumption no longer holds as one dives into the inner part of the NLR where high ionization lines with ionization potentials $\gtrsim 50$ eV arise.

Using only the optical lines, our calculation gives $\rm 12+\log(O/H) \approx 12+\log[(O^{2+}+O^{+})/H^+] = 8.0\pm 0.1$, which is consistent with the value derived by \cite{ubler2023a}.
Despite the uncertainty in the flux of O\,{\sc iii}]$\lambda \lambda 1661,1666$, we also derived an abundance based on the attenuation-corrected flux of the doublet.
Assuming $T_e \approx 1.5\times 10^4$ K and $n_e \sim 10^3 - 10^5~{\rm cm^{-3}}$, we obtained 
$\rm 12+\log(O/H) = 8.19\pm 0.13$\footnote{The uncertainty includes the measurement uncertainty of the flux of O\,{\sc iii}]$\lambda \lambda 1661,1666$ and the uncertainty from the potential range of $A_V$.},
which is marginally larger than the result based on \oiii$\lambda \lambda 4959,5007$ by $1.2\sigma$. 
Even taking into account these small deviations, the results are consistent with a metal-poor NLR in GS\_3073.
We caution that part of the observed O\,{\sc iii}]$\lambda \lambda 1661,1666$ could come from a different zone in the NLR or even the BLR, \redtxt{where} the ICF can significantly deviate from unity.
We further discuss the potential range of ICF($\rm O^{2+}$) for the inner NLR in the next subsection.

\subsection{Nitrogen abundance}
\label{subsec:nitrogen}

In the previous subsections, we have shown that the NLR in GS\_3073 is likely stratified so that the optical lines originate in a less dense zone while the UV lines occupy a denser zone.
In this subsection, we investigate this point in more detail using the nitrogen abundance inferred from different sets of emission lines.

\subsubsection{N/O derived from low-critical density optical lines}

We start with the optical emission lines. Again, we use the measurements from \cite{ubler2023a} as \nii$\lambda \lambda 6548,6583$ is blended with H$\alpha$ in the PRISM spectrum.
The \nii$\lambda 6583$ flux reported in Table~\ref{tab:fluxes} is the sum of the narrow and outflow components, and the narrow component is better constrained as the peak of the narrow line is visible in the G395H spectrum.
In what follows we used the narrow component of \nii$\lambda 6583$ to perform the calculation, which has a flux of $1.07\pm 0.11 \times 10^{-18}~{\rm erg/s/cm^2}$.
In order to calculate N/O, we used the following relation
\begin{equation}
    \rm N/O = \frac{ICF(N^+)}{ICF(O^+)}\frac{N^+}{O^+}.
\end{equation}
We used both attenuation-corrected and uncorrected \nii$\lambda \lambda 6548,6583$/\oii$\lambda \lambda 3726,3729$ to estimate \redtxt{the plausible range of} $\rm N^+/O^+$ assuming an electron temperature of $9.4\pm 0.1\times10^3$ K\footnote{Again, we note the attenuation correction could lead to an \redtxt{underestimation} of $\rm N^+/O^+$ as the attenuation derived from \heii\ could be biased towards the BLR.}. The temperature is derived by scaling the $\rm O^{2+}$ temperature using the relation for NLRs in \cite{dors2020}.
Then, we assumed $\rm \frac{ICF(N^+)}{ICF(O^+)} \approx 0.8$, which is the lower limit quoted in \cite{garnett1990} and \cite{izotov1994} but is more appropriate for NLRs \redtxt{\citep[see e.g., models present in][]{dors2024}}.
\redtxt{Adopting $\rm \frac{ICF(N^+)}{ICF(O^+)} \approx 1$ typically assumed for \hii\ regions would only increase the resulting N/O by 0.1 dex, which does not change our overall conclusion.}
With these assumptions, we derived $\rm \log(N/O) = -1.1 \pm 0.1$, where the uncertainty mainly comes from whether the attenuation correction is applied or not.
As shown in Figure~\ref{fig:no_relation}, the N/O ratio we obtained from optical (forbidden) lines is overall consistent with the N/O versus O/H relation found in the nearby \hii\ regions and SF galaxies (especially GPs that are considered to be local analogues of high-$z$ galaxies; \citealp{kumari2018}).
If \oii$\lambda \lambda 3726,3729$ has a significant outflow component contributing a fractional flux similar to that in the \nii$\lambda 6583$, the resulting N/O would lie above the local relation as indicated by the \redtxt{inverted grey triangle} symbol.

We note that there is also a tentative detection of the auroral line \nii$\lambda 5755$ in the PRISM. 
This line is marginally detected in the G395H spectrum.
If we ignore the potential systematic difference between the PRISM flux and G395H flux and calculate the temperature for $\rm N^+$ using \nii$\lambda 5755$/\nii$\lambda 6583$, the result would be $T_e \gtrsim 3.0\times 10^4$ K at the low density limit, \redtxt{which is inconsistent with the temperature for the low-ionization zone derived above.}
\redtxt{If one assumes instead the regions traced by \niv, \nii, and all the other species have a common density of $n_e \approx 5\times 10^5~{\rm cm^{-3}}$, the $\rm N^+$ temperature is lowered to $T_e = 1.02_{-0.08}^{+0.07}\times 10^4$ K, as \nii$\lambda 6583$ would have been collisionally deexcited significantly.
However, this uniform density model would also lower the $\rm O^{2+}$ temperature significantly and lead to a corresponding $N^{+}$ temperature of $8.6\times 10^3$ K following the scaling relation of \citet{dors2020}, making the whole scenario self-inconsistent.
}
\redtxt{
A potential contaminant for \nii$\lambda 5755$ is [\feii]$\lambda 5747$.
Based on photoionization calculations with \textsc{Cloudy} \citep[c17.03,][]{ferland2017}, we found the flux of [\feii]$\lambda 5747$ would be $\lesssim 10\%$ of that of \nii$\lambda 5755$ for Fe/O values up to solar. Whereas the flux of [\feii]$\lambda 5747$ needs to be roughly 10 times that of \nii$\lambda 5755$ to make the temperatures agree.}
Again, we caution that the above derivation would be affected by the flux difference between PRISM and G395H, the flux measurement of \nii$\lambda 5755$ can be affected by the broad component of the adjacent \hei\ line \redtxt{which is not easy to model} (see Figure~\ref{fig:prism_fe}), \redtxt{the continuum model (which has been assumed to be a smooth power law), and \nii$\lambda 5755$ might still have an unresolved component coming from the high-density region where \nii$\lambda 6583$ has significant collisional deexcitation. Given these sources of systematic uncertainties, we did not use the \nii$\lambda 5755$-based temperature derived above.}

\redtxt{There is another low-ionization line, N\,{\sc ii}]$\lambda 2142$, detected in the UV as listed in Table~\ref{tab:fluxes}. However, since N\,{\sc ii}]$\lambda 2142$ has a critical density of $9.4\times 10^9~{\rm cm^{-3}}$ that is significantly higher than that of \oii, it can have contributions from the BLR and the dense part the NLR. In addition, a large UV attenuation correction would be needed if one uses N\,{\sc ii}]/\oii.
}

\subsubsection{N/O derived from high-critical density UV lines}

Next, we calculated the N/O ratio using UV emission lines. The presence of strong \niii, \niv, and \nv\ lines in the UV indicate a significant fraction of nitrogen is in high ionization states.
The complicating factor of the calculation is the potential contribution from the broad component in the PRISM spectrum and that the \nv\ line is only resolved in the VIMOS spectrum.

To start with, we diagnosed the ionization state of the ionized gas in GS\_3073 using photoionization models.
We used a set of fiducial models for SF regions, NLRs of AGNs, and BLRs of AGNs.
All models are generated by \textsc{Cloudy} \citep[\redtxt{c17.03},][]{ferland2017} and have the same range of metallicity and ionization parameter.
The range of the metallicity is set to $\rm Z/Z_\odot \sim 5\times 10^{-4} - 2$.
The ionization parameter is defined as $U\equiv \Phi _0/n_{\rm H} c$, where $\Phi _0$ is the number flux of photons capable of ionizing hydrogen, $n_{\rm H}$ is the hydrogen density, and $c$ is the speed of light.
The range of the ionization parameter is set to $U \sim 10^{-3.5} - 10^{-1}$.
\redtxt{For the chemical abundance patterns, we started by scaling N/O with O/H following the empirical relation fitted by \citet{groves_model_2004} using a sample of local SF galaxies and \hii\ regions, which can roughly reproduce the relative fluxes of \nii$\lambda 6583$ measured in the Seyfert 2 population from the Sloan Digital Sky Survey IV/Mapping Nearby Galaxies at Apache Point Observatory Survey \citep[SDSS IV/MaNGA,][]{bundy2015,yan2016,ji_bpt_2020}. This relation is given by
\begin{equation}
    \rm N/O = 10^{-1.6} + 10^{2.33+log(O/H)}.
\end{equation}
We then scaled the carbon abundance by the same factor such that $\rm [C/O]=[N/O]$.
The helium abundance was scaled using the relation adopted by \citet{dopita2000}.
We emphasize that the above choices of the scaling relations have little impact on our following diagnostics on the ionization properties.
}

For SF models, we set \redtxt{$n_{\rm H} = 200~{\rm cm^{-3}}$} while adopting an isobaric equation of state (EoS).
We used an spectral energy distribution (SED) from a simple stellar population (SSP) generated by the Binary Population and Spectral Synthesis code \citep[BPASS;][]{stanway2018,byrne2022}, assuming the default parameters and a starburst 1 Myr ago.
We included dust and dust depletion with the default depletion factors given by \textsc{Cloudy}.
\redtxt{For completeness, we also ran SF models with a density of $n_{\rm H} = 5\times 10^5~{\rm cm^{-3}}$, which do not change our conclusions\footnote{The line ratios predicted by the high density SF models are actually more significantly different from the observed ratios.}.}

For NLR models, we set $n_{\rm H} = 5\times 10^5~{\rm cm^{-3}}$.
We note that the diagnostics we show in the following is relatively insensitive to the density.
We also tried $n_{\rm H} = 10^3~{\rm cm^{-3}}$ and our conclusion remains unchanged.
We set the SED as a canonical AGN SED that has an effective big blue bump temperature of $T_{BB} = 10^6$ K, a UV-to-X-ray slope of $-1.4$, a UV slope of $-0.5$, and an X-ray slope of $-1.0$.
\redtxt{The UV spectral shape of GS\_3073 is in rough agreement with the corresponding part of the AGN SED we adopted, although we note that any dust attenuation for the continuum would mean an intrinsically steeper UV slope or a greater stellar light contribution in the UV.
We also checked analytical AGN SED models from \citet{pezzulli2017} with the black hole mass and the range of Eddington ratios similar to those derived by \citet{ubler2023a}. We show in Appendix~\ref{appendix:alternative_models} that our conclusions remain largely unchanged.
}
The rest of the parameters are the same as the SF models.

Finally, we generated the BLR models by setting $n_{\rm H} = 10^{11}~{\rm cm^{-3}}$ and assuming an isopycnic EoS.
We also checked BLR models with $n_{\rm H} = 10^{9}~{\rm cm^{-3}}$ and $n_{\rm H} = 10^{10}~{\rm cm^{-3}}$ and the diagnostic results remain largely the same.
\redtxt{The density range we considered cover the plausible range for BLR clouds showing strong semiforbidden transitions \citep[$10^8~{\rm cm^{-3}}<n_{\rm H}<10^{12}~{\rm cm^{-3}}$,][]{netzer1990}.}
We did not include dust and dust depletion for the BLR models, and we computed two sets of BLR models with hydrogen column densities of $\rm N(H) = 10^{23}~{\rm cm^{-2}}$ and $\rm N(H) = 10^{25}~{\rm cm^{-2}}$, respectively.
\redtxt{The former value of N(H) is typically adopted in photoionization modeling of the BLR emission \citep[e.g.,][]{netzer1990,baldwin2004,temple2021}, the latter is a plausible value for BLRs with Compton-thick clouds, which is a potential explanation for the X-ray weakness of many JWST-identified type-1 AGN including GS\_3073 \citep{maiolino_xrayweak_2024,juodzbalis_absagn_2024}.}
The rest of the parameters are the same as the NLR models.

In the left panel of Figure~\ref{fig:diagrams_n} we plot a hardness diagram composed of line ratios from \niii, \niv, and \nv. The location of any given data point in this diagram is insensitive to the chemical abundances of the gas and it is sensitive primarily to the shape of the ionizing SED and the ionization parameter.
The location in the diagram would also depend, to a lesser extent, on the density of the gas.
One can see AGN models occupy a region very different from the SF models in this diagram due to their overall harder ionizing SED.
We also plot the observed line ratios of GS\_3073.
The fiducial line ratios we adopted (the green star) is a combination of PRISM measurements and VIMOS measurements, where \nv\ and \niv\ are taken from the narrow-line measurements from VIMOS and \niii\ is taken from the PRISM\footnote{The \niii\ fluxes measured from the $0.3^{\prime \prime}$ aperture and the $1^{\prime \prime}$ aperture are very similar as shown by Table~\ref{tab:fluxes}.}.
Overall the observed line ratios are consistent with the fiducial NLR models.
While the \niii\ might have an unresolved broad component, reducing the \niii\ flux would only push the data point towards the upper right location, consistent with a more highly ionized NLR.
We also tried plotting the data point by adopting the \niv\ fluxes measured from the $0.3^{\prime \prime}$-PRISM spectrum and the $1^{\prime \prime}$-PRISM spectrum, the resulting locations (yellow and light green stars) in the diagram are still consistent with the NLR models.
Finally, we tried replacing the \nv\ and \niv\ fluxes with the measurements by \citet{barchiesi2023}, who used a continuum model of a stellar origin to fit the VIMOS spectrum.
It is clear that the resulting location (grey triangle) in the diagram is still more consistent with an AGN origin rather than a stellar origin.

\redtxt{We revisited here several measurement systematics that might impact the above ionization analyses. The first one is matching the intrinsic apertures of VIMOS and PRISM observations. Since the JWST observation has an intrinsically smaller PSF compared to that of the VLT, the aperture matched line ratios should lie between those using the $0.3^{''}$- and $1^{''}$-PRISM extractions. The second one is the contamination from the broad component of \niii. From the left panel of Figure~\ref{fig:diagrams_n}, it is clear that any of such contamination would only bias the data point closer to the SF models. Finally, for the relative flux calibration, since both the Ly$\alpha$ fluxes and the \niv\ fluxes are similar between the (resolution degraded) VIMOS measurements and PRISM measurements (see Section~\ref{subsec:vimos_measurement}), the corresponding systematic bias is unlikely to be significant. Although the broad \nv, similar to the broad \niv, might be lost in the VIMOS spectrum, this does not impact our analyses here.}

In the right panel of Figure~\ref{fig:diagrams_n}, we diagnosed the N/O of GS\_3073 by comparing its UV line ratios with those predicted by photoionization models.
In this \nv/\niv\ versus \niii/\oiiis\ diagram, the $y$ axis traces primarily the hardness of the ionizing SED as well as the ionization parameter, while the $x$ axis traces the abundance ratio of N/O.
The fiducial line ratios we adopted are composed of \nv\ measured from VIMOS, and \niii\ and \oiiis\ measured from PRISM.
One uncertainty therefore comes from the broad-line contribution to \niii\ and \oiiis.
By comparing the BLR and NLR models, one can see at the same N/O, the BLR has a lower \niii/\oiiis.
Therefore, the broad component would only bias the observed \niii/\oiiis\ to a lower value, meaning our measured \niii/\oiiis\ is potentially a lower limit.
Another uncertainty is the flux of \oiiis\ as it is blended with \heii$\lambda 1640$ in PRISM and it is the weaker line of the blend.
Still, we have shown that the fitted flux of \oiiis\ gives a reasonable estimate of the electron temperature.
If we considered a very extreme case where the blend is actually dominated  by \oiiis, which would require \oiiis\ to be blueshifted by $\sim 4000$ km/s and significant dust attenuation, the data would still suggest a super solar N/O as shown by the orange triangle in Figure~\ref{fig:diagrams_n}.

\redtxt{However, we note that the comparison made on the right panel of Figure~\ref{fig:diagrams_n} uses models where N/O is varied accordingly with O/H following \citet{groves_model_2004}.
Since oxygen is a major coolant of the ionized gas and we have constrained the oxygen abundance, we make a more appropriate comparison by fixing the metallicity\footnote{Besides oxygen, the rest of the elements are scaled by the same factor except for nitrogen and carbon. The carbon abundance is set using the value derived later.} to be $\rm 12+log(O/H)=8.0$ while varying N/O in Figure~\ref{fig:diagram_n_no}.
We used a range of N/O from roughly 0.89 dex above solar to 1.29 dex above solar.
By interpolating the model grid linearly in log(N/O) and $\log(U)$ using the \textsc{griddata} function from the \textsc{scipy} package, we found the best-fit NLR model points have $\log(U) = -1.7^{+0.2}_{-0.1}$ and $\rm log(N/O)_{gas} = 0.45^{+0.08}_{-0.13}$.
The best-fit values become $\log(U) = -2.2\pm 0.1$ and $\rm log(N/O)_{gas} = 0.27^{+0.09}_{-0.13}$ if we adopt \citet{barchiesi2023}'s measurements instead.
If no attenuation corrections are applied, the above values become $\log(U) = -2.1\pm 0.1$ and $\rm log(N/O)_{gas} = 0.34^{+0.10}_{-0.15}$ for the higher nebular \nv\ flux, and $\log(U) \approx -2.5$ and $\rm log(N/O)_{gas} = 0.26^{+0.10}_{-0.11}$ for the lower nebular \nv\ flux.
}

\redtxt{From this figure, we can also estimate the potential impact from contamination by SF regions.
Based on our models, low-density SF regions tend to produce significantly lower \nv/\niii\ $\lesssim -3.7$ and slightly higher \niii/\oiiis\ at the same N/O.
Thus, there are two effects. On one hand, the intrinsic N/O would be slightly lower due the contamination in \niii/\oiiis.
On the other hand, the intrinsic \nv/\niii\ would be higher for the NLR, leading to a higher N/O at a fix \niii/\oiiis.
The above effects work in opposite directions and their relative importance depends on the details of the contamination.
As a first order approximation, we added an extra 0.1 dex systematic uncertainty for the model-based N/O in quadrature accounting for these effects.
If we assumed instead the SF contamination is strong in the UV and thus \niv\ has an origin from high-density SF regions with $n_{\rm H} = 5\times 10^5~{\rm cm^{-3}}$, we obtained even lower \nv/\niii\ $\lesssim -4$ and \textit{higher} N/O at fixed \niii/\oiiis, such that our measured \niii/\oiiis\ would correspond to $\rm log(N/O)_{SF;gas} \sim 0.43$ at $\log(U) \sim -1$.
Thus, the above analyses remain largely unchanged.
}

\redtxt{One might also wonder whether the diagnostics involving \nv\ above might be biased by resonant scattering.
While it is difficult to quantify the effect of resonant scattering, its contribution to the observed \nv\ in broad absorption line quasars (BALQSOs) has been examined previously \citep[e.g.,][]{hamann_nv_1996,wang_nv_2010}.
As shown by \citet{hamann_nv_1996}, the theoretical scattered profiles of \nv\ in BALQSOs are relatively broad even compared to the BLR component and the resonant scattering contributes up to $\sim 31\%$ of the flux.
Although narrower profiles are plausible \citep{wang_nv_2010}, they are still considerably broader compared to the BLR component of \target, while what we measured from the VIMOS spectrum is a narrow component.
Thus, in our case, if the nebular \nv\ originates in gas close to AGN, it is not unreasonable to assume little contribution from the resonant scattering to the observed \nv.
We also emphasize that \nv\ is not essential for the abundance determination.
In the right panel of Figure~\ref{fig:diagram_n_no}, we replace \nv/\niii\ with \niv/\niii\ in the $y$ axis. While SF and AGN model grids become closer along the $y$ axis in this diagram, by matching the data point to the AGN model grid, we obtained a similar N/O and $U$ as found in the \nv/\niii\ vs. \niii/\oiiis\ diagram.
}

To crosscheck the N/O inferred from \textsc{Cloudy} models, we used \textsc{PyNeb} to calculate the relative abundances of different ionization species of nitrogen.
We calculated $\rm N^{2+}/O^{2+}$, $\rm N^{3+}/O^{2+}$,  and $\rm N^{4+}/O^{2+}$ from our fiducial flux ratios of \niii/O\,{\sc iii}], \niv/O\,{\sc iii}], and \nv/O\,{\sc iii}] measured in the rest-frame UV, respectively.
We set the density to $n_e = 5\times 10^5~{\rm cm^{-3}}$ as inferred from \niv, and set the temperature to \redtxt{$T_e = 1.45\times 10^4$ K}.
We obtained 
\redtxt{$\rm \log(\frac{N^{4+}+N^{3+}+N^{2+}}{O^{2+}}) = 0.67^{+0.12}_{-0.10}$ with dust attenuation corrections and $\rm \log(\frac{N^{4+}+N^{3+}+N^{2+}}{O^{2+}}) = 0.58^{+0.12}_{-0.09}$ without dust attenuation corrections.}
We note that varying the density in the range of $n_e \sim 10^3 - 10^5 ~{\rm cm^{-3}}$ would vary the result by $\sim 0.01$ dex and varying the temperature in the range of $T_e \sim 10^4 - 2\times 10^4 ~{\rm K}$ would vary the result by $\sim 0.1$ dex.

Now the remaining step is to calculate ICFs to account for the unobserved ions, which depend both on the shape of the ionizing SED as well as the ionization parameter of the ionized gas.
We estimated the ICFs using the photoionization models that match the observed line ratios of GS\_3073 best as shown in Figure~\ref{fig:diagram_n_no}.
\redtxt{As we have discussed, the measured UV line ratios are consistent with a highly ionized NLR, whose ionization parameter is $\log(U) = -2.1\sim -1.7$.}
The ICF for oxygen inferred from the NLR models that match the fiducial line ratios of GS\_3073 is $\rm ICF(O^{2+}) \sim 2$.
Meanwhile, the ICF for nitrogen is $\rm ICF(N^{2+},N^{3+},N^{4+}) \sim 2$, which is similar to $\rm ICF(O^{2+})$.
\redtxt{Applying the ICFs leads to $\rm \log(N/O) \sim 0.57$ with dust attenuation corrections and $\rm \log(N/O) \sim 0.63$ without dust attenuation corrections.}

Last but not least, a major concern with the above \textsc{PyNeb} calculation is the assumption of a single temperature zone, which is not valid even for a single isobaric cloud.
Ideally, the higher ionization species would arise in a zone with a higher electron temperature in the NLR, resulting in a higher emissivity at a fixed abundance.
To correct for the temperature stratification, we again used photoionization models and applied the temperature relations for different ionic species at \redtxt{$\log(U) = -2.1\sim -1.7$}.
The temperatures for different ions predicted by the models lie in a range of \redtxt{1.8-2.1}$\times 10^4$ K.
The temperature-corrected abundance ratio is 
\redtxt{$\rm \log(N/O) = 0.42^{+0.13}_{-0.10}$ when the lines are attenuation corrected,}
about \redtxt{0.15 dex} lower than the estimate based on a uniform temperature.
This result is consistent with the direct comparison with photoionization models in Figure~\ref{fig:diagram_n_no}. 
\redtxt{When no attenuation corrections are applied, the result becomes $\rm \log(N/O) = 0.53^{+0.11}_{-0.10}$.}
In Table~\ref{tab:abundances}, we list N/O measured through different methods as well as the ionic nitrogen abundances. It is clear that the dominant ion of nitrogen is $\rm N^{2+}$.

We note that we did not consider N\,{\sc ii}]$\lambda 2142$ due to the complexity associated with the unknown temperature and density of the low ionization zone as well as the dust attenuation correction.
In general, the inclusion of N\,{\sc ii}]$\lambda 2142$ would make the final N/O even higher if one assumes a lower temperature as appropriate for the low ionization zone. \redtxt{Similar to other semi-forbidden transitions, N\,{\sc ii}]$\lambda 2142$ can have contributions from the BLR due to its high critical density.}

In Figure~\ref{fig:no_relation}, we show the N/O measurements from both the UV lines and optical lines, respectively.
The high N/O derived from UV lines \redtxt{(with attenuation corrections)} clearly place GS\_3073 in the category of the nitrogen-loud AGN.
For comparison, we also plotted several nitrogen-loud targets at high redshifts identified with JWST data \citep{cameron2023,isobe2023,topping2024}.

An extremely interesting result for GS\_3073 is the  different N/O ratios from optical (forbidden) lines \redtxt{with low critical densities} and UV (permitted or semi-forbidden) lines.
\redtxt{One might wonder whether this discrepancy implies any systematic bias in the ionization corrections for optical lines, where only $\rm N^+$ is probed.
However, if the optical \nii\ and \oii\ lines all originate from the same high-density cloud as the UV lines, the ionization correction ratio based on the photoionization models at $\log(U) \sim -2$ becomes $\rm ICF(N^+)/ICF(O^+) \approx 0.26$ for the dense NLR model and $\rm ICF(N^+)/ICF(O^+) \approx 0.35$ for the dense SF model.
Furthermore, since \oii\ has a lower critical density compared to \nii, their emissivity ratio decreases at the high density, leading to a lower $\rm N^+/O^+$ given the observed flux ratio of \nii/\oii.
As a result, the N/O based on the measured flux ratio of \nii/\oii\ is even lower compared to the currently adopted value and the discrepancy remains.
}

\redtxt{An alternative explanation is that the UV line emitting region somehow does not contribute to the low critical density and low ionization lines such as \nii\ and \oii.
This can either be due to the UV line emitting regions are more dust obscured, being density-bounded thereby missing the partially ionized zone, or having a further density stratification such that even \niv\ does not reflect the true average density.
Meanwhile, the UV line emitting region is more nitrogen enhanced compared to the low-density and low-ionization region.
}
While this chemical stratification is not unexpected for AGNs, there has been little direct observational evidence, especially at such high redshifts.
We further discuss the implications of these results in Section~\ref{sec:discussion}.

\subsection{Carbon abundance}

The rich spectral information of GS\_3073 allows the determination of the chemical abundances for elements other than nitrogen and oxygen.
The joint determinations of multiple elemental abundances can better constrain the enrichment mechanisms of the observed region and thus can provide additional clues for the origin of the nitrogen loudness.

In the rest-frame UV spectrum of PRISM, there is detection of \civ\ and \ciii\ lines. Combining these lines with \oiiis, we can roughly estimate the carbon abundance.
\redtxt{We found no detection of C\,{\sc ii}]$\lambda \lambda 2324,2325$ in the UV.}
While there is detection of C\,{\sc ii}$\lambda \lambda 1334,1335$ in the VIMOS spectrum, due to the complex nature of the doublet\footnote{Within the doublet, C\,{\sc ii}$\lambda 1334$ is a resonant line and C\,{\sc ii}$\lambda 1335$ is a fluorescent line.} and the fact that it is unresolved in the spectrum, we did not estimate the carbon abundance using this line.
We note that the observed peak of C\,{\sc ii} is redshifted with respect to the peak of \niv$\lambda \lambda 1483, 1486$ in the VIMOS spectrum, \redtxt{potentially} indicating the presence of an outflow.
Another complicating factor for estimating the total carbon abundance is that the region occupied by $\rm C^{+}$ extends beyond the $\rm H^{+}$ region and goes into the photodissociation region (PDR).
Thus, the fractional contribution from $\rm C^{+}$ has a sensitive dependence on the boundary condition of the ionized cloud.

Figure~\ref{fig:diagram_c} shows a comparison between photoionization models and the observed line ratios of GS\_3073 in the \nv/\niii\ versus \ciii/\oiiis\ diagram.
The fiducial line ratios of GS\_3073 match a range of abundance ratios from $\rm log(C/O)_{gas}\sim -0.32$ to $\rm log(C/O)_{gas}\sim -0.21$ \redtxt{when C/O is scaled with O/H as described in Section~\ref{subsec:nitrogen}}.
\redtxt{When O/H and N/O is fixed and C/O is varied, as shown on the upper right panel of Figure~\ref{fig:diagram_c}, with attenuation corrections, the best-fit NLR parameters are $\log(U) = -1.7^{+0.2}_{-0.1}$ and $\rm log(C/O)_{gas} = -0.15^{+0.13}_{-0.16}$ for the fiducial measurements, and $\log(U) = -2.2\pm 0.1$ and $\rm log(C/O)_{gas} = -0.30^{+0.14}_{-0.15}$ for the measurements based on those of \citet{barchiesi2023}.
The abundances above became $\rm log(C/O)_{gas} = -0.18^{+0.13}_{-0.16}$ (high $\log(U)$) and $\rm log(C/O)_{gas} = -0.26^{+0.15}_{-0.14}$ (low $\log(U)$) if no attenuation corrections were applied.
While the stellar nebular emission contamination as inferred from the SF model predictions would lead to less than 0.1 dex bias in C/O at similar $U$, we included a systematic uncertainty of 0.1 dex in the values above as a conservative estimation.
We note that dielectronic recombination is important for \ciii\ when $T_e > 10^4$ K, which is another source of uncertainty\footnote{In \textsc{Cloudy} simulations, the dielectronic recombination is calculated with rate coefficients from \url{http://amdpp.phys.strath.ac.uk/tamoc/DR/}.}.
Finally, any BLR contribution would lead to underestimation of C/O as \ciii\ is suppressed relative to \oiiis.
Similar to the analyses on N/O, we replaced \nv/\niii\ in the upper right panel of Figure~\ref{fig:diagram_c} with \niv/\niii\ in the bottom panel of Figure~\ref{fig:diagram_c}.
Again, we obtained a similar value of C/O by matching the AGN model grid with the data point.
}

We also estimated C/O based on \textsc{PyNeb}.
To circumvent the $\rm C^{+}$ problem, if we assumed in the ionization zone occupied by $\rm O^{2+}$ that carbon is mainly in $\rm C^{2+}$ and higher ionization species, we obtained 
\redtxt{[($\rm C^{2+}+C^{3+}+C^{4+}+...)/(C^{2+}+C^{3+})$]/[$\rm (O^{2+}+O^{3+}+...)/ O^{2+}$] $\approx$ 0.70-0.72 from the best-fit NLR models, where the range of the value corresponds to $\log(U)=-2.2$\,-\,$-1.7$.
The equation above only corrects for missing ionization species that are more than doubly ionized.
We note that by computing the full ICF into the $\rm C^+$ zone, we obtained a similar value of ICF($\rm C^{2+},C^{3+}$)/ICF($\rm O^{2+}$) $\approx 0.7$, although in this case a significant fraction of C and O comes from singly ionized and neutral species.
This leads to $\rm log(C/O) = -0.38^{+0.13}_{-0.11}$ (high $\log(U)$), and  $\rm log(C/O) = -0.42^{+0.13}_{-0.11}$ (low $\log(U)$). The \textsc{PyNeb} value based on our fiducial measurements with a high $\log(U)$ is lower than the \textsc{Cloudy} value by $1.1\sigma$. The difference might come from the assumption of multiple uniform-temperature zones made in the \textsc{PyNeb} calculation as well as the adopted ICF.
The derived abundances are listed in Table~\ref{tab:abundances}. We note that $\rm C^{2+}$ traced by \ciii\ is significantly more abundant than $\rm C^{3+}$ traced by \civ.
If we only used \ciii/\oiiis\ to get the abundance, we obtained $\rm log(C^{2+}/O^{2+}) = -0.30\pm 0.10$ and ICF($\rm C^{2+}$)/ICF($\rm O^{2+}$) $\approx 1.2$ at $\log(U) = -1.7$. This gives $\rm log(C/O) = -0.24\pm 0.10$, which is more consistent with the model result.
This comparison indicates the observed \civ\ is weaker compared to the model prediction, potentially due to the resonant scattering and self absorption of \civ.
}

Figure~\ref{fig:co_relation} shows the C/O versus O/H relation found in galaxies and stars.
One can see the fiducial carbon abundance of GS\_3073 is only slightly higher compared to other systems with similar metallicities, and the lower limit on the carbon abundance is also compatible with the local relation.
Compared to nitrogen, carbon appears less enriched.
Indeed, normal carbon abundances have been observed in a number of nitrogen-loud AGNs and galaxies.
The AGN candidate at the highest redshift observed by JWST, GN-z11, also shows a lower limit in C/O compatible with the local C/O versus O/H relation \citep{cameron2023}.
In addition, two nitrogen-loud galaxies at $z \sim 6$ with abundances determined by \citet{isobe2023} and \citet{topping2024} also exhibit C/O compatible with local systems, as shown in Figure~\ref{fig:co_relation}.
We further discuss the implication of a normal C/O in a nitrogen-loud system in Section~\ref{sec:discussion}.

\subsection{Iron abundance}
\label{subsec:iron}

\begin{figure}
    \centering\includegraphics[width=0.48\textwidth]{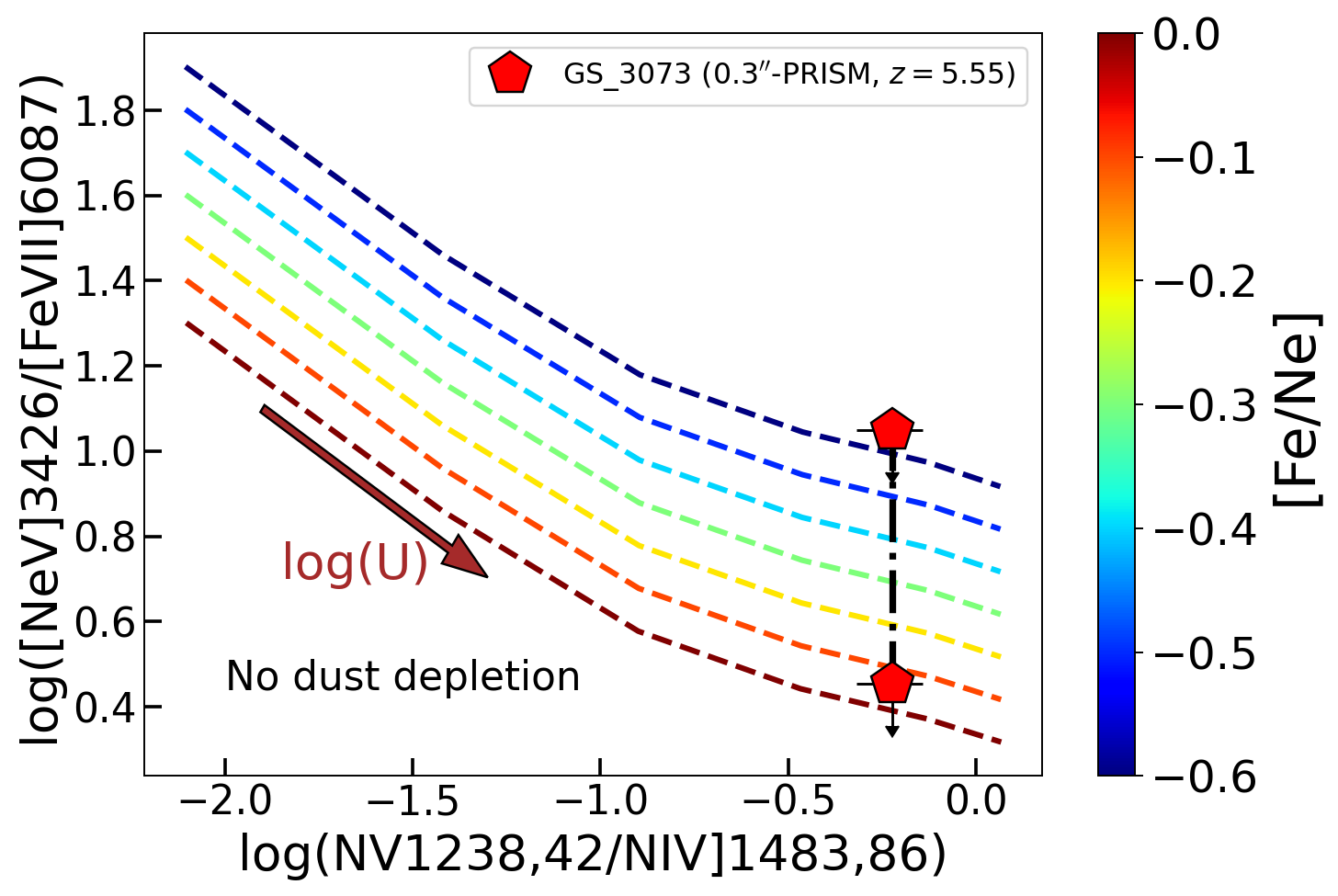}
    
    \caption{Diagnostic diagram composed of \nv/\niv\ and \nev/\fevii.
    The colored lines are photoionization models with parameters similar to the NLR models shown in Figure~\ref{fig:diagrams_n} but having different abundance ratios of F/Ne, \redtxt{where [Fe/Ne] ranges from $-0.6$ dex to 0.}
    All models shown have no dust or dust depletion.
    The red pentagons represent the line ratios measured from the $0.3^{\prime \prime}$-PRISM and VIMOS spectra, with the \nev/\fevii\ being the $3\sigma$ upper limit.
    \redtxt{The higher upper limit was obtained assuming the blended line around 5303 \AA is dominated by \cav, and the lower upper limit was obtained assuming the blended line around 5303 \AA is dominated by \fexiv. The plausible range for the actual upper limit is indicated by ther vertical dashed red line.}
    }
    \label{fig:diagram_fe}
\end{figure}

\begin{figure*}
    \centering\includegraphics[width=0.98\textwidth]{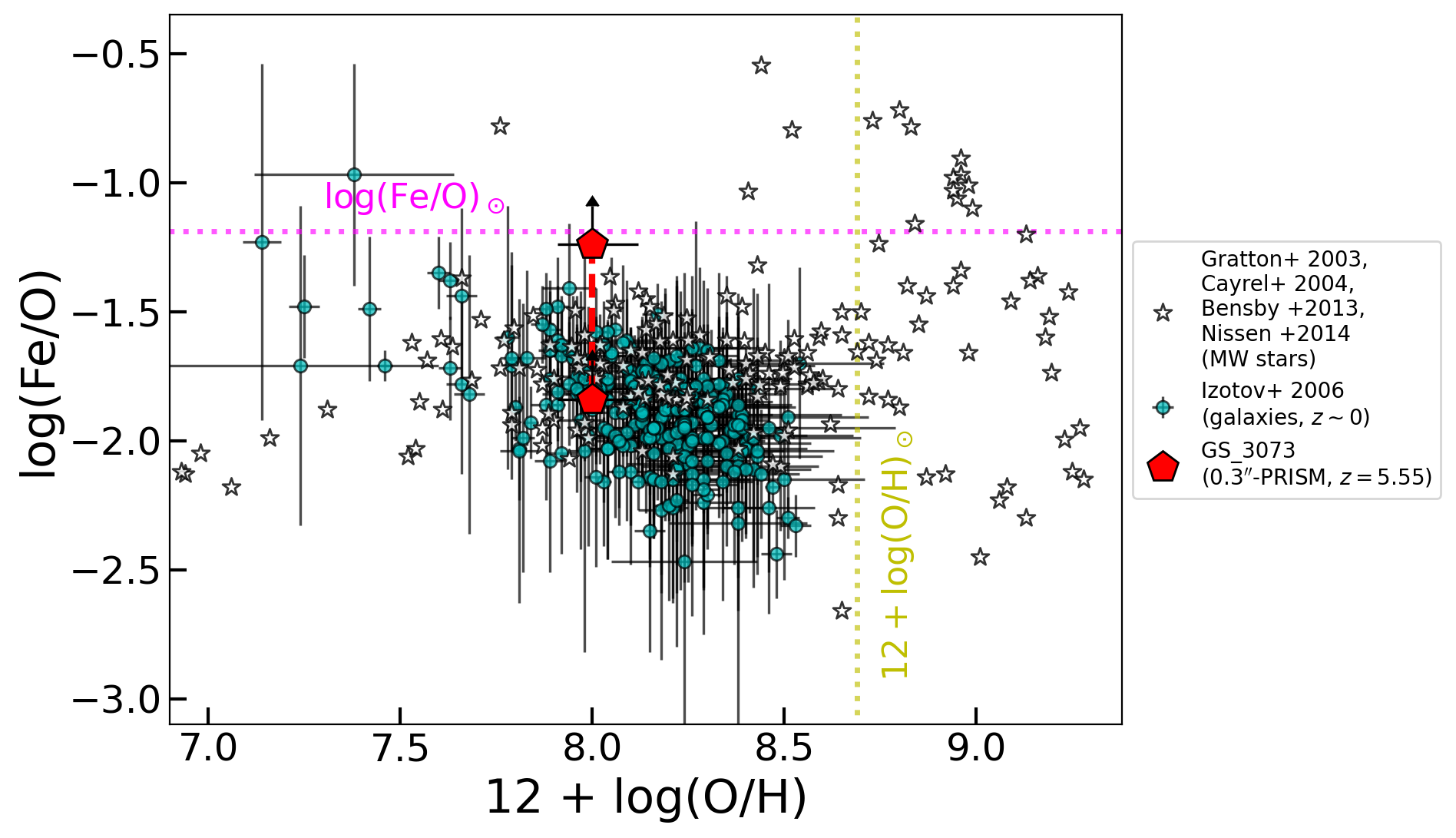}
    
    \caption{Relation between Fe/O and O/H measured from different systems.
    The open stars are MW stars with abundances determined by \citet{gratton2003}, \citet{cayrel2004}, \citet{bensby2013}, and \citet{nissen2014}.
    The cyan circles are local metal-poor galaxies with abundances determined by \citet{izotov2006}.
    The red pentagon represents the abundances we estimated for GS\_3073.
    The dotted lines mark the solar abundances from \citet{grevesse2010}.
    \redtxt{Based on the derived upper limits, the abundance ratio of Fe/O of GS\_3073 can be either similar to or higher than those of the local systems with similar O/H.}
    }
    \label{fig:feo_relation}
\end{figure*}

Iron is another useful element for constraining the enrichment mechanism due to the specific production channels.
Despite being an abundant element, Fe lines are usually difficult to observe due to the significant depletion in dust of Fe in the ISM, which typically leaves only $\sim 1\%$ of Fe in the gas phase \citep{jenkins2009}.
\redtxt{Meanwhile, due to the complicated ICFs for Fe species, the gas-phase Fe abundance is difficult to constrain accurately in observations \citep[e.g.,][]{fe_icf_r05}.
In this subsection, we tried estimating the relative abundance of Fe by using abundance indicators including Fe lines.
}

From Figure~\ref{fig:prism_fit} and Table~\ref{tab:fluxes}, one can see several Fe lines are potentially detected in the PRISM spectrum.
The most significant detection is in \fexiv$\lambda 5303$, which, however, is blended with \cav$\lambda 5309$.
We note that this blend has already been reported as marginally detected in the JWST/NIRSpec G395H spectrum by \citet{ubler2023a}.
\redtxt{The detection of the high ionization line \fexiv$\lambda 5303$ was previously reported in a number of AGN at lower redshift \citep[e.g.,][]{osterbrock_fe14_1981,ward1984,penston1984,alloin_cl_1992,rose2015}. \citet{osterbrock_fe14_1981}, for example, reported detection of the \fexiv$\lambda 5303$+\cav$\lambda 5309$ blend in the Seyfert 1 galaxy III ZW 77. By measuring the centroid of the blended line profile, which is at $5307.4\pm 0.4~\AA$ in the rest frame of III ZW 77, \citet{baldwin1981} concluded that \fexiv$\lambda 5303$ should be present in the spectrum, otherwise the blend should have a centroid around $5309.2~\AA$. We performed a similar check by measuring the line centroid of the blend with \textsc{pPXF} assuming it is a single Gaussian. With 500 times of bootstrap similar to the flux measuring procedure (see Section~\ref{sec:spec_fit}), we obtained a line center at $5306.3^{+3.7}_{-3.6}~\AA$ in the rest frame of other narrow emission lines of GS\_3073. While the line profile is not centering at \cav$\lambda 5309$, the difference is not significant. We also note that the inclusion of a potential nearby line, \fevi$\lambda 5335$, would change the line center. If we include \fevi$\lambda 5335$ and tie its kinematics to those of other narrow lines, the resulting line centroid of the \fexiv$\lambda 5303$+\cav$\lambda 5309$ blend would be at $5300.0^{+3.6}_{-4.3}~\AA$. Regardless, although \fexiv$\lambda 5303$ is potentially present in the spectrum, the S/N and the spectral resolution prevent us from drawing a definite conclusion.}

Due to the high ionization potential of $\rm Fe^{12+}$, \fexiv$\lambda 5303$ is classified as a coronal line that arises from a highly ionized and dense region of the AGN.
Another coronal line frequently observed in AGNs is \fevii$\lambda 6087$.
Due to the similar ionization potentials of $\rm Fe^{5+}$ \redtxt{($98.985\pm 0.015$ eV)} and $\rm Ne^{3+}$ \redtxt{($97.1900\pm 0.0025$ eV)}, \redtxt{$\rm Fe^{6+}$ ($124.9671\pm 0.0025$ eV) and $\rm Ne^{4+}$ ($126.247\pm 0.012$ eV) \citep[from the National Institute of Standards and Technology Atomic Spectra Database,][]{NIST_ASD}}, and the similar critical densities of \fevii$\lambda 6087$ \redtxt{($4.46\times 10^7~{\rm cm^{-3}}$)} and \nev$\lambda 3426$ \redtxt{($1.90\times 10^7~{\rm cm^{-3}}$)}, their ratio is a useful diagnostic of the Fe abundance in AGNs.
In GS\_3073, however, we did not detect \nev$\lambda 3426$\footnote{The non-detection of \nev\ seems common in the current JWST-identified low-luminosity AGN sample, no matter being type-1 or type-2 \citep[e.g.,][]{maiolino2023b,scholtz2023}. While such a lack of \nev\ might not be unexpected based on the analyses of \citet{cleri_n53_2023}, the implication of this result requires more quantitative analyses and is beyond the scope of this work.}. Thus, we calculated a $3\sigma$ upper limit on the flux ratio of \nev$\lambda 3426$/\fevii$\lambda 6087$ and used this to estimate the abundance ratio of Ne/Fe.
\redtxt{A complicating factor here is the contamination from \cav$\lambda 6087$, whose strength is roughly fixed to be 1/5 of \cav$\lambda 5303$ as they have the same upper level. In AGNs with coronal line detections, \cav$\lambda 6087$ is usually much weaker compared to \fevii$\lambda 6087$ \citep[e.g.,][]{rose2015}. This can be understood as Ca is more significantly depleted onto dust grains compared to Fe, as inferred from the observations of the ISM in the Galaxy \citep[e.g.,][]{crinklaw_ca_1994}.
Regardless, we consider below two extreme cases where the 5303 \AA blend is dominated by \fexiv$\lambda 5303$ and \cav$\lambda 5309$, respectively.
In the latter case, \fevii$\lambda 6087$ would have a flux of $1.6\pm 0.9\times 10^{-19}~{\rm erg/s/cm^2}$ for the $0.3^{''}$-PRISM extraction, thus no longer significantly detected. We use the following equation to estimate the $3\sigma$ upper limit of the flux ratio \nev$\lambda 3426$/\fevii$\lambda 6087$
\begin{equation}
    F_{\rm [NeV]}/F_{\rm [Fe VII]} < 3\sqrt{\left(\frac{\sigma _{\rm [NeV]}}{F_{\rm [Fe VII]}}\right)^2+(\frac{\sigma _{\rm [FeVII]}\sigma_{\rm [NeV]}}{F_{\rm [Fe VII]}^2})^2},
\end{equation}
where we have replaced the mean of $F_{\rm [NeV]}$ using the $1\sigma$ formal uncertainty as a conservative estimation, and the fluxes are corrected for dust attenuation as described in Section~\ref{subsec:dust_att}.
}

As shown in Figure~\ref{fig:diagram_fe}, by comparing NLR models with different abundance ratios of Ne/Fe \redtxt{(with O/H, N/O, and C/O fixed to the values derived in Section~\ref{sec:diagnostics})} to the observed line ratios, we can \redtxt{roughly estimate} the Ne/Fe of GS\_3073.
Specifically, based on the less noisy $0.3^{\prime \prime}\times 0.3^{\prime \prime}$ extraction of the PRISM cube, the \redtxt{two upper limits} on the attenuation-corrected flux ratio of \nev$\lambda 3426$/\fevii$\lambda 6087$ implies 
\redtxt{$\rm [Ne/Fe] \lesssim 0.65$ and $\rm [Ne/Fe] \lesssim 0.05$, respectively,}
assuming no dust depletion.
Unlike Fe, Ne is in general not depleted onto dust grains in the ISM.
\redtxt{Thus, if there is dust depletion, the intrinsic abundance ratio of Ne/Fe is even lower.}
While there is clear evidence for the presence of dust based on the observed \heii$\lambda 1640$/\heii$\lambda 4686$, it is possible that the region hosting coronal lines, such as \fevii$\lambda 6087$, has the dust destroyed due to the high temperature and intense ionizing radiation.
\redtxt{We note that the reason we chose NLR models rather than BLR models is in part due to the fact that the critical densities of \nev\ and \fevii\ ($\sim 10^7~{\rm cm^{-3}}$) are lower than the typical densities of broad-line emitting clouds ($\gtrsim 10^8~{\rm cm^{-3}}$; \citealp{netzer1990}).
It is suggested that the coronal-line emitting regions are likely more extended than the BLR but more compact compared to the bulk of the NLR \citep[e.g.,][]{nagao2003}, although there is also observational evidence that they can be more extended \citep{negus_mangacl_2021}.
To check the density, we further computed BLR models with a density of $n_{\rm H} = 10^8~{\rm cm^{-3}}$ and a column density of $N_{\rm H} = 10^{23}~{\rm cm^{-2}}$, and dense NLR models with a density of $n_{\rm H} = 10^7~{\rm cm^{-3}}$.
We found our estimations are largely unchanged even with these high density models.
}

We can convert the constraint on Ne/Fe to Fe/O by estimating Ne/O from emission lines of \neiii\ and \oiii.
The ionization potential of $\rm Ne^{+}$ is close to that of $\rm O^{+}$.
In the optical, the narrow emission lines suggest oxygen is mostly in $\rm O^{2+}$ \citep{ubler2023a}.
Therefore, to a first order approximation, we have $\rm Ne/O \approx Ne^{2+}/O^{2+}$.
This estimation would then only apply to the ionization zone occupied by optical narrow lines.
In the optical, although \neiii$\lambda 3968$ is blended with H$\epsilon$ in PRISM, we can estimate the flux of H$\epsilon$ using the Case B ratio of H$\epsilon$/H$\beta$ modulated by dust attenuation.
From \neiii$\lambda 3968$ and \oiii$\lambda 5007$, we obtained $\rm [Ne/O] \equiv log(Ne/O)-log(Ne/O)_\odot$ to range from $0.22$ to 0 for a range of $n_{\rm H}$ from $10^3~{\rm cm^{-3}}$ to $5\times 10^5~{\rm cm^{-3}}$.
Given that oxygen can be depleted onto dust grains by $\sim 0.22$ dex in the ISM \citep{jenkins2009}, the result is compatible with an intrinsic stellar Ne/O being near-solar or super-solar.
In either case, the resulting lower limit on Fe/O would be near-solar or super-solar.
In Figure~\ref{fig:feo_relation} we compare the Fe/O of GS\_3073 with those in local metal-poor galaxies and MW stars \citep{gratton2003,cayrel2004,izotov2006,bensby2013,nissen2014}.
\redtxt{From this comparison, one can see GS\_3073 has an Fe/O ratio either similar to or higher than those of local stars and galaxies with similar O/H ratios.}

\redtxt{Additional evidence for the plausible value of Fe/O can be found through another significantly detected Fe line, \feiv$\lambda \lambda 2829,2835$ (which are fitted together as a single Gaussian, see Table~\ref{tab:fluxes}). The ionization potentials of $\rm Fe^{2+}$ and $\rm Fe^{3+}$ are $30.651\pm 0.012$ eV and $54.91\pm 0.04$ eV, which are close to those of $\rm O^{+}~(35.12112\pm 0.00006~{\rm eV})$ and $\rm O^{2+}~(54.93554\pm 0.00012~{\rm eV})$ \citep{NIST_ASD}.
Using the attenuation corrected flux ratio of \feiv$\lambda \lambda 2829,2835$/\oiii$\lambda 5007$, we obtained $\rm [O/Fe] \approx 0.5$ (see Appendix~\ref{appendix:b}), consistent with the range derived from \nev$\lambda 3426$/\fevii$\lambda 6087$.
However, one caveat
of using \feiv$\lambda \lambda 2829,2835$/\oiii$\lambda 5007$
is that \feiv$\lambda \lambda 2829,2835$ has a much higher critical density ($\sim 10^8~{\rm cm^{-3}}$) compared to that of \oiii$\lambda 5007$ ($9.39\times 10^5~{\rm cm^{-3}}$), and thus their emitting regions are not necessarily tightly co-spatial.
}

\redtxt{Another} caveat of the above analysis is the coronal-line region might have a different abundance pattern compared to the optical NLR, which we further discuss in the next Section.

\section{Discussion}
\label{sec:discussion}

In this section we discuss the implications of the derived physical conditions for GS\_3073.

\subsection{Enrichment mechanism}

\begin{table*}
	\centering
	\caption{Gas-phase chemical abundances derived from the JWST/NIRSpec and VLT/VIMOS spectra of GS\_3073$^{\rm *}$.}
	{\fontsize{10}{14}\selectfont
	\label{tab:abundances}
	\begin{tabularx}{1.75\columnwidth}{lcccc} 
	    \hline\hline
        Abundance & Value & $n_e$ ($\rm cm^{-3}$) & $T_e$ (K) & $\log(U)$\\
        \hline
        12+log(O/H) & $8.00^{+0.12}_{-0.09}$ & {$ < 10^4$} & $1.45\pm 0.13 \times 10^4$ & -\\
        log(N/O)$_{\rm Optical}^{\rm a}$ & $-1.10^{+0.18}_{-0.20}$ (w/o outflow) & {$< 10^4$} & $9.4\pm 0.1 \times 10^3$ & -\\
        & $-0.58^{+0.18}_{-0.20}$ (w/ outflow) & {$< 10^4$} & $9.4\pm 0.1 \times 10^3$ & -\\
        log(N/O)$_{\rm UV;~Low}^{\rm b}$ & 
        \redtxt{$>-0.21^{\rm c}$} 
        & $5.0\pm 1.0\times 10^5$ & $1.8$\,-\,$2.1\times 10^4$ & $-1.7^{+0.2}_{-0.1}$\\
         log(N/O)$_{\rm UV;~Fiducial}^{\rm b}$ & \redtxt{$0.42^{+0.13}_{-0.10}$} & $5.0\pm 1.0\times 10^5$ & $1.8$\,-\,$2.1\times 10^4$ & $-1.7^{+0.2}_{-0.1}$\\
         \redtxt{$\rm log(N^{2+}/O^{2+})$} & \redtxt{$0.28\pm 0.10$} & $5.0\pm 1.0\times 10^5$ & $1.8\times 10^4$ & -\\
         \redtxt{$\rm log(N^{3+}/O^{2+})$} & \redtxt{$0.02\pm 0.10$} & $5.0\pm 1.0\times 10^5$ & $2.0\times 10^4$ & -\\
         \redtxt{$\rm log(N^{4+}/O^{2+})$} & \redtxt{$-0.46\pm 0.14$} & $5.0\pm 1.0\times 10^5$ & $2.1\times 10^4$ & -\\
        log(C/O)$_{\rm UV;~Low}$ & 
        \redtxt{$>-1.01^{\rm c}$} 
        & $5.0\pm 1.0\times 10^5$ & $1.7$\,-\,$2.0\times 10^4$ & $-1.7^{+0.2}_{-0.1}$\\
         log(C/O)$_{\rm UV;~Fiducial}$ &  \redtxt{$-0.38^{+0.13}_{-0.11}$} & $5.0\pm 1.0\times 10^5$ & $1.7$\,-\,$2.0\times 10^4$ & $-1.7^{+0.2}_{-0.1}$\\
         \redtxt{$\rm log(C^{2+}/O^{2+})$} & \redtxt{$-0.30\pm 0.10$} & $5.0\pm 1.0\times 10^5$ & $1.7\times 10^4$ & -\\
         \redtxt{$\rm log(C^{3+}/O^{2+})$} & \redtxt{$-1.13\pm 0.11$} & $5.0\pm 1.0\times 10^5$ & $2.0\times 10^4$ & -\\
        $\rm log(Fe^{6+}/Ne^{4+})$ & \redtxt{$>-1.08$\,-\,$-0.48$} & $5.0\pm 1.0\times 10^5$ & $2.0\times 10^4$ & $-1.7^{+0.2}_{-0.1}$ \\
        log(Fe/O) & \redtxt{$>-1.84$\,-\,$-1.24$ (\fevii)} & $5.0\pm 1.0\times 10^5$ & - & -\\
         & \redtxt{$-1.74\pm 0.05$ (\feiv)} & $5.0\pm 1.0\times 10^5$ & - & -\\
	    \hline
	\end{tabularx}
        \begin{tablenotes}
        \small
        \item $\bf Notes.$
        \item $^{\rm *}$ Assuming the physical conditions from AGN NLR models. See Section~\ref{sec:diagnostics} for discussions on the impact of different ionized regions.
        \item $^{\rm a}$ Calculated based on \nii/\oii\ in the optical.
        \item $^{\rm b}$ Calculated based on \nv, \niv, \niii, and \oiiis\ in the UV.
        \item $^{\rm c}$ Assuming \oiiis\ has the flux of the whole \heii+\oiiis\ blend (see Section~\ref{sec:diagnostics}).
        \end{tablenotes}
	}
\end{table*}

\subsubsection{Enrichment of nitrogen and carbon}

In Table~\ref{tab:abundances}, we summarize the chemical abundances that we have derived for GS\_3073.
The chemical abundance pattern of GS\_3073 shows a exceptionally enriched nitrogen abundance with respect to the oxygen abundance.
While AGB stars can preferentially raise the abundance ratio of N/O in the ISM \citep[e.g.,][and references therein]{nomoto2013}, their enrichment usually lags behind that of CCSNe, which preferentially boost the relative abundance of oxygen in the ISM.
As a result, the abundance pattern in typical galaxies would follow the trend shown by local \hii\ regions and SF galaxies in Figure~\ref{fig:no_relation}.

As noted by \citet{isobe2023}, the relative abundances of carbon and nitrogen in the nitrogen-loud high-redshift systems confirmed by JWST are similar to those of nitrogen-enhanced metal-poor (NEMP) stars and metal-poor globular cluster (GC) stars.
Thus, \citet{isobe2023} suggest these JWST systems are potential progenitors of GCs. Still, the actual enrichment mechanism is uncertain and people have suggested enrichment by WR stars \citep{kobayashi2024,senchyna2023,watanabe2024}, 
ejecta from AGB stars combined with inflows of pristine gas
\citep{dantona2023},
VMSs with stellar masses of $100~M_\odot < M < 1000~M_\odot$ \citep{vink2023},
SMSs with stellar masses of $M \gtrsim 10^3~M_\odot$ \citep{charbonnel2023,nagele2023}, and TDEs \citep{cameron2023}.
In all the above cases, the nitrogen-enriched (and potentially carbon-enriched) outermost layers of stars are stripped via stellar winds or interactions with black holes to enrich the surrounding ISM.

While the above enrichment mechanisms can all produce a high N/O in the early Universe in a metal-poor environment, the predictions on the abundance ratio of N/C differ among them.
As shown by the chemical enrichment models of \citet{isobe2023} and \citet{watanabe2024}, while SMSs preferentially lead to an abundance pattern with $\rm log(N/O)\gtrsim -0.2$ and $\rm log(C/O)\gtrsim -0.8$, WR stars and TDEs produce a wide range of abundance patterns with $\rm log(N/O)\lesssim 0.6$ and $\rm log(C/O)\lesssim -0.3$.
Purely judging from the abundance ratios of N/O and C/O of GS\_3073, the enrichment 
by SMSs
or WR stars is plausible \citep{limongi2018,nagele2023,watanabe2024}.
For the enrichment purely by AGB ejecta, the equilibrium abundance ratios of N/O and C/O from the CNO cycle are generally higher than the JWST-confirmed nitrogen-loud sources including GS\_3073 ($\rm log(N/O)_{CNO~ equil.} \sim 1$, $\rm log(C/O)_{CNO~ equil.} \sim -0.5$--$-0.3$; \citealp{maeder2015}).
However, if the gas is initially slightly enriched in oxygen and is later mixed with pristine gas, the observed range of N/O and C/O of the JWST-confirmed nitrogen-loud sources can be reproduced by AGB enrichment models \citep{dantona2023}.
Finally, for VMSs, calculations by \citet{vink2023} show they are capable of reproducing the range of N/O inferred for GN-z11 but full chemical evolution models including VMSs remain to be developed.

An important question is the physical scale as well as the time scale of the enrichment required to explain the observations.
Take the highest-redshift AGN candidate, GN-z11, as an example. It is possible that the nitrogen-loudness is limited to the BLR of the AGN, which would then only require a few stars to enrich a physical region of $\sim 10^{-2}$ pc \citep{maiolino2023}.
However, confirmations of broad-line AGNs become increasing difficult at early times due to the decreasing masses of SMBHs \citep{maiolino2023b,maiolino2023}.
Another nitrogen-loud galaxy, CEERS\_01019 at $z=8.679$, is also classified as an AGN based on the presence of high ionization lines \citep{isobe2023}.
In the case of the nitrogen-loud galaxy, RXC J2248-ID, at $z=6.106$, \citet{topping2024} argue the ionization is SF-dominated and thus the enrichment is likely galactic-wide.
While there is a clear broad component underlying H$\alpha$ in the spectrum of RXC J2248-ID, \citet{topping2024} argue for an outflow origin as a broad component is also found in \oiii$\lambda 5007$, although weaker and with a different profile. In addition, the strength of \heii$\lambda 1640$ is weaker than the predictions of AGN photoionization models.
Finally, for the last high-redshift nitrogen-loud target shown in Figure~\ref{fig:diagrams_n}, GLASS\_150008 at $z = 6.229$, \citet{isobe2023} argue it is not an AGN due to the absence of high ionization lines.
{It is not clear whether the enriched regions in these sources are limited to small physical regions illuminated by AGNs or spread out across the entire systems.}
In the case of GS\_3073, the physical scale of nitrogen enhancement likely reaches the high-ionization part of the NLR but not the entire NLR according to our analyses in Section~\ref{sec:diagnostics}.
This points towards an enrichment site no larger than a few hundred parsecs or even a few tens of parsecs (possible for the size of a dense NLR; \citealp{peterson2013})
but not confined within the sub-parsec scale of the BLR.

The time scale of the enrichment, on the other hand, is closely related to the rate of the occurrence of the nitrogen loudness.
The resulting N/O from WR star enrichment, for example, would rapidly drop $\sim 10^{6.5}$ yr after an initial burst of star formation, which can be extended to $\sim 10^{6.9}$ yr if direct collapse of massive WR stars is considered to suppress the enrichment from CCSNe \citep{limongi2018,watanabe2024}.
\redtxt{Still, we note that binary evolution can further extend the WR phase \citep{eldridge2009} and its impact on the chemical enrichment remains to be investigated in chemical evolution models.}
The AGB enrichment, on the other hand, tends to build up gradually $\sim 10^8 - 10^{9}$ yr after an initial burst of star formation, although the first AGB develops as early as 40 Myr after the onset of star formation.
Due to the potential presence of the AGN continuum, the star-formation history (SFH) of GS\_3073 cannot be well constrained.
The only time constraint is the age of the Universe at $z=5.55$, which is $\sim 10^9$ yr after the Big Bang, compatible with all aforementioned scenarios.
Still, 
\redtxt{without binary evolution,}
the plausible timing for the enrichment scenario with WR stars might be strict in order
\redtxt{have the WR wind enrichment}
while still being ahead of the subsequent CCSNe enrichment that lowers the overall N/O.

Since GS\_3073 is a confirmed Type-1 AGN at a high redshift, its chemical abundance pattern might provide new clues also for the puzzling class of nitrogen-loud AGNs.
It remains a matter of debate whether the nitrogen-loud quasars are overall chemically enriched with high oxygen abundances as well, or whether they are particularly enriched only in nitrogen \citep{araki2012,batra2014,jiang2008,matsuoka2011,matsuoka2017}.
Similarly, whether the enrichment is limited to a small region surrounding the AGN, or rather occurs galaxy-wide, remains an open question. \redtxt{\citep{collin1999,baldwin2003,wang2011,araki2012,cantiello2021,huang_agndiscsf_2023}.}
It is noteworthy that AGN activity is not necessary directly related to the nitrogen enrichment in nitrogen-loud galaxies, but it could be the key to revealing nitrogen-loud regions due to the high luminosities of AGNs \citep{baldwin2003}.

According to the enrichment track for nitrogen-loud quasars mainly through AGB stars as adopted by \citet{hamann1993}, for example, a low abundance ratio of $\rm log(N/C) \lesssim -1$ is generally expected at subsolar metallicities.
This does not match our measurements, where 
\redtxt{$\rm log(N/C) \sim$ 0.7-0.8.}
Another AGB-enrichment scenario proposed by \citet{dantona2023} to explain the nitrogen and carbon abundances of GN-z11 is able to reach a high N/O and a low C/O at a metallicity of $\rm 12+\log(O/H) = 8.04$.
In the scenario of \citet{dantona2023}, ejecta from intermediate-mass AGB stars (with yields computed by \citealp{ventura2013}) from a massive GC or a nuclear star cluster (NSC, which is defined as a massive, bright, and compact stellar system usually within the central 50 pc of the host galaxy; \citealp{neumayer2020}) enrich the central region of GN-z11.
Meanwhile, certain dilution of the chemical abundances of the ISM through the inflow of chemically pristine gas might be needed to maintain a low metallicity.
In the case of GS\_3073, the abundance pattern could result from enrichment by AGB stars in an NSC with an age of $\sim 200-300$ Myr \citep{ventura2013,dantona2023}.
It is noteworthy that the physical sizes of NSCs are comparable with that of the enrichment site we speculated for CS\_3073 and the dense environments of NSCs can potentially lead to formations of early SMBHs \citep[][and references therein]{neumayer2020}.

As another possibility, the enrichment by VMSs or SMSs is also compatible with the enrichment sites of nitrogen-loud AGNs, if they are indeed spatially confined in NSCs.
In the case of GS\_3073, our analyses point towards a scenario where nitrogen is clearly enriched at a low oxygen abundance and the enrichment seems limited to a dense and highly ionized zone traced by rest-frame UV lines.
This indicates chemical enrichment in a dense environment where stellar encounters might occur and lead to production of SMSs \citep{ebisuzaki2001,portegieszwart2004,gieles2018}.
We note that the dense star forming regions are not necessarily limited to NSCs but can arise from the self-gravitating part of the accretion disks of AGNs \citep[e.g.,][]{wang2011,cantiello2021}.

As speculated by the early studies of nitrogen-loud quasars, the nitrogen-loud feature might be a result of the co-evolution of AGNs and nuclear star formation \citep[e.g.,][]{hamann1993,baldwin2003}.
According to this scenario, the AGN is initially highly obscured by the nuclear star formation, and the BLR only starts to be visible after the gas supply is nearly exhausted and the star formation declines.
With this setup, the AGN has a higher chance to shine upon the gas recently enriched with nitrogen by NSCs and exhibit the nitrogen-loud feature.
Interestingly, based on the previous observation of [C\,{\sc ii}]$\lambda 158~{\rm \mu m}$ of GS\_3073 in the ALMA Large
Program to INvestigate C+ at Early Times survey \citep[ALPINE;][]{lefevre2020}, \citet{Dessauges-Zavadsky2020} inferred a \redtxt{potentially} low molecular gas mass fraction of $f_{\rm molgas}\equiv M_{\rm molgas}/(M_{\rm molgas}+M_{*}) = 0.10^{+0.13}_{-0.06}$.
This indicates a significant reduction of the cold gas through rapid star formation or AGN feedback.
In the early Universe, with the star formation becomes more bursty and compact, the resulting ionization might play an important role in the nuclear region as well.
This might explain the seemingly more frequent discoveries of nitrogen-loud galaxies at early times by JWST, if in these galaxies, the AGN activities and intense star formation are preferentially observed in a later phase of rapidly evolving NSCs or dense star formation environments.
\redtxt{
Still, we caution that the conversion from the luminosity of [C\,{\sc ii}]$\lambda 158~{\rm \mu m}$ to the mass of the molecular gas has various uncertainties including the dependence on the metallicity \citep{zanella_c2_2018}.
Finally, it is worth noting that according to \citet{schaerer_nloud_2024}, most of the nitrogen enhanced galaxies confirmed by JWST thus far are relatively compact and likely have dense SF environments.}

\subsubsection{Enrichment of iron}

An additional constraint on the enrichment mechanism comes from the abundance of iron.
The typical enrichment channel for Fe is by the ejecta of Type-Ia supernovae, which has a delay time of $\sim 10^{8.5}-10^9$ yr, compatible with the age of the Universe at $z=5.55$, although prompt production channels as short as a few 10 Myr have been suggested \citep{matteucci2006}.
\redtxt{The iron abundance of GS\_3073 is relatively uncertain, as discussed in Section~\ref{subsec:iron}, especially given the potential presence of regions with different densities and chemical enrichment.
While the most plausible range of the gas-phase Fe/O as constrained by high ionization lines including \fevii\ and \feiv\ is consistent with those of local stars and metal-poor galaxies, if Fe and O has been depleted onto dust grains, the intrinsic Fe/O can be significantly higher.
}

\redtxt{If \target\ has an intrinsically low Fe/O, it would be consistent with the scenario where the Type-Ia supernova enrichment has not made a significant contribution to the overall enrichment, similar to the local metal-poor galaxies.
Compared to, for example, the CCSNe enrichment models computed by \citet{watanabe2024} with yields based on \citet{nomoto2013}, $\rm [Fe/O] \approx -0.5$ can be achieved with enrichment purely from CCSNe.
As a result, there would be not enough time for significant nitrogen and carbon enrichment through the normal AGB enrichment unless the time is exactly between the plausible delay times of the AGB enrichment and the Type-Ia supernova enrichment.
}
\redtxt{On the other hand, if \target\ has an intrinsic Fe/O value} of near-solar or even super-solar at a metallicity as low as $\rm 12+log(O/H) \sim 8.0$, one would need an enrichment history that does not over-enrich oxygen.
This again requires the suppression of CCSNe enrichment from previous star formation. 
Alternatively, the overabundance of Fe can be produced by the enrichment of metal-poor hypernovae (HNe).
As shown by \citet{watanabe2024}, enrichment by HNe from massive stars alone at $\sim 20\%$ solar abundances (corresponding to $\rm 12+log(O/H) \sim 8.0$) are able to produced near-solar Fe/O and Ne/O, and a slightly super-solar Fe/Ne.

In the local Universe, super-solar iron abundances up to $\rm Fe/Fe_\odot \sim 5$ were often found in the accretion disks of AGNs based on X-ray observations \citep[e.g.,][]{jiang2019}.
Super-solar abundance ratios of Fe/O have also been found in some local extremely metal-poor galaxies (EMPGs), which, however, have low abundance ratios of N/O at the same time \citep{kojima2021}.
According to \citet{kojima2021} and \citet{isobe2022}, the preferred enrichment mechanism for the high Fe/O in EMPGs is by NHe and/or CCSNe from VMSs with stellar masses of $M_{*} \gtrsim 30-300~M_\odot$, which can potentially form by the infall of pristine gas or mergers of stars in dense star clusters.
If the enrichment of Fe occurred before the enrichment of N, it could also enhance the strength of stellar winds that carry N \citep{vink2023}.

Combining the constraints from abundance ratios of N/O, C/O, and Fe/O, the enrichment can be driven by intermediate-mass AGB stars, VMSs/SMSs, \redtxt{or classical WR stars} in the NSC of GS\_3073 combined with \redtxt{CCSNe (Fe-poor), or} Type-Ia supernovae or HNe \redtxt{(Fe-rich)}.

\subsection{A stratified galaxy}
\label{subsec:stratification}

An interesting characteristic of GS\_3073 is the different gas-phase properties inferred from different sets of emission lines.

First of all, the electron densities inferred from the narrow lines of \sii$\lambda \lambda 6716,6731$, \ariv$\lambda \lambda 4711, 4740$, and \niv$\lambda \lambda 1483,1486$ increase from $\sim 10^3~{\rm cm^{-3}}$ to $\sim 10^5~{\rm cm^{-3}}$.
For Ar and S, $\rm Ar^{2+}$ has a much higher ionization potential compared to $\rm S^{0}$ and \ariv$\lambda \lambda 4711, 4740$ is more sensitive to the density at the high-density regime.
Meanwhile, there is evidence in local \hii\ regions that \sii$\lambda \lambda 6716,6731$ is biased towards low density regions \citep{mendez-delgado2023}.
For N, while the ionization potential of $\rm N^{2+}$ is only slightly higher than $\rm Ar^{2+}$, the higher critical density of \niv$\lambda \lambda 1483,1486$ makes it more likely to arise from a denser region when there is density stratification.
A similar high density from \niv$\lambda \lambda 1483,1486$ has also been reported for the nitrogen-rich galaxy at $z = 6.106$ by \citet{topping2024}.
\redtxt{In the local nitrogen enhanced galaxy Mrk 996, \citet{james2009} found a density stratification where the high density gas reaches a density of $\log(n_e) = 7.25^{+1.25}_{-0.75}~{\rm cm^{-3}}$ and is nitrogen enhanced.
The low density gas has $n_e \approx 10$-170 $\rm cm^{-3}$ is nitrogen-normal compared to local galaxies.
}
\redtxt{In another higher redshift and lensed galaxy, the Sunburst Arc at $z = 2.37$, \citet{pascale2023} found evidence for high-pressure, high-density, and nitrogen enhanced clouds surrounding a young super star cluster.
The high-density clouds have densities of $n_e \sim 10^5~{\rm cm^{-3}}$ and are likely surrounded by more extended low-density and nitrogen-normal clouds with densities of $n_e \sim 10^3$-$10^4~{\rm cm^{-3}}$.
}
Finally, it is noteworthy that we only used the narrow component of \niv$\lambda \lambda 1483,1486$, meaning the density stratification is within the NLR.
The density stratification of NLRs has already been noted in previous observations of AGNs \citep[e.g.,][]{binette1999}.

An even more interesting type of stratification in the NLR of GS\_3073 is the ionization stratification.
The ICF for oxygen as inferred from the optical \heii\ and \hei\ lines implies an ionization state where most oxygen is in $\rm O^{2+}$.
However, the ionization condition for UV lines is clearly different.
The presence of strong \niv\ and \nv\ lines as well as the evidence of O\,{\sc vi} in the Lyman continuum regime requires a non-negligible amount of high ionization species.
Based on the ionic species of UV lines, we inferred there is likely more than 1/4 of oxygen in ionization stages higher than $\rm O^{2+}$ within the UV-line dominated zone.
This result suggests the UV-line dominated zone is not only denser but also more highly ionized.
Besides these UV lines, there is also \redtxt{potential} detection of optical coronal lines that have ionization potentials $\gtrsim 100$ eV, which likely arise from an even more central and highly ionized zone \citep{ferguson1997,murayama1998,rose2015}.

The \redtxt{potential} presence of \fevii$\lambda 6087$ and \fexiv$\lambda 5303$, \redtxt{if confirmed, might} suggest little Fe depletion in the coronal-line region \redtxt{\citep{nagao2003,mckaig_cl_2024}}.
Meanwhile, dust attenuation is clearly present given the observed \heii\ UV-to-optical ratio.
While it remains unclear how much dust depletion would be sufficient to produce the moderate amount of attenuation we observed, these results point to a possibility that the dust distribution is also stratified.
In such a scenario, dust grains are destroyed in the coronal-line region by the intense radiation field, but survive in less ionized and outer zones.
While the stratification of dust distribution has long been considered as part of the standard AGN models, the exact location at which dust starts to emerge is debatable \citep[e.g.,][]{shields2010}.
For GS\_3073, our measurements indicate the destruction of dust beyond the BLR into the NLR, where forbidden lines with high critical densities can still arise.
This is consistent with the result of \citet{nagao2006}, where the line ratios from the high-ionization part of the NLRs in a sample of X-ray selected intermediate-redshift AGNs are more aligned with predictions from dust-free photoionization models.

Finally, perhaps the most intriguing type of stratification we found in GS\_3073 is the chemical abundance stratification, \redtxt{as also found by \citet{james2009} and \citet{pascale2023} in Mrk 996 and the Sunburst Arc, respectively}.
The abundance ratio of N/O inferred from \redtxt{the low-critical density and low-ionization lines including \oii\ and \nii} in the PRISM and G395H spectra appears only slightly higher than the N/O of local galaxies with similar metallicities.
In contrast, the N/O inferred from \redtxt{the high-critical density and high-ionization} UV lines is significantly higher than local galaxies, and also the highest among the nitrogen-loud galaxies confirmed by JWST, as shown in Figure~\ref{fig:no_relation}.
While the potential outflow component in the optical lines and the blending of \heii\ and \oiiis\ could bring the N/O ratios estimated from \redtxt{the two sets of} lines closer, the lower limit on the UV N/O is still significantly higher than the highest N/O we estimated \redtxt{from \nii/\oii}.
This result suggests the nitrogen is more enriched in the UV-line dominated zone, which is a denser and more highly ionized zone compared to the \redtxt{low-density and low-ionization zone where lines such as \oii\ and \nii\ originate in}.
Presumably, the UV-line dominated zone lies closer to the accretion disk of the central SMBH and thus is more directly affected by the abundance pattern potentially resulting from dense nuclear star formation.
This would then suggest the nitrogen-loudness is confined to the BLR and the inner NLR and would reduce the total mass of stars required to build the observed abundances.
Still, with the current observations, we cannot constrain precisely the physical scale of the enriched region.

Also, it is possible that the optical narrow lines are contaminated by SF regions in GS\_3073.
The narrow optical line ratios of GS\_3073, as shown by \citet{ubler2023a}, lies close to the boundary of SF regions and AGN regions in the BPT diagram.
While this behavior can be simply explained by a metal-poor NLR \citep{scholtz2023}, we cannot fully rule out the contribution from metal-poor SF regions.
If the peculiar enrichment of nitrogen in GS\_3073 is confined to the nuclear region as we speculated, then the non-nuclear region in GS\_3073 can simply show ``normal'' abundances.
As a result, the contamination from non-nuclear region for optical lines would lead to an estimation of a less extreme N/O.

In summary, we found evidence for stratification in the abundance pattern of nitrogen in GS\_3073, which appears to correlate with the stratification in the electron density and the ionization state of the gas.
We thereby speculate that the enrichment is confined to the dense and highly ionized central region of GS\_3073.
Our results suggest that the nitrogen abundance is enriched locally but at a scale larger than the BLR as traced by the highly ionized part of the NLR.
This is consistent with the enrichment by NSCs we discussed in the last subsection.

\section{Conclusions}
\label{sec:conclude}

In this work we investigated the chemical abundance pattern of GS\_3073, a galaxy at $z = 5.55$, hosting a Type-1.8 AGN, by using JWST/NIRSpec IFS observations and VLT/VIMOS data. The presence of an AGN illuminating the gas in the nuclear region and the rest of the galaxy, combined with the sensitivity of JWST-NIRSpec and VLT-VIMOS, allows the detection of about 40 emission lines in the rest-frame UV and optical wavelength range, hence enabling an unprecedented investigation in terms of chemical enrichment and physical properties of the gas.

We summarize our findings in the following.
\begin{itemize}
    \item The flux ratios between the (semi-forbidden) rest-frame UV emission lines of GS\_3073 imply an ionization condition consistent with the inner region of the galaxy (inner part of the Narrow Line Region, or a combination of an NLR and a BLR), characterized by high densities  of $n_e\sim 10^5~{\rm cm}^{-3}$ and with an ionization parameter of \redtxt{$\rm \log U \sim -1.5$\,-\,$-2$}.
    This is consistent with the interpretations based on the rest-frame optical permitted 
    lines, where hydrogen Balmer lines and helium lines show both narrow and broad components.
    \item The rest-frame UV emission-line flux ratios of GS\_3073 indicate that, in such dense gas, nitrogen is highly enriched with respect to oxygen (N/O about 20 times higher than solar), carbon is not as enriched as nitrogen (C/O about solar \redtxt{or slightly subsolar}), and Fe/O is \redtxt{subsolar}.
    In particular, the relative abundance of nitrogen
    is clearly above the N/O versus O/H relation observed in low-redshift galaxies.
    The N/O and C/O abundance patterns of GS\_3073 are similar to those of nitrogen-loud quasars at low redshifts as well as several nitrogen-loud galaxies recently revealed by JWST, and GS\_3073 is actually among the most nitrogen-rich of them all. Therefore, with its $\sim40$ lines detected, this system provides an excellent test bench for investigating nitrogen-enriched systems in the early Universe. In particular, the constraints on the iron abundance provide new insights to be compared with chemical evolution models.
    \item The abundance ratio of N/O inferred from the rest-frame optical {\it forbidden} emission lines \redtxt{including \nii\ and \oii} (tracing much lower gas densities) is significantly lower than that inferred from the rest-frame UV (semi-forbidden) emission lines.
    We interpret this difference as an indication of a stratification in the chemical abundances, where the inner, denser, and more highly ionized region of the galaxy is more chemically enriched.
    The chemical stratification is associated with an ionization stratification and a density stratification, where the UV lines generally show higher ionization states and a higher electron density compared to \redtxt{the low-ionization and low-critical density} optical emission lines. Importantly, the extreme nitrogen enrichment is confined to the central, high-density region, not to the bulk of the host galaxy.
    \item While the flux ratio between the UV and optical \heii\ lines is consistent with a moderate amount of dust attenuation with $A_V \sim 0.39-0.49$, \redtxt{the potential} detection of the coronal line \fevii$\lambda 6087$ indicates destruction of dust grains in the inner part of the galaxy, likely due to the strong AGN radiation field.
    \item The chemical abundance pattern of GS\_3073 can be explained by a variety of enrichment mechanisms.
    It is possible to reproduce the observed N/O and C/O with enrichment by AGB ejecta, 
    winds from Very Massive Stars (VMSs), Super Massive Stars (SMSs), \redtxt{or classical WR stars} in a dense Nuclear Star Cluster (NSC).
    To reproduce the observed Fe/Ne and Ne/O abundance ratios, enrichment by a combination of Core Collapse SNe and Type-Ia supernovae, or Hypernovae is \redtxt{plausible}.
    \item Given the observed stratification in the density, ionization state, and chemical abundances, we speculate the extreme enrichment pattern seen in GS\_3073 preferentially occurred in a dense NSC, meaning the enrichment is spatially confined within the central few parsecs.
    Such an environment would also allow production of VMSs and SMSs through frequent stellar encounters.
    \item The potential lack of molecular gas in GS\_3073 suggests it might have reached a later stage of star formation due to the consumption of gas and feedback.
    Therefore, the reason that GS\_3073 shows nitrogen-loudness might be similar to the AGN-SF co-evolution scenario proposed for nitrogen-loud quasars, where the AGN becomes less obscured but still active close to the end of a rapid star formation in a dense nuclear environment.
\end{itemize}

As star formation becomes more bursty and intense in the metal-poor and dense environments in the early Universe, the chance 
of observing galaxies in the nitrogen-loud phase of their chemical evolution might increase.
This potentially explains the discoveries of several nitrogen-loud galaxies by JWST, as well as the fact that not all of them exhibit clear evidence for AGN activity, although AGN illumination of the central, dense region facilitates their discovery.
With the currently observed number of early galaxies increasing thanks to JWST, we might start to be able to study the statistical behavior of nitrogen enrichment at high redshifts.
The statistical analyses on the chemical abundance patterns would be vital for understanding the general enrichment channel at early times.
In addition, as already noted by many current studies, new methods for distinguishing AGN ionization from SF ionization at high redshifts are urgently needed.
This would help us understand how the (peculiar) abundance patterns might be associated with AGN or SF activities and further test previous theories for chemical enrichment of quasars and dense star clusters.
Finally, how the metals are distributed differently in the early galaxies remain poorly constrained in observations.
The chemical stratification in GS\_3073 implies that the early enrichment in galaxies may be highly inhomogeneous.
Future explorations of deep IFS data of high-$z$ galaxies will be useful for understanding the differential enrichment of metals in young galaxies in the early Universe.

\section*{Acknowledgements}
\redtxt{We thank the anonymous referee for thoughtful suggestions that improve the clarity of this manuscript.}
\redtxt{We thank Gary Ferland and Cristina Ramos Almeida for helpful discussions.
We thank Rosa Valiante for kindly providing AGN SED tables.
}
We thank Andrea Grazian for kindly providing the 1D spectrum of GS\_3073 reduced from the previous observation by VLT/VIMOS.
We thank Georges Meynet and Francesca D'Antona for kindly providing insightful comments on a draft of the manuscript.
XJ, RM, and FDE acknowledge ERC Advanced Grant 695671 “QUENCH” and support by the Science and Technology Facilities Council (STFC) and by the UKRI Frontier Research grant RISEandFALL.
RM also acknowledges funding from a research professorship from the Royal Society.
H{\"U} gratefully acknowledges support by the Isaac Newton Trust and by the Kavli Foundation through a Newton-Kavli Junior Fellowship.
SA, MP, and BRdP acknowledge grant PID2021-127718NB-I00 funded by the Spanish Ministry of Science and Innovation/State Agency of Research (MICIN/AEI/10.13039/501100011033).
AJB acknowledges funding from the ``FirstGalaxies'' Advanced Grant from the European Research Council (ERC) under the European Union's Horizon 2020 research and innovation program (Grant agreement No. 789056).
GC acknowledges the support of the INAF Large Grant 2022 ``The metal circle: a new sharp view of the baryon
cycle up to Cosmic Dawn with the latest generation IFU facilities''.
IL acknowledges support from PID2022-140483NB-C22 funded by AEI 10.13039/501100011033 and BDC 20221289 funded by MCIN by the Recovery, Transformation and Resilience Plan from the Spanish State, and by NextGenerationEU from the European Union through the Recovery and Resilience Facility.
\redtxt{CHIANTI is a collaborative project involving George Mason University, the University of Michigan (USA), University of Cambridge (UK) and NASA Goddard Space Flight Center (USA).}

\section*{Data Availability}

The JWST/NIRSpec IFS data used in this paper are available through the MAST portal.
The VLT/VIMOS data used in this paper are available from the ESO Phase 3 Archive.
All analysis results of this paper will be shared on reasonable request to the corresponding author.

%
\bibliographystyle{mnras} 
\bibliography{ref} 
%

\appendix
\section{JWST/NIRCam imaging of GS\_3073}
\label{appendix:a}

\begin{figure*}
    \centering\includegraphics[width=0.32\textwidth]{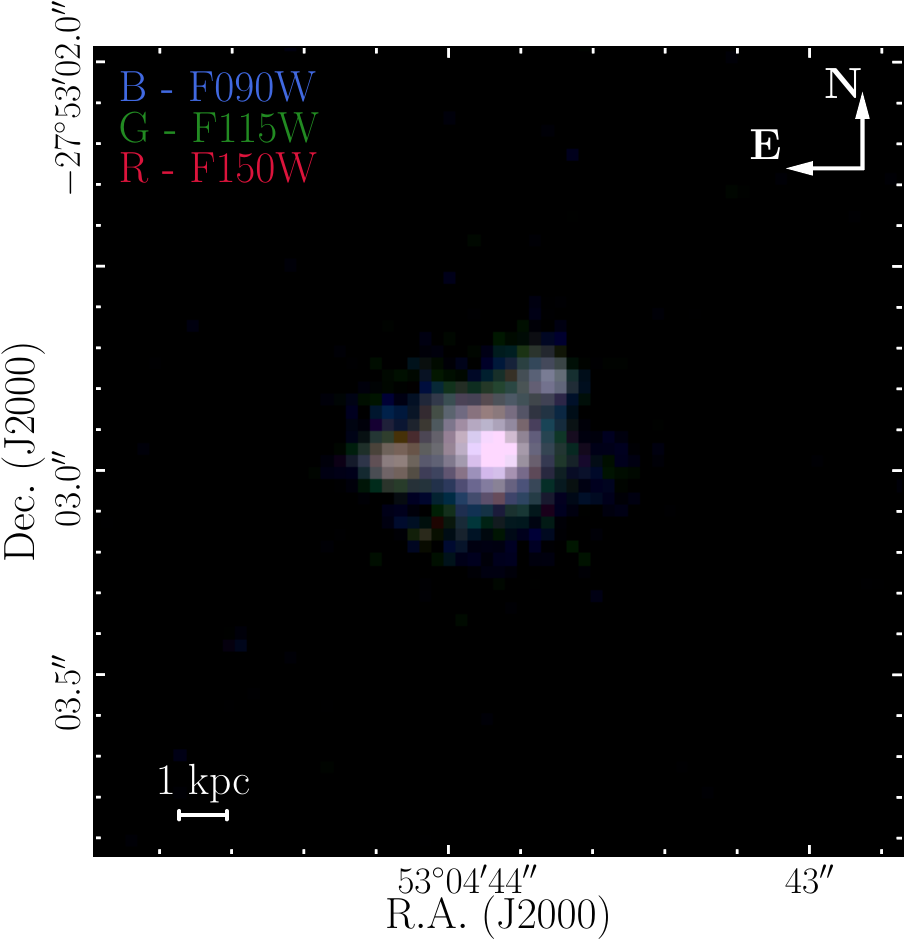}
    \includegraphics[width=0.32\textwidth]{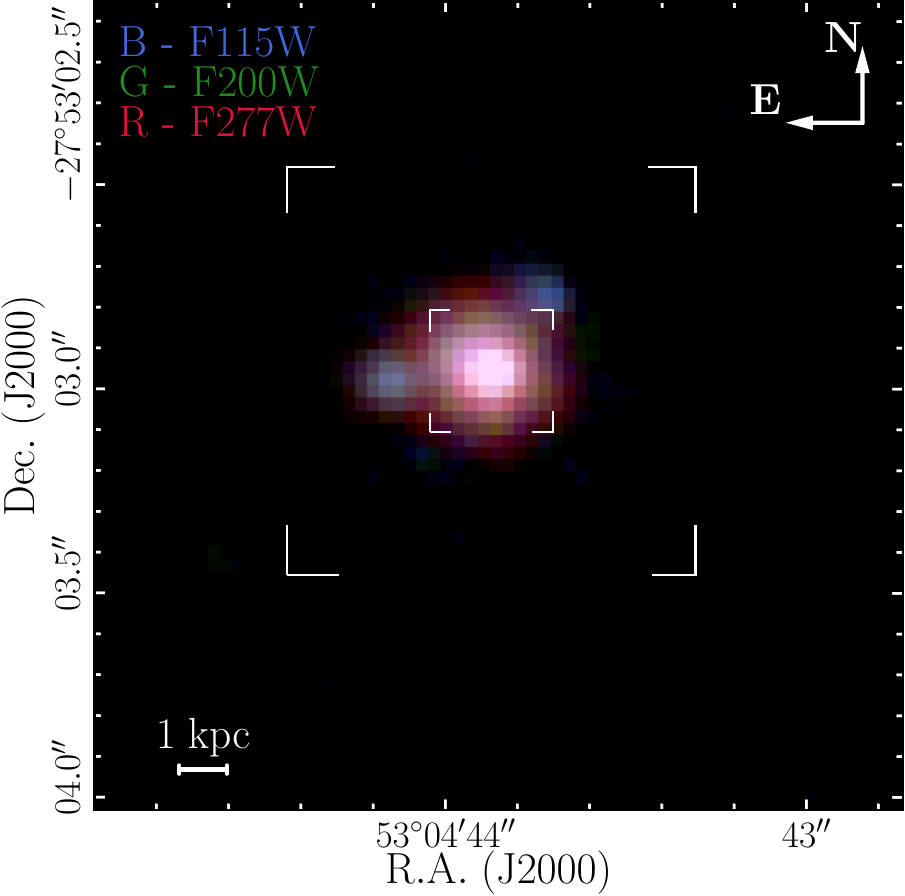}
    \includegraphics[width=0.32\textwidth]{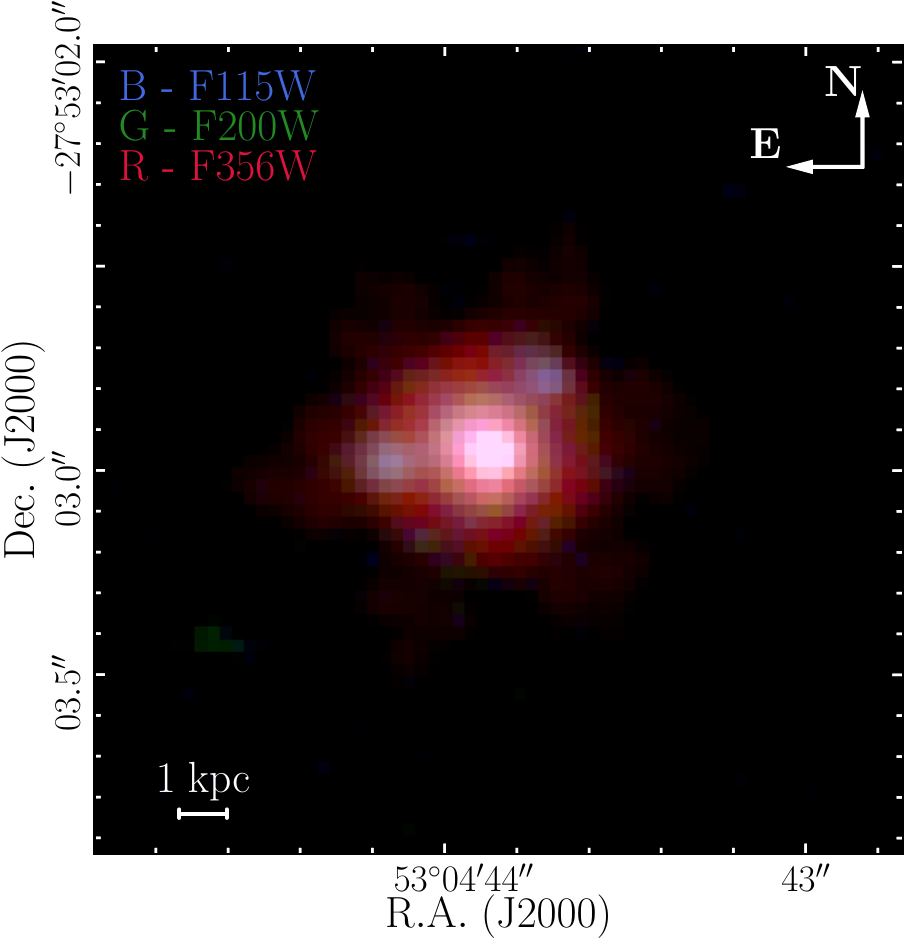}
    
    \caption{Wide-band colour-composite images constructed from JWST/NIRCam observations of GS\_3073 \citep{rieke2023}.
    \redtxt{In the middle panel, we show also the $0.3^{''}$-aperture and the $1^{''}$-aperture we used for extracting NIRSpec/PRISM IFU data.}
    }
    \label{fig:nircam_img}
\end{figure*}

In this Appendix we present JWST/NIRCam images of GS\_3073, obtained as part the JWST Advanced Deep Extragalactic Survey \citep[JADES;][]{eisenstein2023}, and publicly available as part of the JADES data release 1 \citep{rieke2023}.
In addition to the continuum, the different filters capture nebular line emission from \nv\ (F090W), \heii--\ciii\ (F115W) and \oiii\ (F356W); Ly$\alpha$ and \civ\ fall under the low-sensitivity region of F090W and F115W, respectively. F150W and F200W probe mostly the rest-UV continuum, without high equivalent-width line emission.
Figure~\ref{fig:nircam_img} show the color-composite images constructed from the above bands.

The images show two distinct sources in addition to GS\_3073, as reported by \citet{ubler2023a} based on NIRSpec PRISM data. Leveraging the superior spatial resolution of NIRCam at 0.9--1.5~$\mu$m, we can clearly resolve the east and north-west sources as being spatially separated from GS\_3073.
\redtxt{From the middle panel of Figure~\ref{fig:nircam_img}, one can see a $0.3^{''}$-extraction aperture as we adopted for our fiducial measurements would avoid light contamination from the two additional sources.
In comparison, a $1^{''}$-extraction aperture as adopted for generating an alternative PRISM spectrum as well as a $1^{''}$-slit used for extracting the VIMOS spectrum would include the two additional sources.
}
We note that neither source displays an excess in F356W-F277W or F410M-F444W, implying weak line emission in these bands.
Both sources are undetected in HST/ACS F435W and F606W, consistent with $z>5$. Both sources appear significantly bluer in colour than GS\_3073, which could indicate either bright nebular emission in F090W and F115W, or a steep continuum.
\redtxt{To further understand the potential contamination from these sources in the $1^{''}$-spectra, we extract a $0.1^{''}\times0.1^{''}$ box centering at each source in the PRISM IFU data.
We found that the integrated light centering at the source to the east of \target\ contributes an \oiii$\lambda 5007$ flux roughly 10\% of that from the central $0.3^{''}$-extraction, and an \niv$\lambda 1483,1486$ flux less than 5\% of that from the central $0.3^{''}$-extraction.
In comparison, the integrated light centering at the source to the northwest of \target\ contributes an \oiii$\lambda 5007$ flux roughly 1\% of that from the central $0.3^{''}$-extraction, and an \niv$\lambda 1483,1486$ flux up to 1\% of that from the central $0.3^{''}$-extraction.
In conclusion, these additional sources unlikely contaminate the nebular emission of the central source significantly and we emphasize that our main conclusions are based on the central $0.3^{''}$-extraction with little contamination.
}

Assuming both sources to be at the same redshift as GS\_3073, they could be interpreted as low-mass satellites dominated by blue stellar populations.
Intriguingly, similar sources are found near other broad-line AGN at redshifts $z\sim5$ \citep{maiolino2023b,matthee2024}, suggesting a possible connection between strong AGN activity and minor mergers.

\section{Nebular attenuation curve}
\label{appendix:dust}

\begin{figure}
    \centering\includegraphics[width=\linewidth]{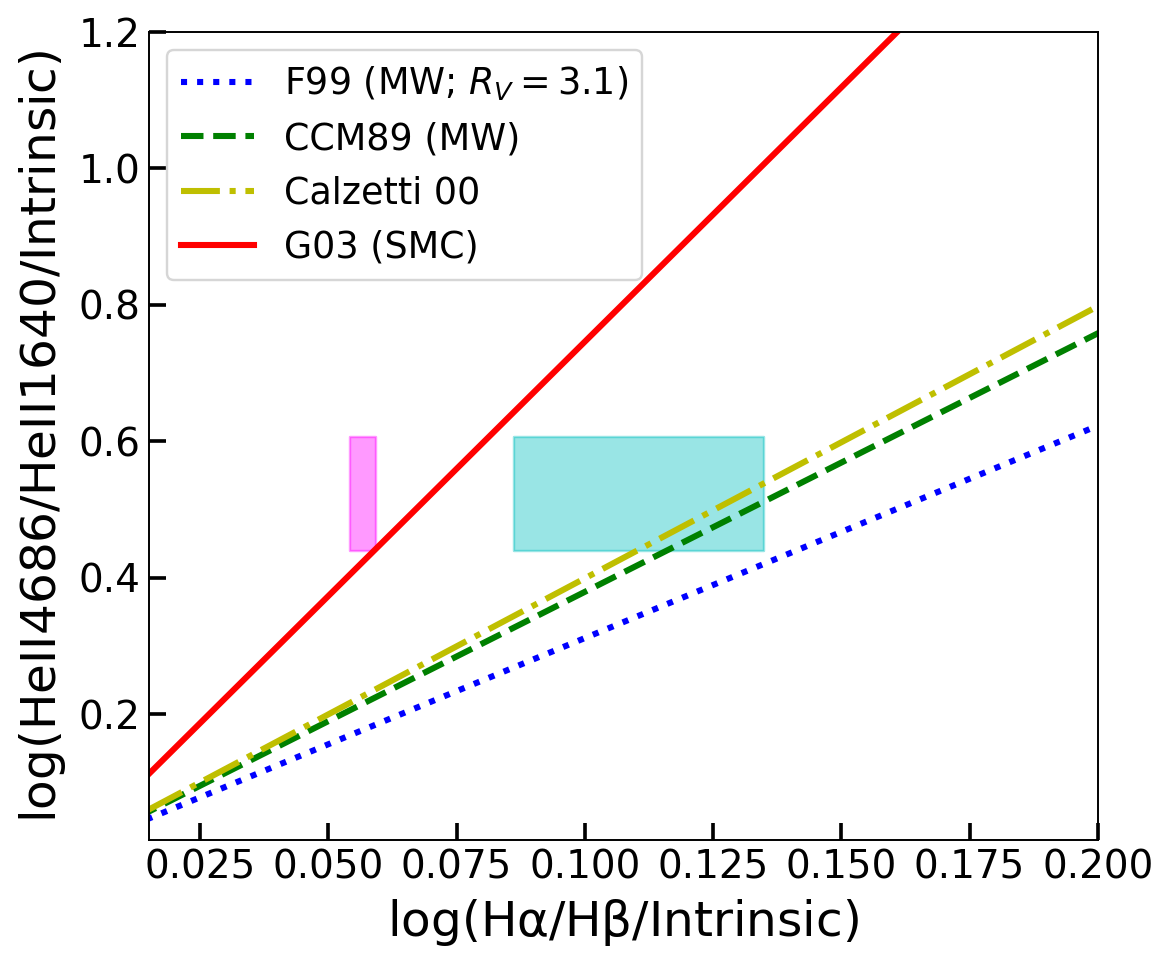}
    
    \caption{\redtxt{Diagnostic diagram constraining the shape of the UV-to-optical attenuation curve. The lines displayed correspond to the predicted \heii\ flux ratio versus the Balmer decrement for extinction curves from \citet{cardelli1989}, \citet{fitzpatrick1999}, \citet{calzetti2000}, and \citet{gordon2003}. The magenta shaded region corresponds to the plausible range of line ratios measured from the $0.3^{''}$-PRISM spectrum. For the cyan shaded region, we replace the Balmer decrement with that of the broad component measured in the R2700 spectrum by \citet{ubler2023a}.}
    }
    \label{fig:uv_o_curve}
\end{figure}

\begin{figure*}
    \centering\includegraphics[width=\columnwidth]{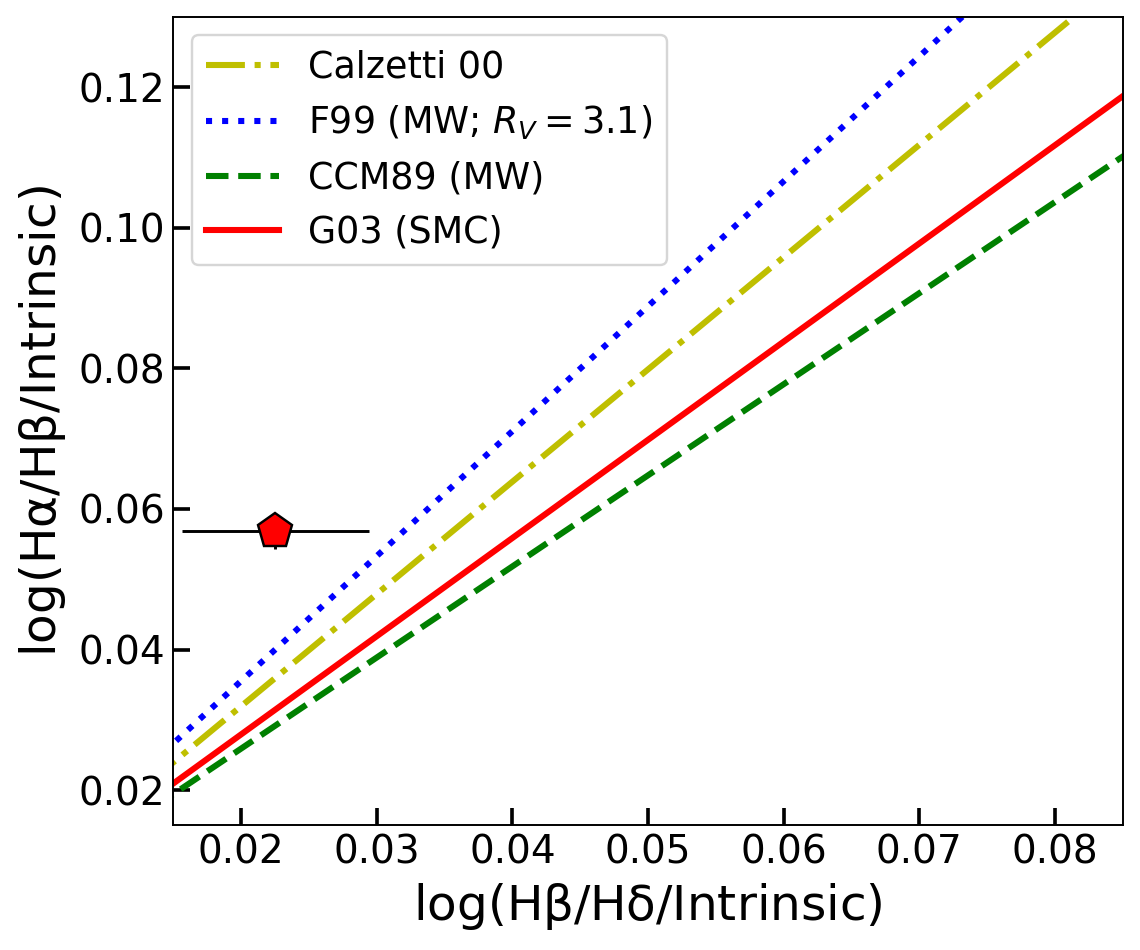}
    \centering\includegraphics[width=\columnwidth]{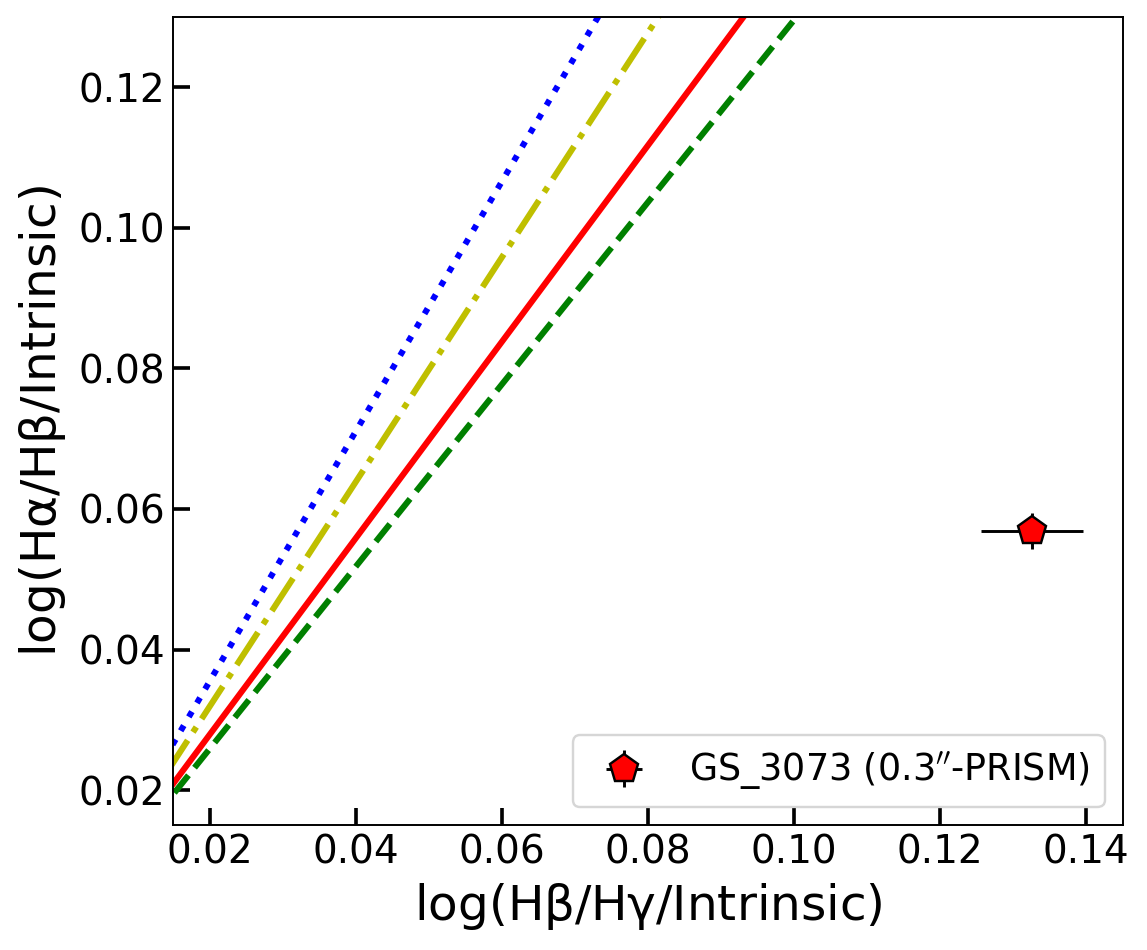}
    
    \caption{\redtxt{Diagnostic diagrams constraining the shape of optical attenuation curve. The extinction curves are the same as the ones shown in Figure~\ref{fig:uv_o_curve}. The red pentagons represent the values we measured from the $0.3^{''}$-PRISM spectrum.}
    }
    \label{fig:o_curve}
\end{figure*}

\redtxt{In Section~\ref{subsec:dust_att}, we have discussed the dust attenuation probed by \heii\ line ratios and the hydrogen Balmer lines, where the attenuation curve is assumed to follow the shape of an average SMC extinction curve \citep{gordon2003}.
In this Appendix we investigate the shape of the attenuation curve in greater detail.
}

\redtxt{Figures~\ref{fig:uv_o_curve} and \ref{fig:o_curve} show comparisons between the predicted line ratios from some of the extinction curves in the literature and those measured from the NIRSpec spectra of GS\_3073.
All line ratios are normalized by the Case B values from \citet{draine11_book}.
The total line flux ratios measured from the PRISM spectrum prefers steep attenuation in the UV closer to the predictions by the SMC curve. If we replace the Balmer derement with the value measured from the broad lines in the R2700 NIRSpec/grating spectrum, the observed line ratios move closer to the predictions by the curve from either \citet{cardelli1989} or \citet{calzetti2000}. The caveat, however, is that we cannot compare it with the flux ratio from the broad \heii\ lines since they are not resolved in PRISM. Also, it is likely that the BLR is more obscured and thus the shape of the attenuation curve for the broad lines is also close to that of the SMC extinction curve.
For the optical part, the left panel of Figure~\ref{fig:o_curve} suggests it is close to the MW extinction curve from \citet{fitzpatrick1999}, although being slightly steeper redwards to H$\beta$. However, from Figure~\ref{fig:o_curve}, it appears H$\gamma$ is more attenuated compared to H$\delta$. We caution that the attenuation from H$\gamma$ is potentially affected by blending with \oiii$\lambda 4363$ in the PRISM spectrum\footnote{We note that in the main text we did not use \oiii$\lambda 4363$ measured in PRISM during our analyses but took the value measured in G395H.} (e.g., a $\sim 20$\% underestimation for H$\gamma$ would make H$\beta$/H$\gamma$ roughly consistent with the attenuation curves shown in Figure~\ref{fig:o_curve}).
}

\redtxt{Although in the main text we have assumed the SMC extinction curve when dereddening for example \nii/\oii, assuming other extinction curves only change the correction by 0.02 dex at most.
In most cases, the SMC correction in the optical regime serves as an upper limit.
When using \nii/\oii, we also considered the case where there is no extinction for these narrow lines, thereby including various possibilities.
Again, we would like to point out the spatial distribution of dust in GS\_3073 is likely complex, with the nuclear region being more obscured. This is supported by the significantly different Balmer decrements from broad emission lines and narrow emission lines in the R2700 NIRSpec/grating spectrum by \citet{ubler2023a}.}
\redtxt{We will explore more systematically shapes of the attenuation curves in JWST identified type-1 AGN in future work.}

\section{Comparisons with additional photoionization models}
\label{appendix:alternative_models}

\begin{figure*}
    \centering\includegraphics[width=\linewidth]{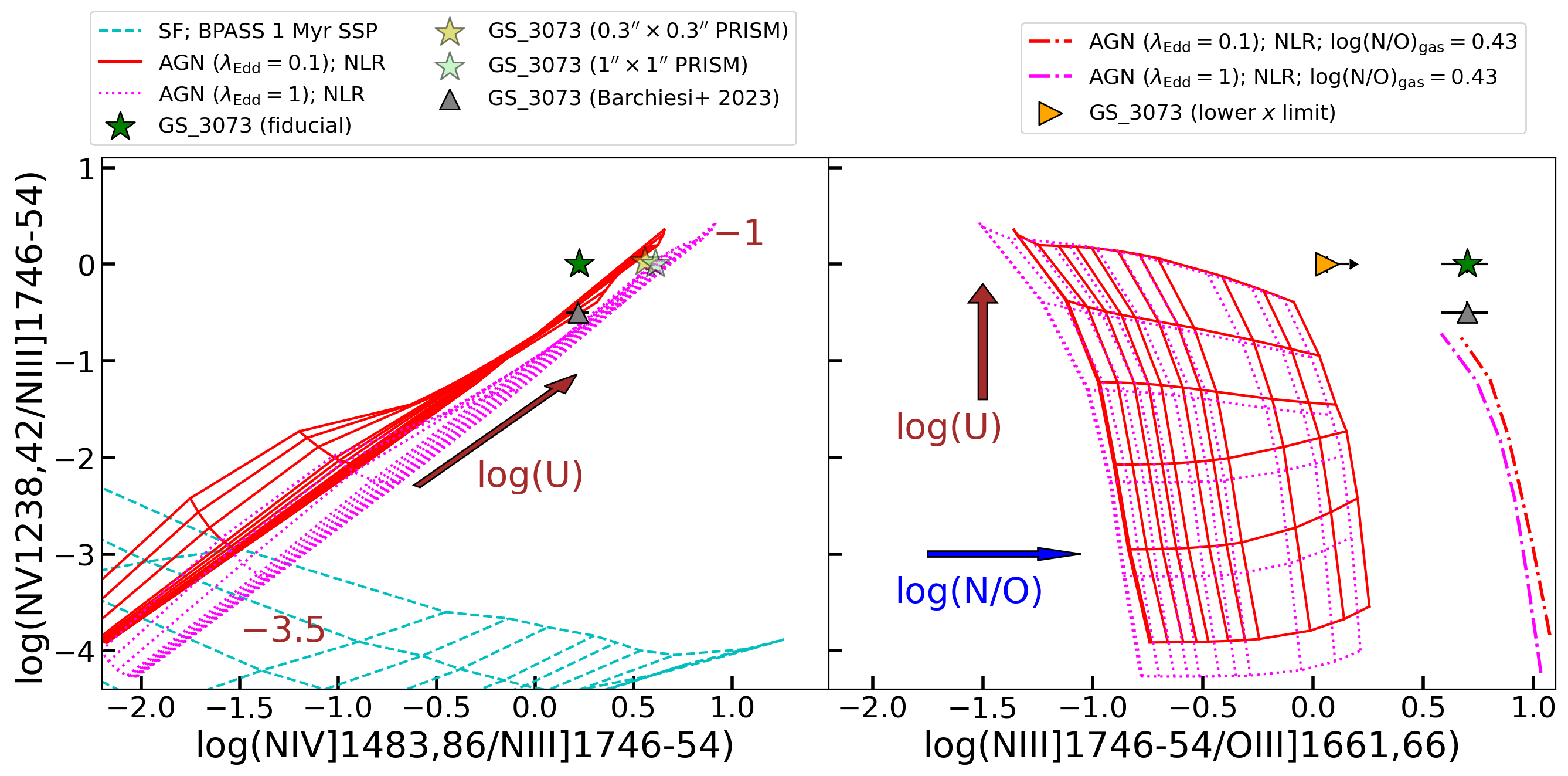}
    
    \caption{Comparisons between observed line ratios of \target\ and those predicted by two additional AGN NLR models.
    These new models assuming a functional form of the SED parametrized by the black hole mass and the Eddington ratio as adopted by
    \citet{pezzulli2017}.
    }
    \label{fig:other_seds}
\end{figure*}

\redtxt{In Section~\ref{sec:diagnostics}, we have compared the measured line ratios of \target\ with a fiducial set of photoionization models.
Here we show further comparisons with models under different assumptions.}

\redtxt{In Figure~\ref{fig:other_seds}, we show two additional NLR models. Different from our fiducial NLR model, these new models have their incident SEDs generated by the analytical AGN SED described in \citet{pezzulli2017}, which is a function of the black hole mass and the Eddington ratio.
We considered two SEDs with a black hole mass of $M_{\rm BH} = 10^8~M_{\odot}$ and Eddington ratios of $\lambda _{\rm Edd} = 0.1$ and 1, matching the range of parameters estimated by \citet{ubler2023a} for \target.
We set the rest of the model parameters the same as those in the fiducial model.
}

\redtxt{Overall these additional models do not change our conclusions. The measured nebular line ratios are still more consistent with ionization by the AGN continuum emission, although the best-fit ionization parameter becomes $\log U \approx -1.5\sim -1$. From the right panel of Figure~\ref{fig:other_seds}, one can see the gas-phase N/O indicated by the NLR models is still roughly 0.43 dex.
}

\section{Gas-phase Fe/O based on [OIII]/[FeIV]}
\label{appendix:b}

\begin{figure}
    \centering\includegraphics[width=\linewidth]{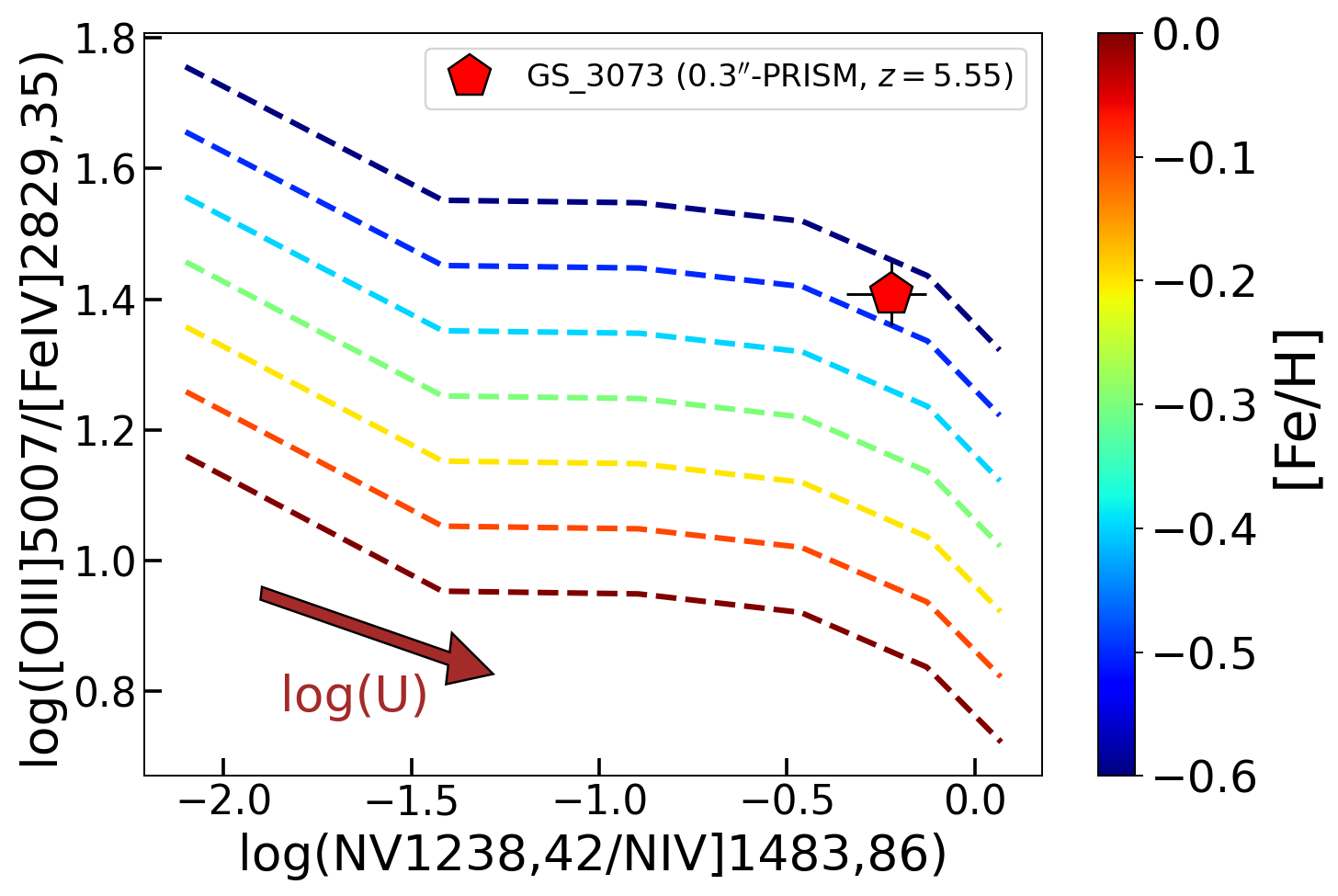}
    
    \caption{\redtxt{Diagnostic diagram composed of \nv/\niv\ and \oiii/\feiv.
    The colored lines are photoionization models with parameters similar to the NLR models shown in Figure~\ref{fig:diagram_fe}, having different Fe abundances in a range of $-0.6 \leq {\rm [Fe/H]} \leq 0$, where ${\rm [Fe/H] \equiv log(Fe/H)-log(Fe/H)_\odot}$.}
    }
    \label{fig:diagram_fe4o3}
\end{figure}

\redtxt{In the main text we used the upper limit on the flux ratio of \nev/\fevii\ to constrain the gas-phase abundance of Fe. In this Appendix we investigate the Fe abundance constrained by the flux ratio of \oiii$\lambda 5007$/\feiv$\lambda \lambda 2829,2835$.
In Figure~\ref{fig:diagram_fe4o3}, we compare the measured \oiii$\lambda 5007$/\feiv$\lambda \lambda 2829,2835$ versus \nv$\lambda \lambda 1238,1242$/\niv$\lambda \lambda 1483,1486$ from the $0.3^{''}$-PRISM spectrum of GS\_3073 to those predicted by photoionization models for NLRs of AGN.
We have corrected the flux ratio of \oiii$\lambda 5007$/\feiv$\lambda \lambda 2829,2835$ for dust attenuation assuming $A_V = 0.44$ and using the SMC extinction curve (see Section~\ref{subsec:dust_att}).
The photoionization models have the same parameters as those in Figure~\ref{fig:diagram_fe}. Specifically, the hydrogen density is assumed to be $n_{\rm H} = 5\times 10^5~{\rm cm^{-3}}$, the oxygen abundance is set to 20\% solar (although the nitrogen and carbon abundances are not relevant here, they are set to follow the constraints in Figure~\ref{fig:diagrams_n} and Figure~\ref{fig:diagram_c}), neither dust nor dust depletion is included, the ionization parameter is varied in a range of $-3.5\leq \log(U) \leq -1$, and the Fe abundance is varied in a range of $-0.6 \leq {\rm [Fe/H]} \leq 0$.}

\redtxt{From Figure~\ref{fig:diagram_fe4o3}, the best matched Fe abundance is roughly 0.5-0.6 dex below solar, consistent with the lower limits constrained by \nev/\fevii, which ranges from 0.6 dex below solar to solar.
If Fe is depleted onto dust grains in the \feiv\ emitting region, the intrinsic Fe abundance should be higher.
Regardless, as mentioned in the main text, one needs to be cautious about this simple model-based derivation given the very different critical densities of \feiv\ and \oiii.
}

\bsp	
\label{lastpage}
\end{document}